\documentclass{jfm}
\usepackage{graphicx}
\usepackage{amsmath}
\usepackage{booktabs}
\usepackage{mathtools}
\usepackage{subcaption}
\usepackage{multirow}
\usepackage{empheq}
\usepackage{natbib}
\usepackage{enumitem}

\ifCUPmtlplainloaded \else
  \checkfont{eurm10}
  \iffontfound
    \IfFileExists{upmath.sty}
      {\typeout{^^JFound AMS Euler Roman fonts on the system,
                   using the 'upmath' package.^^J}%
       \usepackage{upmath}}
      {\typeout{^^JFound AMS Euler Roman fonts on the system, but you
                   dont seem to have the}%
       \typeout{'upmath' package installed. JFM.cls can take advantage
                 of these fonts,^^Jif you use 'upmath' package.^^J}%
      }
  \else
  \fi
\fi

\ifCUPmtlplainloaded \else
  \checkfont{msam10}
  \iffontfound
    \IfFileExists{amssymb.sty}
      {\typeout{^^JFound AMS Symbol fonts on the system, using the
                'amssymb' package.^^J}%
       \usepackage{amssymb}%
         \let\leq=\leqslant
         \let\geq=\geqslant
      }{}
  \fi
\fi

\ifCUPmtlplainloaded \else
  \IfFileExists{amsbsy.sty}
    {\typeout{^^JFound the 'amsbsy' package on the system, using it.^^J}%
     \usepackage{amsbsy}}
    {\providecommand\boldsymbol[1]{\mbox{\boldmath $##1$}}}
\fi

\newcommand\Rey{\mbox{\textit{Re}}}  % Reynolds number
\newcommand\karman{K\'{a}rm\'{a}n~}

\newcommand{\dd}{\, \mathrm{d}}  % roman d for integral

\begin{document}

\title[Fish swimming efficiency]{Optimal undulating swimming for a
  single fish-like body and for a pair of interacting swimmers}
%\author{Audrey P. Maertens, Michael S. Triantafyllou}

\author[Audrey P. Maertens, Amy Gao, and Michael S. Triantafyllou]%
       {Audrey P. Maertens \thanks{Current affiliation: EPFL, LMH, Avenue de Cour 33 bis, 1007 Lausanne, Switzerland. Email address for correspondence:
           audrey.maertens@epfl.ch}, Amy Gao, and Michael S. Triantafyllou}

% NOTE: A full address must be provided: department, university/institution, town/city, zipcode/postcode, country.
\affiliation{Center for Ocean Engineering, Massachusetts Institute of 
Technology,\\77 Massachusetts Avenue, Cambridge 02139, USA
}

%\ead{maertens@alum.mit.edu}
%\date{\today}
%\ead{mistetri@mit.edu}
\maketitle

%\address[mit]{Massachusetts Institute of Technology, 77 %Massachusetts Avenue, Cambridge 02139, USA}

\begin{abstract}

We establish through numerical simulation conditions for optimal undulatory propulsion for a single fish, and for a pair of hydrodynamically interacting fish, accounting for linear and angular recoil.  We first employ systematic 2D simulations to identify conditions for minimal propulsive power of a self-propelled fish, and continue with targeted 3D simulations for a danio-like fish.  We find that the Strouhal number, phase angle between heave and pitch at the trailing edge, and angle of attack are principal parameters.  Angular recoil has significant impact on efficiency, while optimized body bending requires maximum bending amplitude upstream of the trailing edge. For 2D simulations, imposing a deformation based on measured displacement for carangiform swimming provides efficiency of 40\%, which increases for an optimized profile to 57\%; for a 3D fish, the corresponding increase is from 22\% to 35\%; all at Reynolds number 5000.

Next, we turn to 2D simulation of two hydrodynamically interacting fish.  We find that the upstream fish benefits energetically only for small distances.  In contrast, the downstream fish can benefit at any position that allows interaction with the upstream wake, provided its body motion is timed appropriately with respect to the oncoming vortices.  For an in-line configuration, one body length apart, the optimal efficiency of the downstream fish can increase to 66\%; for an offset arrangement it can reach 81\%.  This proves that in groups of fish, energy savings can be achieved for downstream fish through interaction with oncoming vortices, even when the downstream fish lies directly inside the jet-like flow of an upstream fish.

\end{abstract}

% keywords are entered during the submission process and should not be included in the manuscript
%\begin{keywords}
%biological fluid dynamics, propulsion, swimming, vortex interactions
%\end{keywords}

%\begin{keywords}
%lateral line \sep object identification \sep viscous effects 
%\sep boundary layer \sep Orr-Sommerfeld \sep convective instability
%\end{keywords}

%\end{frontmatter}

%%%%%%%%%%%%%%%%%%%%%%%%%%%%%%%%%%%%%%%%%%%%%%%%%%%%%%%%%%%%%%%%%% 
\section{Introduction}

The grace and agility of swimming fish and marine mammals have excited
the curiosity of scientists for a long time.  For example, the the
Northern pike (\emph{Enox Lucius}) can reach accelerations up to $25g$
\citep{harper_fast-start_1990}; the European eel (\emph{Anguilla
  Anguilla}) annually swims over $5000$~km across the Atlantic Ocean
while fasting \citep{ginneken_eel_2005}; fish employing body
undulation as their primary means of propulsion greatly surpass all
engineered vehicles in terms of fast-starting and maneuvering
capabilities. In the hope of shedding light to the fluid mechanisms
behind the aquatic animals' extraordinary performance, biologists,
hydrodynamicists and engineers have observed fish swimming
\citep{gray_studies_1933, videler_fast_1984,
  tytell_hydrodynamics_2004}, measured their metabolic rates
\citep{bainbridge_problems_1961, webb_swimming_1971}, proposed
hydrodynamic principles and scaling laws
\citep{gero_hydrodynamic_1952, lighthill_note_1960,
  triantafyllou_wake_1991, gazzola_scaling_2014,
  van_weerden_meta-analysis_2014}, and built robots replicating the
function of fish \citep{triantafyllou_efficient_1995,
  stefanini_novel_2012,sefati_mutually_2013,
  ijspeert_biorobotics_2014}.

With the increase in computational power, computational fluid dynamics
(CFD) provides an attractive alternative means of studying fish
swimming because of the detailed flow images it can
convey\citep{deng_numerical_2013}. Since the viscous simulations of a
two-dimensional self-propelled anguilliform swimmer by
\citet{carling_self-propelled_1998}, a variety of methods have been
developed to simulate fish swimming. These methods range from
arbitrary Eulerian-Lagrangian methods with deformable mesh
\citep{kern_simulations_2006}, to immersed boundary methods
\citep{borazjani_numerical_2008, shirgaonkar_new_2009, liu_flow_2011,
  bergmann_effect_2014}, to multiparticle collision dynamics methods
\citep{reid_fluid_2012} and viscous vortex particle methods
\citep{eldredge_numerical_2006}. CFD is a unique complement to
experiments on live fish that can potentially give access to full
three-dimensional flow structures as well as local forces and
power. The application of CFD to the study of fish swimming is still
in its infancy, while a number of modelling decisions also need to be
made. Once these modelling and numerical questions are resolved, CFD
becomes a very powerful tool providing unmatched detail of the flow
properties, while allowing wide parametric searches through systematic
changes in the body geometry or the swimming kinematics. As a result,
there has recently been a number of publications reporting efforts in
optimizing fish shape and/or swimming motion
\citep{kern_simulations_2006, van_rees_optimal_2013, eloy_best_2013,
  tokic_optimal_2012}. In this paper we first present a methodology
for simulating fish swimming in which the impact of modelling choices
are carefully quantified. We use this methodology to investigate
efficient swimming for an undulating body.

In addition to optimizing their self-generated flow structures, fish
may be able to use the flow patterns from another swimming fish to
save energy. Whether energy saving is an important reason for
schooling has long been a matter of
discussion. \citet{weihs_hydromechanics_1973} is one of the few papers
proposing a hydrodynamic theory of schooling, viz.~that fish can save
energy by swimming in a `diamond' configuration, taking advantage of
areas of reduced average oncoming velocity that form between adjacent
propulsive wakes.  \citet{partridge_evidence_1979} later commented
that saithe, herring and cod do not swim in the diamond pattern, which
led \citet{pitcher_functions_1986} to write that ``no valid evidence
of hydrodynamic advantage has been produced, and existing evidence
contradicts most aspects of the only quantitative testable theory
published.'' Yet, as pointed out by \citet{abrahams_fish_1987}, such
conclusions may be premature because they ignore the potential
trade-offs involved in school functions. Indeed, despite the
difficulty of assessing the importance of energy saving in schooling
due to the dynamic nature of schools, there has been experimental
evidence that fish located in the rear part of a school spend less
energy than those in the front \citep{killen_aerobic_2012}. A recent
paper suggests that in a fish school, individuals in every position
have reduced costs of swimming, compared to when they swim at the same
speed but alone \citep{marras_fish_2014}. Further, the recent finding
that ibises in a flock position themselves and phase their motion such
that they can take advantage of the vortices left by the ibis in front
of them, suggests that analogous mechanisms might be found for fish
schools as well \citep{portugal_upwash_2014}. In this paper we
investigate the mechanisms by which two fish swimming as a pair can
save energy.

By optimizing fish-like swimming kinematics and comparing them with
the parameters observed for various fish species, we can shed light on
the processes that led to the development of the swimming
characteristics of each species, while from an engineering point of
view we can derive new design principles for propulsion, inspired by
efficient living organisms, potentially even exceeding their
performance \citep{van_rees_optimal_2013}.  By investigating
strategies for fish-to-fish hydrodynamic interaction, we can shed
light on fish school formation and assess its potential hydrodynamic
benefits.  In \S\,\ref{sec:modelling}, we discuss modelling
considerations for the simulation of fish swimming and briefly present
the governing equations numerical details specific to fish swimming
simulations. We use the model and numerical method to optimize the
gait of an undulating fish-like foil in open-water
(\S\,\ref{sec:1fish}) and the positioning and timing for a pair of
undulating fish-like foils (\S\,\ref{sec:2fish}).

%%%%%%%%%%%%%%%%%%%%%%%%%%%%%%%%%%%%%%%%%%%%%%%%%%%%%%%%%%%%%%%%%%

\section{Methods: modelling and simulations \label{sec:modelling}}

Aquatic animals exhibit a wide variety of designs and propulsion
modes. However, most fish and cetaceans generate thrust by bending
their bodies into a backward-traveling wave that extends to the caudal
fin, a type of swimming often classified as body and/or caudal fin
(BCF) locomotion \citep{sfakiotakis_review_1999}. In the present
paper, we investigate the efficiency of BCF propulsion, with
particular examples drawn from eels that undulate their whole body
(anguilliform motion), as well as saithe and mackerel that only
undulate the aft third of their body (carangiform motion)
\citep{breder_locomotion_1926}. For expedient calculations and since
the swimming motion bends the body as a plate, i.e.~it is
quasi-two-dimensional, we first use two-dimensional simulations to
investigate the impact of various kinematic parameters and then
compare the results with three-dimensional simulations of a
three-dimensional, danio-shaped body.

%%%%%%%%%%%%%%%%%%%%%
\subsection{Fish shape and swimming motion}

\begin{figure}
 \centerline{\includegraphics[width=0.5\textwidth]{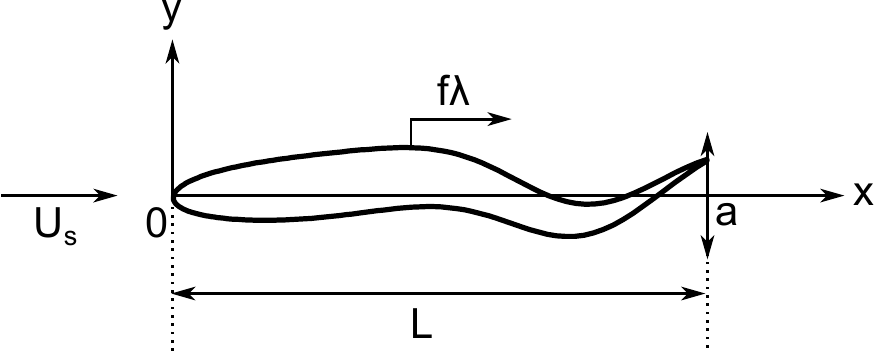}}
  \caption[Schematic showing the fish model parameters.]{Schematic
    showing the fish model parameters. An elongated body of length $L$
    undulates in a flow of speed $U_s$ with a wave traveling backward
    at speed $f\lambda$ and amplitude $a$ at the trailing edge.}
\label{fig:fish_schematic}
\end{figure}

\begin{figure}
 \centerline{\includegraphics[width=0.75\textwidth]{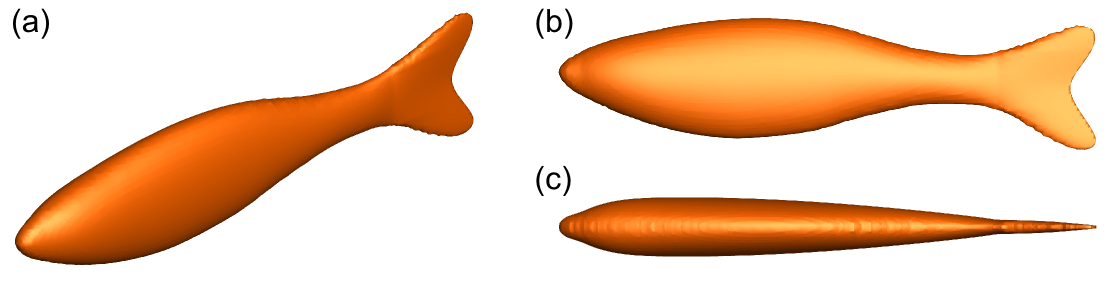}}
  \caption{Three-dimensional fish geometry based on a giant danio.}
\label{fig:danio}
\end{figure}

In order to capture the main parameters of BCF swimming while keeping
the problem complexity manageable, we model the main body of the fish
and its caudal fin but not the other fins or details on the body such
as scales, finlets, and other protrusions. We represent a swimming
fish by a neutrally buoyant undulating body of length $L=1$, as
illustrated in figure \ref{fig:fish_schematic}. For the
two-dimensional simulations, a NACA0012 shape is chosen at rest,
whereas a danio-shaped body, shown in figure \ref{fig:danio}, is used
for the 3D simulations. The body propels itself at average speed $U_s$
in a fluid of kinematic viscosity $\nu$ and density $\rho$ by
oscillating its mid-line in the transverse direction $y$. The leading
edge of the body at rest is located at $x=0$ and its trailing edge at
$x=1$.

In 2D simulations, we will refer to the body as a `fish' rather than a
flexible foil to avoid confusion with the caudal fin, and with
non-self-propelled flexible foils used as propulsors.

We employ traveling wave kinematics including recoil that resemble
those observed in fish according to either carangiform or anguilliform
swimming. The lateral displacement, $h$, of a point located at $x$
along the foil is given at time $t$ by:
 \begin{align}
 \label{eq:motion}
 h(x,t) &= h_0(x,t)+B(x,t)+y_1(x)  \nonumber \\
 &=a_0 A(x)\sin \big( 2\pi ( x/\lambda - ft +\phi)\big) + B(x,t) +y_1(x) \nonumber \\
 &=g(x)\sin \big( 2\pi (ft + \psi(x) )\big) +y_1(x)
\end{align}  
where $A(x)$, with $A(1)=1$, is the envelope of the qprescribed backward
traveling wave of wavelength $\lambda$ and frequency $f$,
\begin{equation}
B(x,t) = \left(a_r+b_rx \right) \sin \big( 2\pi (ft +\phi _r)\big)
\end{equation}
is the recoil term due to the hydrodynamic forces on the fish, and
\begin{equation} \label{eq:camber}
y_1(x) = C(x^2+\gamma x + \beta )
\end{equation}
can be used for steering (see Appendix\,\ref{sec:PID}) by adding
camber to the fish, while $\gamma$ and $\beta$ ensure that linear and
angular momentum are conserved through the deformation. $y_1$ is
necessary to ensure stability but, in steady regime, $y_1 \ll a_0$.

The parameter $a_0$ determines the amplitude of the deformation $h_0$
at the trailing edge. It is adjusted through a feedback control loop
to ensure that the average net drag on the foil is $0$, as described
in Appendix\,\ref{sec:PID}. $h_0(x,t)$ can be used without the recoil
and steering terms in order to prescribe the full kinematics of the
swimmer, in which case $h(x,t)=h_0(x,t)$. For a freely moving body
with prescribed deformation, the recoil is computed from the
hydrodynamic forces on the body. In the latter case, the envelope of
the actual displacement is given by $g(x)$, with peak to peak
amplitude at the trailing edge given by $a=2g(1)$.

\begin{figure}
	\centering
\begin{subfigure}[b]{0.4\textwidth}
	\centering
	\includegraphics[width=\textwidth]{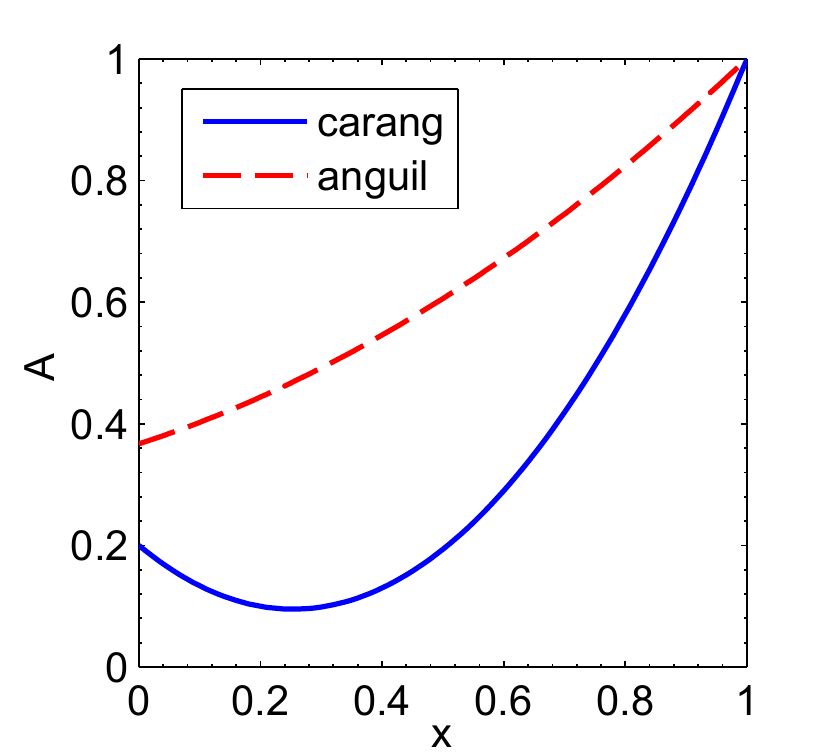}
	\caption{Prescribed amplitude envelopes}
	\label{fig:envelop}
\end{subfigure}
\hspace{0.05\textwidth}
\begin{subfigure}[b]{0.42\textwidth}
	\centering
	\includegraphics[width=\textwidth]{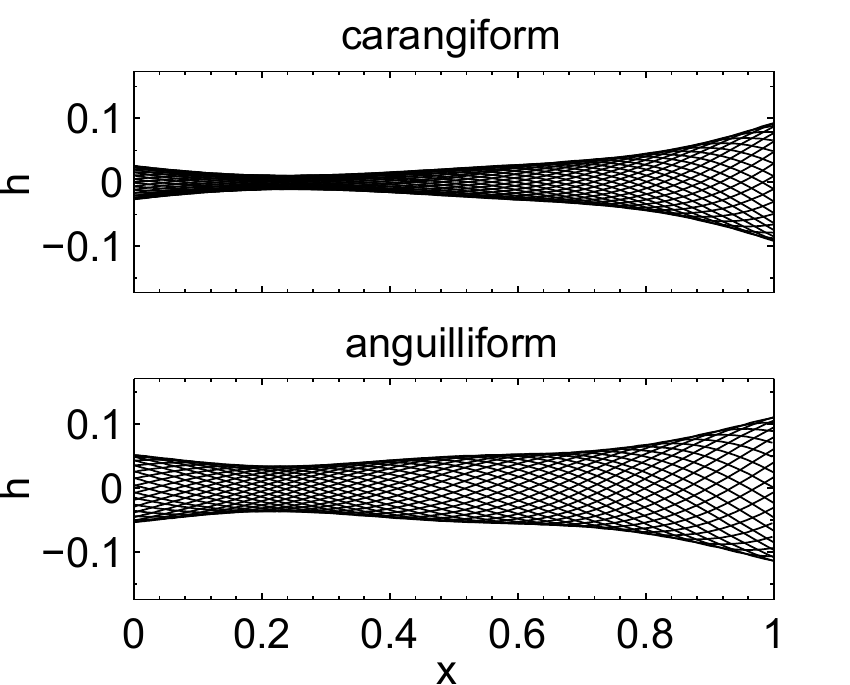}
	\caption{Mid-line displacement}
	\label{fig:motion}
\end{subfigure}
\caption{Carangiform and anguilliform motion for $f=1.8$ and $a_0
  =0.1$ at Reynolds number $\Rey = 5000$ with recoil.}
\end{figure}

The prescribed kinematics of a carangiform swimmer, based on the
experimental observation of steadily swimming saithe
\citep{videler_fast_1984, videler_fish_1993}, is often modeled as:
\begin{equation} \label{eq:carangiform}
a_0 = 0.1,\quad A(x)=1-0.825(x-1)+1.625(x^2-1), \quad
\lambda = 1,\quad B(x,t) = 0.
\end{equation}
This motion is for example used in \citet{borazjani_numerical_2008,
  dong_characteristics_2007} and, in the rest of the paper, will be
referred to as the carangiform gait. Experimental observations of
American eels \citep{tytell_hydrodynamics_2004-1} provide that
anguilliform motion can be represented by:
\begin{equation} \label{eq:anguilliform}
a_0 = 0.1,\quad A(x)=1+0.323(x-1)+0.310(x^2-1), \quad
\lambda = 1,\quad B(x,t) = 0.
\end{equation}
Figure \ref{fig:envelop} shows the prescribed envelope $A(x)$ for the
carangiform (resp. anguilliform) swimmer defined in equation
\ref{eq:carangiform} (resp.~equation \ref{eq:anguilliform}). Figure
\ref{fig:motion} illustrates the resulting mid-line displacement in
the presence of the recoil term.

%%%%%%%%%%%%%%%%
\subsection{Kinematic parameters}

The goal of this paper is to identify kinematic parameters that
minimize the self-propelled swimming power $P_{in}$ for a given speed
(Reynolds number) and body shape. In order to quantify the fitness of
each motion, the quasi-propulsive efficiency $\eta_{QP}$ is used,
which compares $P_{in}$ to the useful power, i.e.~the resistance $R$
of the rigid-straight towed body at the same speed $U_s$ times the
speed: $\eta_{QP}=RU_s/\overline{P_{in}}$. Indeed, as discussed in
\cite{maertens_efficiency_2015}, the Froude propulsive efficiency is
not appropriate here as it is zero for a self-propelled body.

For rigid flapping foils, the parameters that characterize the motion
and its performance have been extensively studied
\citep{anderson_oscillating_1998, read_forces_2003}). The principal
kinematic parameters are the Strouhal number and the maximum nominal
angle of attack, and, to a lesser degree, the heave amplitude to chord
ratio, and the phase angle between heave and pitch; all, typically,
measured at $25\%$ of the chord. The Strouhal number is a wake
parameter, since it characterizes the dynamics of the (unstable) wake
\citep{triantafyllou_wake_1991, triantafyllou_optimal_1993}; hence the
width of the wake must, in principle, be used as the characteristic
length.  However, the width of the wake is unavailable beforehand, so
this characteristic length is approximated typically by the peak to
peak motion of the trailing edge. Hence, for an undulating flexible
foil, we define the Strouhal number, heave amplitude, pitch angle and
nominal angle of attack at the trailing edge. These parameters and
others used throughout this paper are summarized in table
\ref{tab:phys_param}.  While the motion cannot be characterized by
these parameters alone, they play an important role in determining the
swimming efficiency. Changing the amplitude of motion and Strouhal
number can be achieved through parameters like $a_0$ and $f$ (though,
for a given motion and average velocity, there is a unique amplitude
that ensures a steady velocity), but the pitch amplitude
$\theta_{\max}$ and maximum angle of attack $\alpha_{\max}$ cannot be
directly controlled. Therefore, when optimizing the swimming gait, it
is important to choose a parametrization that allows to adjust the
pitch and angle of attack amplitudes independently of the heave
amplitude and Strouhal number.  This is best done by changing
parameters that control the derivative of the prescribed envelope
$A(x)$ at the trailing edge.

In this study, the lateral flexing motion (i.e.~the lateral motion
after the linear and angular recoil are subtracted) is characterized
by four parameters: the frequency, amplitude, and two parameters
controlling the shape of $A(x)$, the envelope of the unsteady bending
motion. With prescribed frequency, the amplitude is adjusted to ensure
self-propulsion, while the two remaining parameters are varied
systematically in order to identify the values that minimize the
swimming power (equivalently, maximize the quasi-propulsive efficiency
$\eta_{QP}$) for a given Reynolds number and body flexing shape. In
order to explore a wide range of motions, two different
parametrizations are used for $A(x)$: a quadratic envelope and a
Gaussian envelope (see details in \S\,\ref{sec:1fish}). The wavelength
of the traveling wave is fixed, equal to the body length, and the
maximum amplitude is adjusted to ensure the average swimming speed of
the self-propelled fish results in Reynolds number $\Rey=5000$. The
linear and angular recoil terms are computed by integration of the
hydrodynamic forces.
 
\begin{table}
  \begin{center}
  \begin{tabular*}{0.8\textwidth}{@{\extracolsep{\fill}} ccc}
  \toprule
   Name & Symbol & Expression  \\
     \midrule
     Power coefficient & $C_P$ & $2P_{in}/(\rho U_s^3L)$ \\
     Quasi-propulsive efficiency & $\eta_{QP}$ & $RU_s/\overline{P_{in}}$ \\
     Reynolds number & \Rey & $U_sL/\nu$ \\
     Strouhal number & $St$ & $fa/U_s$ \\
     Pitch angle & $\theta$ & \\
     Angle of attack & $\alpha$ & \\
     Peak to peak heave & $a$ \\
     heave-pitch phase angle & $\psi$ \\
       \bottomrule  
  \end{tabular*}
  \caption{Kinematic and other dimensionless parameters.}
  \label{tab:phys_param}
  \end{center}
\end{table}

%%%%%%%%%%%
\subsection{Governing equations and numerical implementation \label{sec:numerical}}

In a self-propelled swimming body, its motion is determined by the
coupled fluid-body dynamics.  The physical parameters are
non-dimensionalized by the fish body length $L$, its intended average
cruising speed $U_s$, and the density of water $\rho$.

\begin{figure}
 \centerline{\includegraphics[width=0.7\textwidth]{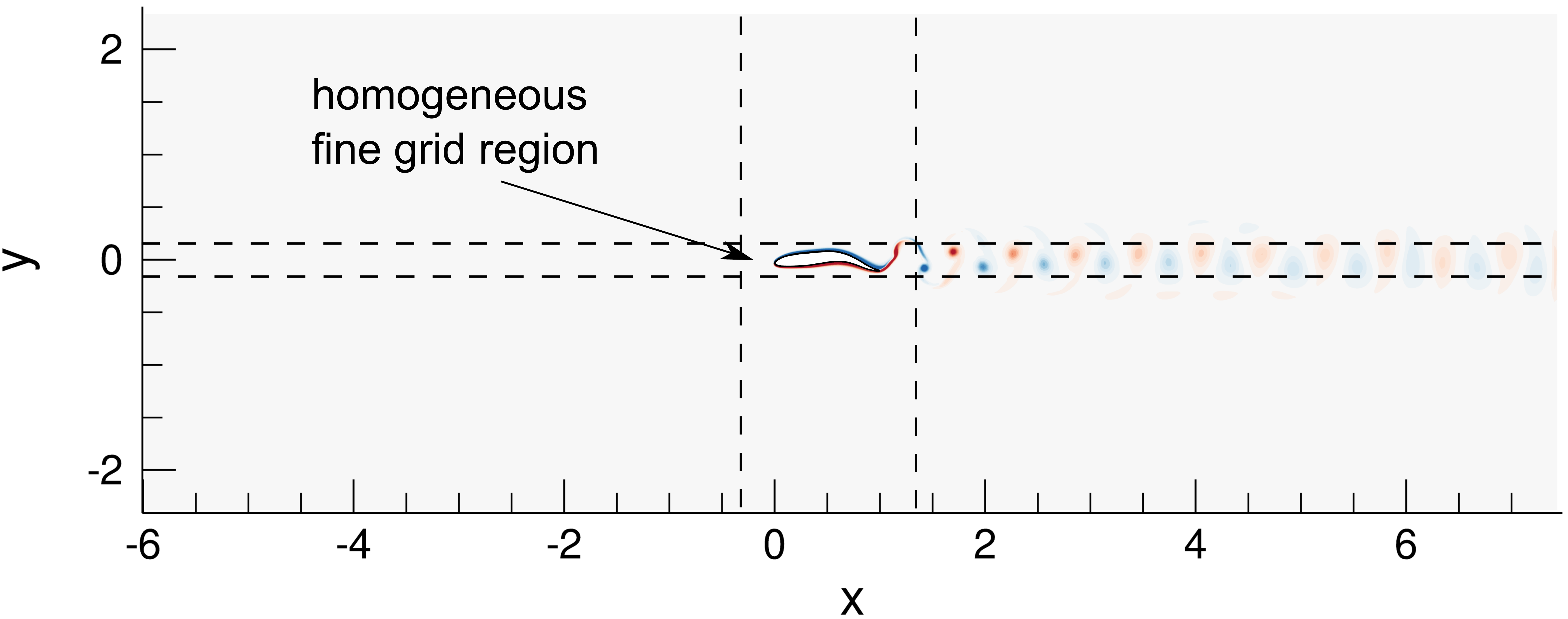}}
  \caption{Flow configuration for the undulating NACA0012
    simulations. The vorticity field for the carangiform motion with
    $f=1.8$ and zero mean drag is shown as an example.}
\label{fig:geometry}
\end{figure}

In order to solve the coupled fluid/body problem described above, we
adapted the \emph{2nd order} boundary data immersion method (BDIM)
presented in \citet{maertens_accurate_2015}. The validation of the
numerical method presented in this section for simulating
self-propelled undulating bodies is presented in Appendix
\ref{sec:valid}. For the two-dimensional simulations, constant
velocity $\vec{u}=\vec{U}_s$ is used on the inlet ($x=-6$), periodic
boundary conditions on the upper and lower boundaries ($y= \pm 2.4$),
and a zero gradient exit condition with global flux correction
($x=7$). The Cartesian grid is uniform near the fish with grid size
$\dd x=\dd y=1/160$ and uses a $2\%$ geometric expansion ratio for the
spacing in the far-field, as illustrated in figure \ref{fig:geometry}.
The three-dimensional simulations are run on a $6 \times 3 \times 3$
domain with constant velocity $\vec{u}=\vec{U}_s$ on the inlet, a zero
gradient exit condition with with global flux correction and periodic
boundary conditions along $y$ and $z$ boundaries. The Cartesian grid
is uniform near the fish with grid size $\dd x = \dd y = \dd z =
1/100$ and uses a $4\%$ geometric expansion ratio for the spacing in
the far-field.

The fluid and body equations are integrated over the fluid and body
domains, respectively, $\Omega_f$ and $\Omega_b$, with a kernel of
radius $\epsilon = 2\dd x$.  The BDIM equations for the smoothed
velocity field $\vec{u}_{\epsilon}$ are valid over the complete domain
$\Omega = \Omega_f \cup \Omega_b$ and enforce the no-slip boundary
condition at the interface.  These equations, integrated from time $t$
to time $t_+=t+\Delta t$, are:
\begin{subequations}\label{eq:BDIM}
\begin{empheq}[left=\empheqlbrace]{align}
&  \vec{u}_{\epsilon}(t_+)  = \; \vec{v}(t_+) +
 \left(\mu_0^{\epsilon}(d)  +\mu_1^{\epsilon}(d)\frac{\partial}{\partial n}
 \right)\big(\vec{u}_{\epsilon}(t)- \vec{v}(t_+) 
 +\vec{R}_{\Delta t}-\vec{\partial P}_{\Delta t}\big) \\
&  \vec{\nabla} \cdot \vec{u}_{\epsilon}(t_+) =0 
\end{empheq}
\end{subequations}
where $\vec{v}$ is the velocity field associated with the closest
body, $\hat{n}$ the unitary normal to the closest fluid/solid boundary
(pointing toward the fluid), and $d$ the signed distance to the
closest boundary ($d>0$ within the fluid, $d<0$ inside a
body). $\mu_0^{\epsilon}$ and $\mu_1^{\epsilon}$ are respectively the
zeroth and first central moments of the smooth delta kernel (see
\cite{maertens_accurate_2015} for more details). The pressure impulse
$\vec{\partial P}_{ \Delta t}$ and $\vec{R}_{\Delta t}$ accounting for
all the non-pressure terms are defined as:
\begin{equation}
\vec{R}_{\Delta t}(\vec{u}) = \int_{t_0}^{t_0+\Delta t}\left[-\left(\vec{u}\cdot \vec{\nabla}  \right) \vec{u} +\nu\nabla ^2\vec{u} \right]\ \text{d}t 
\text{ ,} \qquad
\vec{\partial P}_{\Delta t} =\int_{t_0}^{t_0+\Delta t}\frac{1}{\rho}\vec{\nabla}p \ \text{d}t.
\end{equation}
In order to simplify the equations of motion, we consider motion
within the $(x,\,y)$ plane, such that the translational velocity of
the body center of mass (COM), $\vec{v}_c$, is a two-dimensional
vector $(v_c^x,\,v_c^y)$, and its rotation velocity is
$\omega_b=\omega_b^z$.  We then define the generalized velocity
$\boldsymbol{V}$, location $\boldsymbol{X}$, and force
$\boldsymbol{F}$ vectors, as well as the generalized mass matrix
$\mathbf{M}$:
\begin{equation}
\boldsymbol{V} = 
\begin{pmatrix}
v_c^x\\
v_c^y\\
\omega_b
\end{pmatrix},
\quad
\boldsymbol{X} = \frac{\text{d}\boldsymbol{V}}{\text{d}t},
\quad
\boldsymbol{F} = 
\begin{pmatrix}
F^x_h\\
F^y_h\\
M_c^z
\end{pmatrix},
\quad
\mathbf{M} = 
\begin{pmatrix}
m & 0 & 0 \\
0 & m & 0 \\
0 & 0 & I_c
\end{pmatrix},
\end{equation}
where $\vec{F}_h$ is the hydrodynamic force on the body, $m$ is the
mass of the body which has density $\rho_b=\rho$ and $I_c$ its moment
of inertia with respect to the COM. The motion of the body is governed
by:
\begin{equation}  \label{eq:body}
\frac{\text{d}}{\text{d} t} \left( \mathbf{M} \boldsymbol{V} \right) =
\boldsymbol{F}.
\end{equation}
The coupled dynamic equations are discretized using a sequentially
staggered Euler explicit integration scheme with Heun's
corrector. Sequentially staggered schemes are computationally
efficient, but for large added mass they become unconditionally
unstable \citep{forster_artificial_2007}, regardless of the particular
scheme used. In order to stabilize the numerical scheme, we introduce
the virtual added mass matrix $\mathbf{M_a}$.

The virtual added mass, which is used in an implicit added mass scheme
\citep{connell_flapping_2007, zhu_propulsion_2008, peng_energy_2009},
can eliminate the instability due to large added mass, but its exact
value will not affect the results.  In the case of an undulating fish,
the coefficients of the matrix can be estimated from the added mass of
the fish at zero angle of attack, or heuristically tuned to avoid
instability. In the present simulations, the virtual added mass is a
diagonal matrix with value $[0\ 11m\ 13m]$.
 
We also define the total mass as:
\begin{equation}
\mathbf{M_T} = \mathbf{M} + \mathbf{M_a}.
\end{equation}
With these new definitions, we integrate equation \ref{eq:body} over a
time-step $\Delta t$ in the form:
\begin{equation} \label{eq:body2}
\boldsymbol{V}(t+\Delta t)= \boldsymbol{V}(t) +
\mathbf{M_T}^{-1}\int_t^{t+\Delta t} \left[ \boldsymbol{F} + 
\mathbf{M_a} \frac{\text{d} \boldsymbol{V} }{\text{d} \tau} \right]  
\text{d} \tau .
\end{equation}
At each time step $t_n$, the fluid and body velocities, $\vec{u}_n =
\vec{u}_{\epsilon}(t_n)$ and $\vec{v}_n = \vec{v}(t_n)$ respectively,
are calculated from the velocities and forces at the previous time
steps according to equations \ref{eq:BDIM} and \ref{eq:body2}.

%%%%%%%
\subsection{The importance of recoil}

Equation \ref{eq:body2} determines the recoil $B(x,t)$ resulting from
the prescribed motion $h_0(x,t)$ and the related hydrodynamic forces
$\boldsymbol{F}$. Due to the significant added complexity incurred by
the recoil term, most of the earlier simulation studies neglected it
\citep{borazjani_numerical_2008, dong_characteristics_2007}. However,
the amplitude of this term, and its impact on the estimated swimming
power are substantial \citep{reid_fluid_2012}, as illustrated below.

\begin{figure}
	\centering
\begin{subfigure}[b]{0.44\textwidth}
	\centering
	\includegraphics[width=\textwidth]{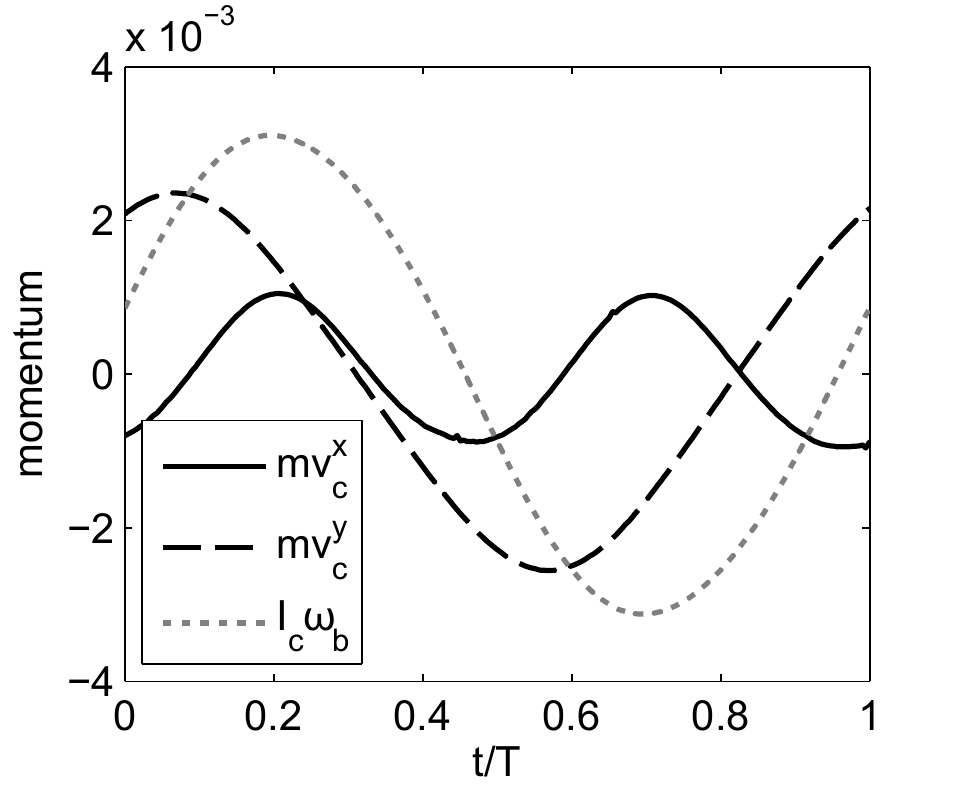}
	\caption{Linear and angular momentum}
	\label{fig:momentum}
\end{subfigure}
\hspace{0.05\textwidth}
\begin{subfigure}[b]{0.45\textwidth}
	\centering
	\includegraphics[width=\textwidth]{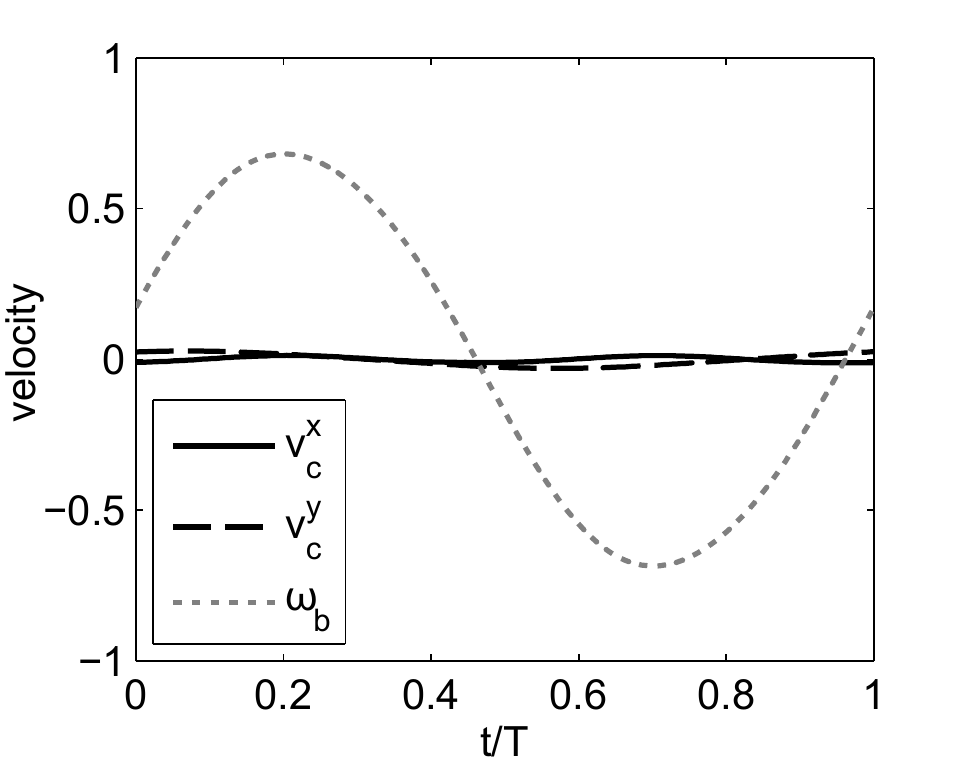}
	\caption{Velocity and rotation rate}
	\label{fig:velocity}
\end{subfigure}
\caption{(a) Linear and angular momentum and (b) corresponding
  velocities for a neutrally buoyant self-propelled NACA0012 with
  carangiform motion at frequency $f=1/T=2.1$.}
\end{figure}

We consider first the carangiform motion of Eq. \ref{eq:carangiform}
with frequency $f=2.1$.  Figure \ref{fig:momentum} shows the
dimensionless linear and angular momentum for the self-propelled fish,
including the recoil, as determined by the hydrodynamic forces and
adaptive amplitude $a_0$. The angular and transverse momentum are
larger than the longitudinal momentum, but the three amplitudes are
comparable. However, the non-dimensional moment of inertia of the fish
is much smaller than its mass:
\begin{equation}
m=0.081,\, \qquad I_c=0.0045 ,
\end{equation}
where the mass and moment of inertia are non-dimensionalized by the
length $L$ and density $\rho$. Therefore, whereas the linear momentum
results in velocities smaller than $3\%$ of the free-stream $U_s$, the
rotation of the fish generates velocities at the trailing edge up to
$40\%$ of the free-stream, as shown in Figure \ref{fig:velocity}. This
observation suggests that, whereas the longitudinal motion of the fish
might be negligible, the transverse motion, and specifically the
motion due to the free-rotation, are important.

\begin{figure}
 \centerline{\includegraphics[width=0.56\textwidth]{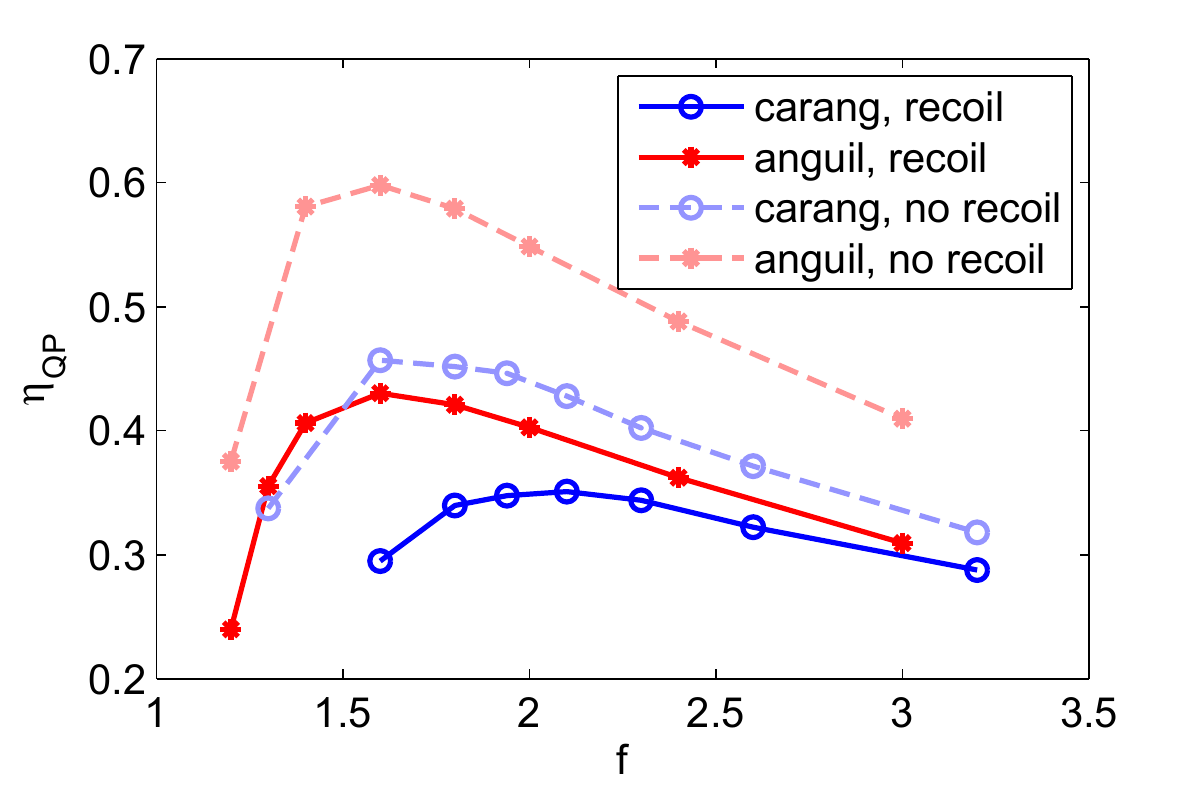}}
  \caption{Quasi-propulsive efficiency as a function of frequency for
    the carangiform and anguilliform motions with and without recoil.}
\label{fig:recoil}
\end{figure}

In order to further illustrate this result, figure \ref{fig:recoil}
shows the quasi-propulsive efficiency as a function of frequency for
the carangiform and anguilliform motions with and without recoil. The
figure shows that, at all frequencies, the undulation with recoil
requires more power than the undulation without recoil. Therefore,
simulations that do not allow for recoil are likely to underestimate
the swimming power, as discussed in \citet{reid_flow_2009}. The figure
also shows that the optimal frequency without recoil might differ from
the optimal frequency with recoil. In the cases studied here, the
optimal frequency for the carangiform undulation without recoil is
around $f=1.6$, while with recoil it is around $f=2.1$.

In summary, we have shown here that the impact of the pitch motion of
the fish on swimming performance is significant. In order to estimate
meaningful values of fish swimming efficiency, it is critical to allow
for recoil.

%%%%%%%
\subsection{Gait optimization procedure}

\citet{eloy_best_2013} and \citet{tokic_optimal_2012} combined an
evolutionary algorithm with Lighthill's potential flow slender-body
model to simultaneously optimize the shape and kinematics, using,
respectively, $22$ and $9$ parameters. In this paper we employ viscous
simulations that are far more demanding computationally, hence we
parameterize the amplitude $A(x)$ using only two parameters, because
the small number of parameters allows us to find an optimum with a
reduced number of evaluations, and it also facilitates the
visualization and interpretation of the results.

For a given kinematic parametrization and frequency, the envelope
$A(x)$ is optimized using derivative-free optimization
\citep{rios_derivative-free_2013}. We apply the BOBYQA algorithm that
performs bound-constrained optimization using an iteratively
constructed quadratic approximation for the objective function
\citep{powell_bobyqa_2009}. For each set of parameters, the viscous
simulation is run for $15$ non-dimensional time units, and the average
power coefficient $\overline{C_P}$ across the last $10$ undulation
periods is calculated. Based on the values of $\overline{C_P}$, the
implementation of BOBYQA provided by the NLopt free C library
\citep{nlopt} interfaced with Matlab computes the next set of
parameters. In order to avoid finding a local minimum due to numerical
noise, after the algorithm has converged, it is run again using the
previously found minimum as a starting point.

%%%%%%%%%%%%%%%%%%%%%%%%%%%%%%%%%%%%%%%%%%%%%%%%%%%%%%%%%%%%%%%%%%  
\section{Efficiency of swimming in open-water}

The goal in this section is to identify undulatory gaits that require
the minimum amount of power ($\overline{P_{in}}$) to drive an
elongated body at speed $U_s$, such that the Reynolds number is
$\Rey=5000$. In other words, we want to maximize the quasi-propulsive
efficiency $\eta_{QP}$ of the self-propelled undulating body and
identify the key parameters under the constraints of fixed body size
and shape, as well as Reynolds number. We first consider a
two-dimensional NACA0012-shaped fish and then apply the results to a
three-dimensional danio-shaped body.

%%%%%%%%%%%%%%%%%%%%%%%%%%%%%
\subsection{Gait optimization for a two-dimensional foil \label{sec:1fish}}

\begin{figure}
 \centerline{\includegraphics[width=0.7\textwidth]{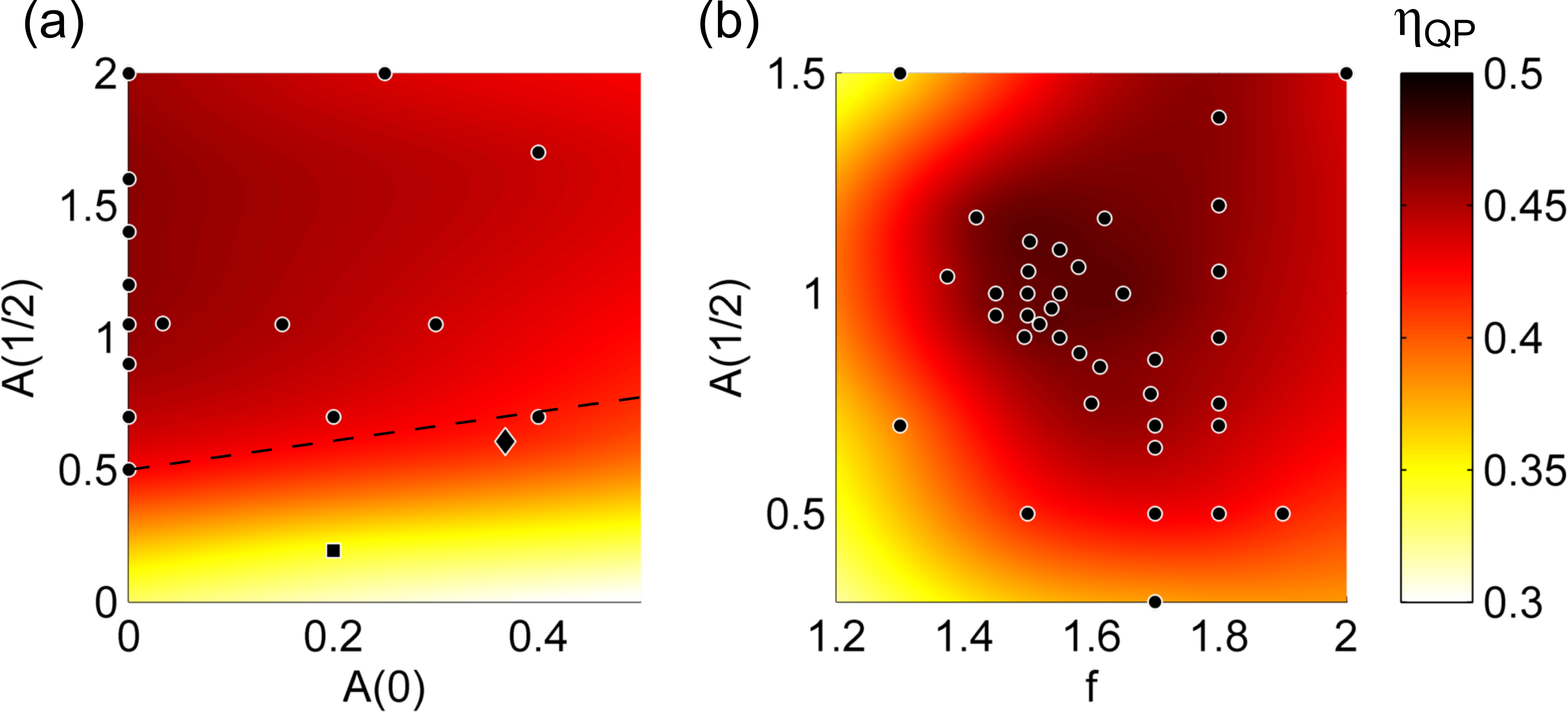}}
  \caption[$\eta _{QP}$ as a function of $A(0)$ or $f$ and $A(1/2)$
    for quadratic envelopes.]{$\eta _{QP}$ as a function of $A(0)$ or
    $f$ and $A(1/2)$ for quadratic envelopes. The black dots show the
    location of the points that have been used to build the thin-plate
    smoothing spline ($\mathsf{tpaps}$ function in Matlab with
    smoothing parameter $p=0.999$) represented in color. (a): fixed
    frequency $f=1.8$. The carangiform and anguilliform motions are
    respectively denoted by a black square and a black diamond, and a
    dashed line shows the location of linear envelopes (points below
    this line correspond to convex envelopes, while above it the
    envelopes are concave). (b): fixed leading edge value $A(0)=0$.}
\label{fig:optim_poly}
\end{figure}

For several values of undulation frequency, we optimize the
deformation envelope $A(x)$. $A(x)$ has been traditionally modeled by
a quadratic function, of the form:
\begin{equation}
A(x)=1+c_1(x-1)+c_2(x^2-1).
\end{equation}
In the figures we parametrize each envelope by $A(0)$ and $A(1/2)$,
the envelope amplitude at the leading edge and mid-chord respectively
(the amplitude at the trailing edge being constrained to
$A(1)=1$). Indeed, $A(0)$ and $A(1/2)$ are more meaningful than $c_1$
and $c_2$ and can easily be restricted to a rectangle. First, we fix
the undulation frequency to $f=1.8$ and optimize the quadratic
envelope $A(x)$, restricting $A(0)$ to positive values. Figure
\ref{fig:optim_poly}a shows the efficiency as a function of $A(0)$ and
$A(1/2)$. The carangiform envelope used in previous sections is
denoted by a black square, and the anguilliform gait through a
diamond. It is clear that the envelopes above the dashed line, which
are concave envelopes with a peak upstream from the trailing edge,
have good efficiency. The efficiency decreases very quickly below the
dashed line, as the envelope becomes convex with an increasing
amplitude at the trailing edge. Therefore, the envelope traditionally
used to model carangiform swimming is inefficient, whereas the
anguilliform envelope, which is closer to a straight line, is much
more efficient. Among the concave envelopes, $A(0)=0$ seems best,
together with $1 \leq A(1/2) \leq 1.7$, where the efficiency reaches a
value of $48 \%$. Since the optimal quadratic gait saturates the
constraint $A(0) \geq 0$, we then fix the leading edge amplitude to
$A(0)=0$ and optimize the undulation frequency $f$ and the second
envelope parameter $A(1/2)$. Figure \ref{fig:optim_poly}b shows the
efficiency as a function of $f$ and $A(1/2)$. Here again, around the
optimal point, the efficiency is not very sensitive to the exact value
of $f$ and $A(1/2)$. The optimal quadratic envelope ($A(0)=0$,
$A(1/2)=1$, $A(1)=1$) has a maximum amplitude at $x=3/4$ and reaches
an efficiency of $\eta_{QP}=49\%$ around $f=1.6$.

\begin{figure}
 \centerline{\includegraphics[width=0.7\textwidth]{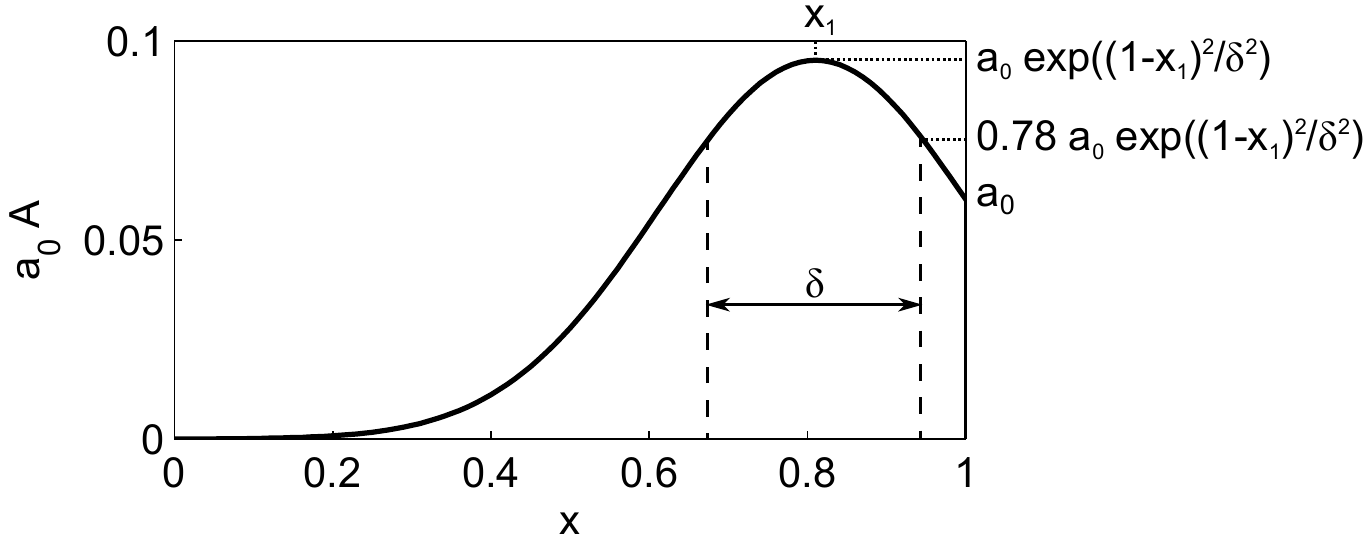}}
  \caption{Definition of the parameters for a Gaussian envelope.}
\label{fig:envelop_params}
\end{figure}

A quadratic envelope has been traditionally used to describe the
displacement envelope of undulating fish which is maximum at the
trailing edge, but the envelope of the curvature amplitude in saithe
and mackerel has a distinctive peak around the peduncle section
\cite{videler_fast_1984}. The results from figure \ref{fig:optim_poly}
also suggest that the efficiency is higher if the deformation is
largest upstream of the trailing edge rather than at the trailing edge
itself. Such envelopes can be better modeled by a Gaussian function of
the form:
\begin{equation}
A(x)=\exp\left({-\left(\frac{x-x_1}{\delta}\right)^2+\left(\frac{1-x_1} {\delta}\right)^2}\right),
\end{equation}
where $x_1$ parametrizes the location of the peak and $\delta$ its
width, as shown in figure \ref{fig:envelop_params}. With the Gaussian
function, it is easy to change the pitch and angle of attack
amplitudes at the tail by adjusting the location and width of the
peak. Since the Gaussian envelope is always positive, the entire
$(x_1,\, \delta)$ space can be used to search for an optimal gait
without running into degenerate gaits.

\begin{figure}
 \centerline{\includegraphics[width=0.85\textwidth]{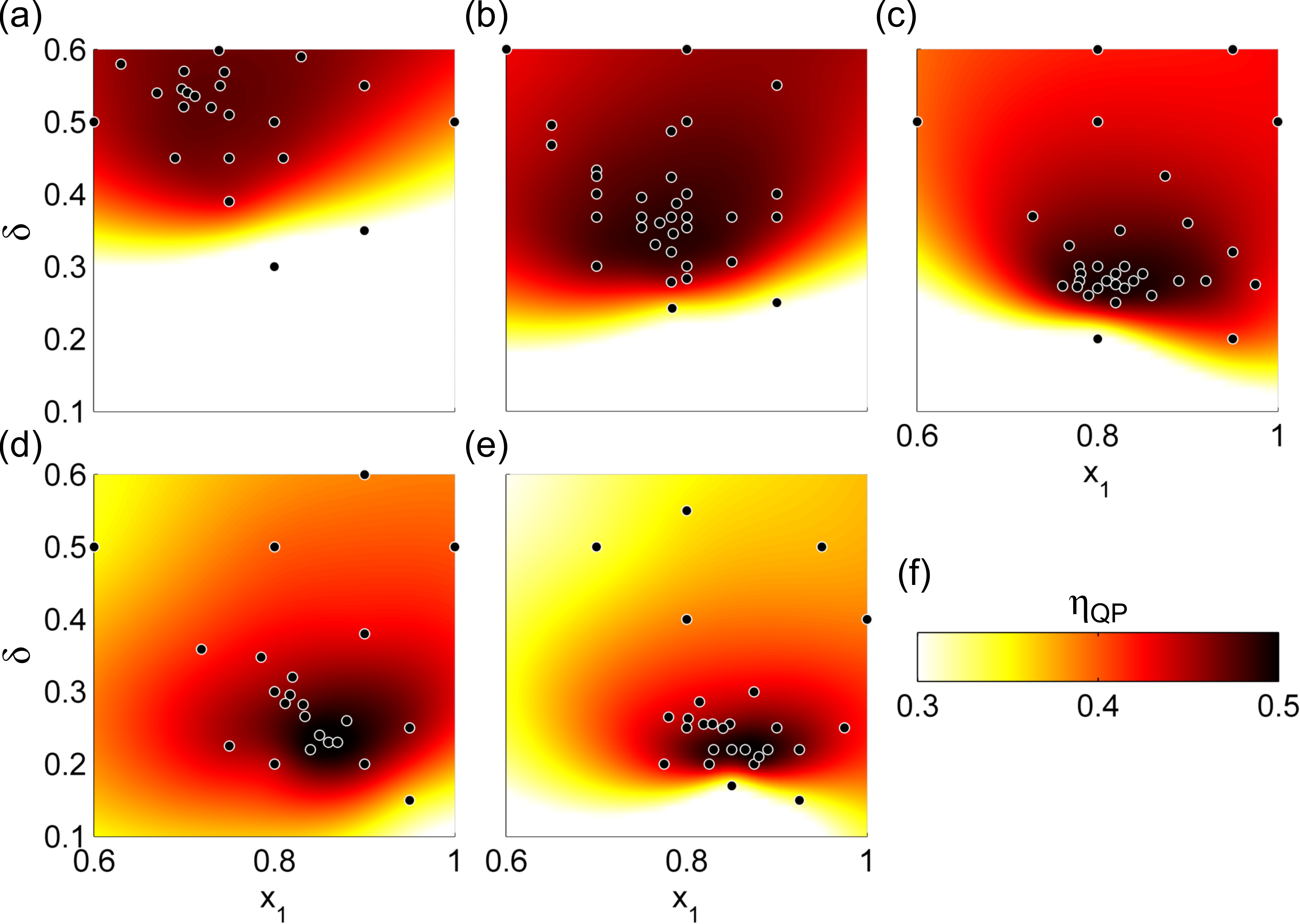}}
  \caption[$\eta _{QP}$ as a function of $x_1$ and $\delta$ near the
    optimum for Gaussian envelopes.]{$\eta _{QP}$ as a function of
    $x_1$ and $\delta$ near the optimum for Gaussian envelopes. (a):
    $f=1.5$, (b): $f=1.8$, (c): $f=2.1$, (d): $f=2.4$, (e): $f=2.7$,
    (f): colorbar. The black dots show the location of the points that
    have been used to build the thin-plate smoothing spline
    ($\mathsf{tpaps}$ function in Matlab with smoothing parameter
    $p=0.999$) represented in color.}
\label{fig:optim_gauss}
\end{figure}

Figure \ref{fig:optim_gauss} shows the efficiency as a function of
$x_1$ and $\delta$ in the neighborhood of the optimal envelope for
$\lambda=1$ and five frequencies ranging from $f=1.5$ to $f=2.7$. For
all frequencies, the efficiency decreases very rapidly as $\delta$ is
decreased below its optimal value, while the efficiency is much less
sensitive to increases above this optimal value. Moreover, while for
all frequencies it is possible to find a region in the
$(x_1,\,\delta)$ space that reaches an efficiency of $50\%$ (see table
\ref{tab:opt_gauss} for details), the optimal envelope clearly depends
on the frequency.

\begin{figure}
 \centerline{\includegraphics[width=0.85\textwidth]{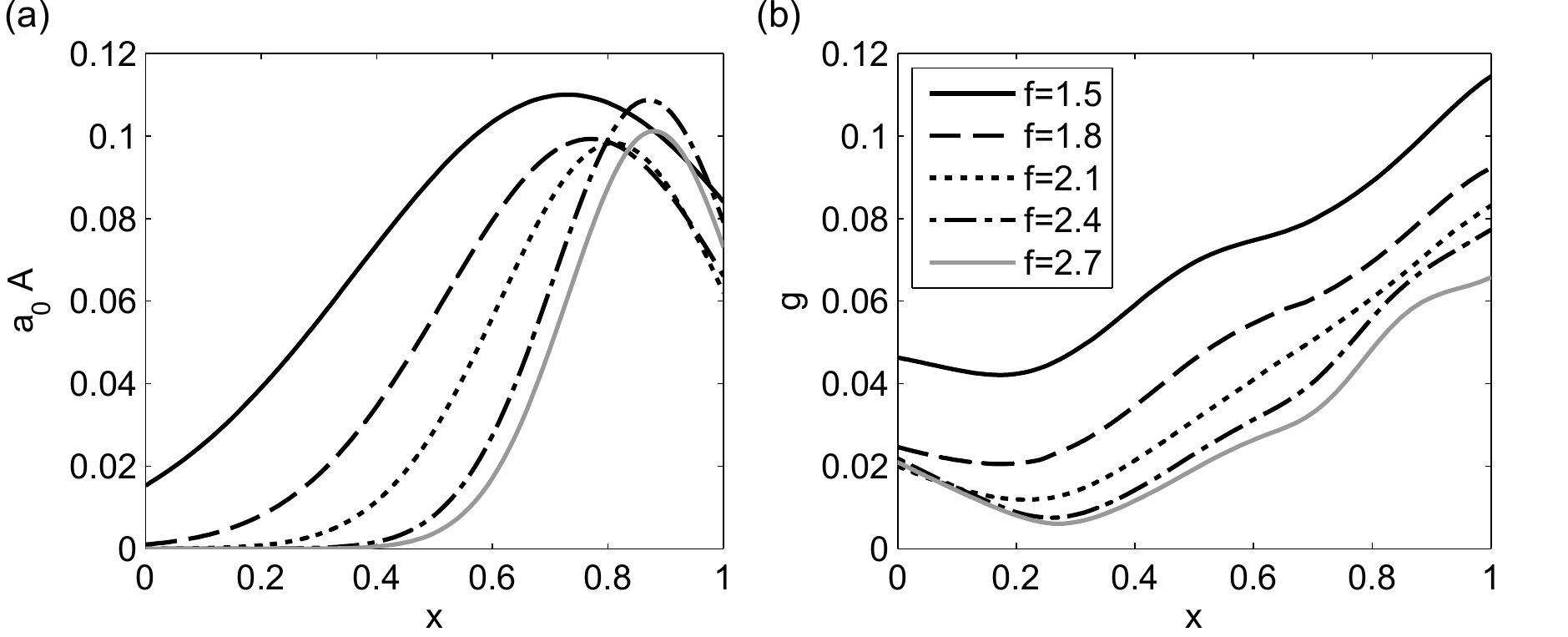}}
  \caption[Optimized (a) prescribed deformation envelopes and (b)
    displacement envelopes for the Gaussian
    parameterization.]{Optimized (a) prescribed deformation envelopes
    and (b) displacement envelopes for the Gaussian
    parametrization. $\lambda=1$ and
    $f=[1.5,\,1.8,\,2.1,\,2.4,\,2.7]$.}
\label{fig:opt_gauss_A}
\end{figure}

At low frequency, gaits with undulations of the entire body
($x_1=0.73$ and $\delta=0.52$ at $f=1.5$) are most efficient, while at
high frequency, the undulations should be restricted to a narrow
region ($\delta=0.21$ at $f=2.7$) located around $25\%$ of the
trailing edge ($x_1=0.88$ at $f=2.7$). However, for all frequencies,
the optimized deformation envelope $A(x)$, shown in figure
\ref{fig:opt_gauss_A}a, is qualitatively similar to the curvature
envelope from \citet{videler_fast_1984}, with a small amplitude at the
leading edge, a peak $10$ to $30\%$ from the trailing edge, and a
sharp decrease in amplitude at the trailing edge. Moreover, for all
frequencies, the amplitude of the peak is very close to $0.1$.

The corresponding displacement envelopes $g(x)$ are shown in figure
\ref{fig:opt_gauss_A}b. The displacement envelopes are qualitatively
similar to the carangiform displacement envelope from
\citet{videler_fast_1984}, with a minimum amplitude around $x=0.25$
and a maximum amplitude at the trailing edge. While the amplitude at
the leading edge decreases by a factor of two from $f=1.5$ to $f=1.8$,
it remains almost constant for $f$ from $2.1$ to $2.7$ with a value
$g(0)=0.02$ very close to that of \citet{videler_fast_1984}.

It is interesting to note that, since quadratic envelopes can only
result in functions with a wide peak, they can reach the same
efficiency as the wide peak Gaussian envelopes at low frequency
($f=1.5$), but not at high frequency ($f=2$) where a sharp peak is
advantageous.

%%%%%%%%%%%%%%%%%%%%%
\subsection{Characterization of efficient swimming gaits for a two-dimensional fish}

We showed in the previous section that, by changing the location and
width of the peak in a Gaussian deformation envelope, a very efficient
gait can be designed for a large range of undulation frequencies.

\begin{figure}
 \centerline{\includegraphics[width=1\textwidth]{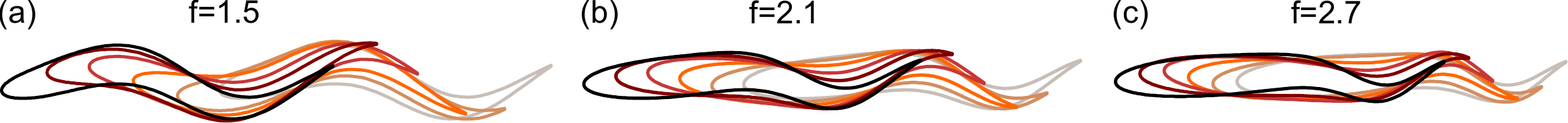}}
  \caption{Superimposed body outlines over one undulation period for
    three frequencies.}
\label{fig:body_motion_gauss}
\end{figure}

Figure \ref{fig:body_motion_gauss} shows the deformed fish for three
optimized gaits. As expected from figure \ref{fig:opt_gauss_A}, at
$f=1.5$ the entire length of the fish undergoes noticeable deformation
and displacement, resulting in a swimming motion that is similar to
that of an anguilliform swimmer, with a moderate curvature along the
entire body: the deformation of the fish matches the large wavelength
trajectory of the trailing edge and thus avoids the efficiency loss
associated with a large angle of attack. At higher frequency, the
front half of the fish undergoes virtually no deformation, resulting
in a swimming motion very similar to a carangiform ($f=2.1$) or even a
thuniform ($f=2.7$) swimmer. At this frequency, the region that would
correspond to the peduncle deforms with a very large curvature caused
by the sharp peak in the Gaussian envelope. This allows the
deformation of the fish to match the small wavelength trajectory of
the trailing edge and thus avoid the efficiency loss associated with a
large angle of attack.
%%$$

\begin{figure}
 \centerline{\includegraphics[width=1\textwidth]{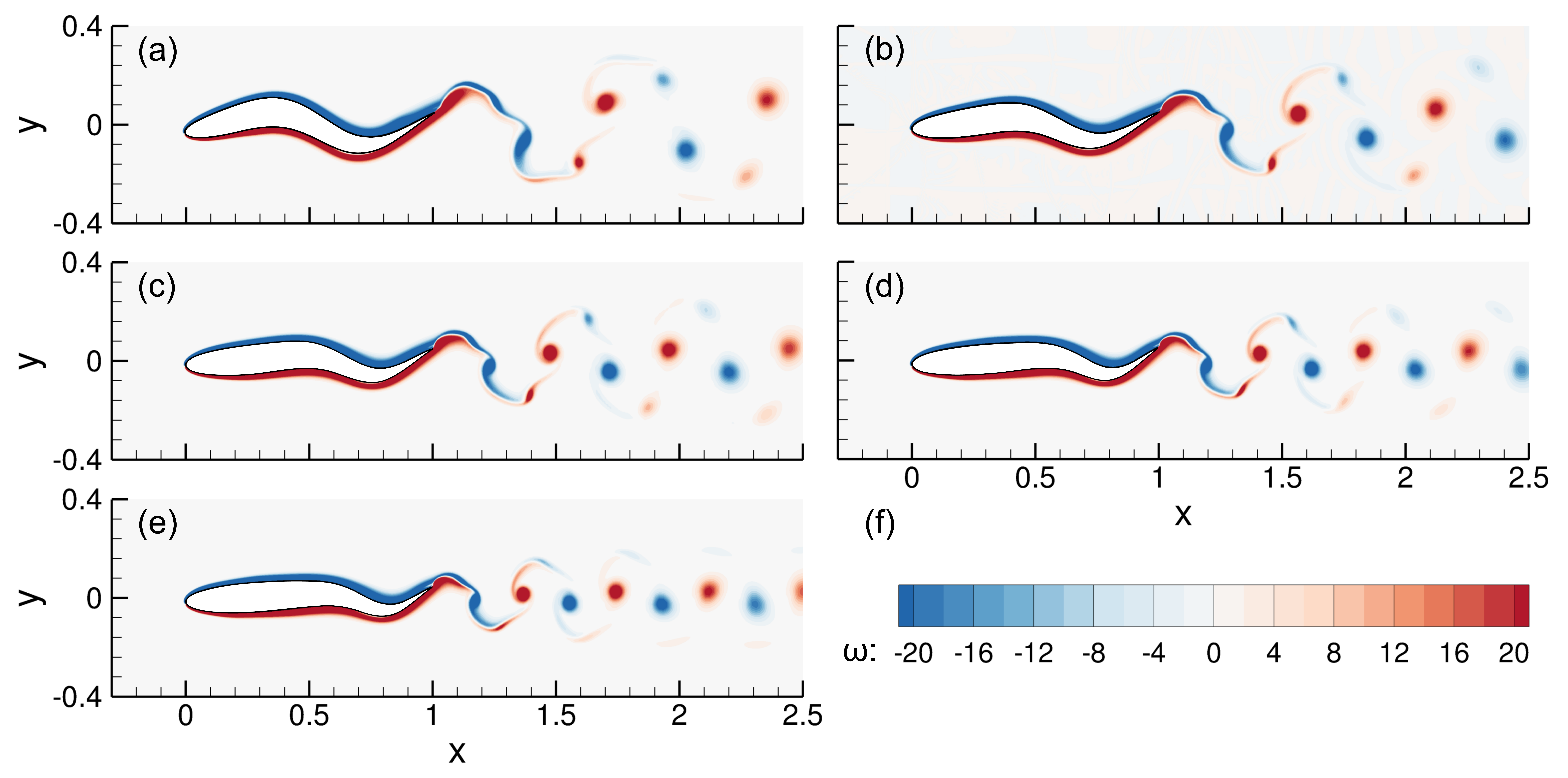}}
  \caption[Snapshots of vorticity for optimized gaits at
    $t/T=0\ (\text{mod } 1)$.]{Snapshots of vorticity for optimized
    gaits at $t/T=0\ (\text{mod } 1)$. (a): $f=1.5$, (b): $f=1.8$,
    (c): $f=2.1$, (d): $f=2.4$, (e): $f=2.7$, (f): colorbar.}
\label{fig:opt_gauss_vort}
\end{figure}

Figure \ref{fig:opt_gauss_vort} shows the deformed fish and vorticity
snapshots for the five optimized gaits at $t/T=0\ (\text{mod } 1)$,
where $T=1/f$ is the undulation period.  With a Gaussian deformation
envelope, a peak width specifically tailored to the undulation
frequency allows for reduced angle of attack at all frequencies. This
helps the boundary layer remain attached, as previously observed for
waves traveling faster than the free stream \citep{taneda_visual_1977,
  shen_turbulent_2003}. As for thrust-producing flapping foils, a
reverse \karman vortex street forms in the wake. The width and
wavelength of the reverse \karman vortex street decreases with
increasing undulation frequency, and secondary small vortices develop
at low frequency.

Table \ref{tab:opt_gauss} summarizes the parameters and properties of
the five optimized gaits. The quasi-propulsive efficiency $\eta_{QP}$
of these undulatory gaits is of prime interest. The efficiency reaches
$57\%$ for $f=2.7$, whereas the least efficient frequency, $f=1.5$
reaches $\eta_{QP}=49\%$. An other important parameter is the Strouhal
number, which is close to $St=0.35$. The consistency of the Strouhal
number for the optimized envelopes across frequencies suggests that,
for a given Reynolds number, there exists an optimal Strouhal number
that can be reached with a large range of frequencies. Like the
Strouhal number, the maximum pitch angle $\theta_{\max}$ and maximum
angle of attack $\alpha_{\max}$ are almost constant across the five
optimized gaits, with a value close to $\theta_{\max}=31^{\circ}$ and
$\alpha_{\max}=17^{\circ}$. The corresponding phase angle between the
heave and pitch of the trailing edge is $\psi = 82^{\circ}$. The
results from this optimization show that, as in rigid flapping foils,
the efficiency of undulating fish is primarily driven by the Strouhal
number, angle of attack, heave motion (or pitch motion), and
heave-pitch phase angle, all at the trailing edge. The optimal
Strouhal number, pitch angle, and angle of attack can be attained by
tuning the envelope peak for each frequency.

\begin{table}
    \begin{subtable}[t]{\textwidth}
        \centering
  \begin{tabular*}{\textwidth}{@{\extracolsep{\fill}} ccccccccccc}
  \toprule
      $f$  & $x_1$   & $\delta$ & $a_0$ & $a$ & $\theta_{\max}(^{\circ})$ &  $\alpha_{\max}(^{\circ})$ & $\psi(^{\circ})$ & $St$ & $\overline{C_P}$ & $\eta_{QP}$\\
      \midrule
       $1.5$  & $0.73$ & $0.52$ & $0.084$ & $0.23$ & $31$ &  $17$ & $82$ & $0.34$ & $0.093$ & $0.49$ \\
       $1.8$  & $0.77$ & $0.36$ & $0.066$ & $0.18$ & $28$ &  $19$ & $82$ & $0.33$ & $0.089$ & $0.52$ \\
       $2.1$  & $0.81$ & $0.28$ & $0.062$ & $0.16$ & $29$ &  $20$ & $81$ & $0.35$ & $0.087$ & $0.53$ \\
       $2.4$  & $0.87$ & $0.23$ & $0.079$ & $0.15$ & $35$ &  $16$ & $82$ & $0.37$ & $0.083$ & $0.56$ \\
       $2.7$  & $0.88$ & $0.21$ & $0.073$ & $0.13$ & $34$ &  $15$ & $84$ & $0.36$ & $0.081$ & $0.57$ \\
       \bottomrule  
  \end{tabular*}
  \caption{Optimized envelopes at several frequencies.}
  \label{tab:opt_gauss}
    \end{subtable}
    
    \begin{subtable}[b]{\textwidth}
        \centering
        \vspace{0.35cm}
  \begin{tabular*}{1\textwidth}{@{\extracolsep{\fill}} ccccccccccc}
  \toprule
     $f$  & $x_1$   & $\delta$ & $a_0$ & $a$ & $\theta_{\max}(^{\circ})$ &  $\alpha_{\max}(^{\circ})$ & $\psi(^{\circ})$ & $St$ & $\overline{C_P}$ & $\eta_{QP}$\\
      \midrule
       $1.8$ & $0.65$ & $0.50$ & $0.050$ & $0.17$ & $22$ & $22$ & $82$ & $0.30$ & $0.097$ & $0.45$ \\
       $1.8$ & $0.78$ & $0.49$ & $0.068$ & $0.18$ & $26$ & $21$ & $78$ & $0.32$ & $0.093$ & $0.47$ \\
       $1.8$ & $0.80$ & $0.29$ & $0.082$ & $0.22$ & $36$ & $17$ & $84$ & $0.40$ & $0.096$ & $0.46$ \\
       $1.8$ & $0.85$ & $0.31$ & $0.101$ & $0.22$ & $37$ & $16$ & $82$ & $0.40$ & $0.095$ & $0.46$ \\
       $1.8$ & $0.90$ & $0.37$ & $0.103$ & $0.21$ & $35$ & $19$ & $78$ & $0.38$ & $0.094$ & $0.46$ \\
       $1.8$ & $0.90$ & $0.25$ & $0.186$ & $0.32$ & $53$ & $10$ & $87$ & $0.57$ & $0.140$ & $0.31$ \\
       \bottomrule  
  \end{tabular*}
  \caption{Examples of envelopes around the optimal gait at $f=1.8$.}
  \label{tab:nonopt_gauss}
    \end{subtable}
    \caption[Parameters and properties of gaits with Gaussian
      envelopes.]{Parameters and properties of gaits with Gaussian
      envelopes. Motion parameters are the frequency $f$, peak
      location $x_1$, peak width $\delta$ and amplitude
      $a_0$. Properties are the peak to peak displacement amplitude at
      the trailing edge $a$, maximum pitch angle at the trailing edge
      $\theta_{\max}$, maximum angle of attack $\alpha_{\max}$, heave
      and pitch phase angle $\psi$, Strouhal number $St$,
      time-averaged power coefficient $\overline{C_P}$ and the
      quasi-propulsive efficiency $\eta_{QP}$.}
\end{table}

In order to better understand the impact of $x_1$ and $\delta$ on the
gait properties, table \ref{tab:nonopt_gauss} summarizes these
properties for several values of $x_1$ and $\delta$ near the optimum
for $f=1.8$. As the location of the peak moves aft and its width
decreases, the portion of the fish undergoing significant deformation
reduces, therefore a larger amplitude is necessary to ensure that
enough thrust is produced. As a result, the Strouhal number and
maximum pitch angle increase. This observation also allows us to
interpret the optimization results. For a fixed envelope $A(x)$, the
Strouhal number of a self-propelled undulating fish increases with
decreasing frequency. In order to mitigate this effect, an envelope
with widespread undulations (small $x_1$ and large $\delta$) that can
produce the same thrust with smaller amplitude makes it possible to
reach the optimal Strouhal number even at low frequency. Similarly, at
high Reynolds number, a large $x_1$ and a small $\delta$ make it
possible to generate the required thrust at the optimal Strouhal
number. When the undulation frequency reaches about $2.5$, the optimal
envelope parameters reach a plateau at $x_1 \approx 0.9$ and $\delta
\approx 0.2$.

\begin{figure}
 \centerline{\includegraphics[width=1\textwidth]{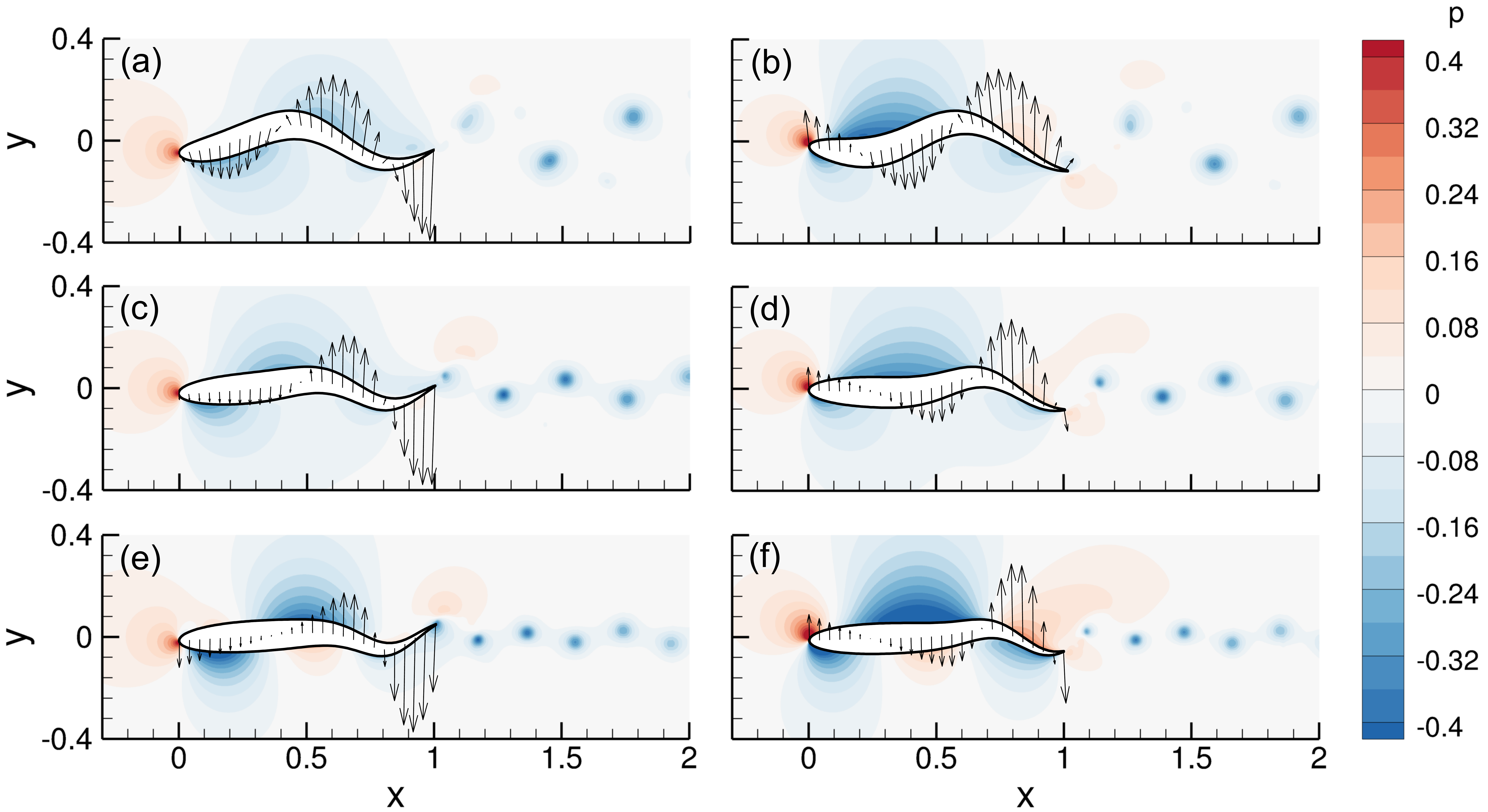}}
  \caption[Snapshots of pressure field with arrows showing the body
    velocity.]{Snapshots of pressure field with arrows showing the
    body velocity. (a, b): optimized Gaussian envelope at $f=1.5$; (c,
    d): optimized Gaussian envelope at $f=2.1$; (e, f): optimized
    Gaussian envelope at $f=2.7$. (a, c): $t/T=0.12\ (\text{mod } 1)$
    (minimum power for $f=1.5$ and $f=2.1$); (e): $t/T=0\ (\text{mod }
    1)$ (minimum power for $f=2.7$); (b, d): $t/T=0.34\ (\text{mod }
    1)$ (maximum power for $f=1.5$ and $f=2.1$; (f):
    $t/T=0.29\ (\text{mod } 1)$ (maximum power for $f=2.7$).}
\label{fig:opt_gauss_press}
\end{figure}

Figure \ref{fig:opt_gauss_press} shows the pressure field and body
velocity for the optimized envelopes with frequency
$f=[1.5,\,2.1,\,2.7]$ at their respective time of minimum and maximum
power. For $f=1.5$ (figures \ref{fig:opt_gauss_press}a,b) and $f=2.7$
(figures \ref{fig:opt_gauss_press}e,f), there are three distinct
sections along the upper side of the fish: high pressure near the
leading edge, low pressure in the middle and high pressure near the
trailing edge (and the opposite on the other side).  In figures
\ref{fig:opt_gauss_press}b,f, these sections almost exactly match
transverse velocity of respectively positive, negative, and positive
sign, resulting in very large instantaneous swimming
power. Conversely, in figures \ref{fig:opt_gauss_press}a,e, the sign
of the transverse velocity is reversed, resulting in a significant
negative swimming power. For $f=2.1$, the pressure changes along the
fish are smaller, and the pressure is close to zero along a large
portion of the fish. Moreover, unlike for $f=2.7$, the sign changes in
pressure do not match the sign changes in transverse velocity. For
instance, at $t/T=0$, the pressure along the bottom side of the fish
near the trailing edge is positive (not shown here), which would
result in a positive swimming power. Therefore, the minimum power is
reached at a later time $t/T=0.12$, at which point the amplitude is
largest in areas where the pressure is close to zero, resulting in a
very small power. Similarly, the maximum power reached at $t/T=0.34$
is not as large as for $f=2.7$ because the sections of high pressure
do not exactly match the sections of large transverse velocity.

It must be pointed out that there are additional parameters affecting
the efficiency of an undulating fish, since the efficiency ranges from
$\eta_{QP}=0.49$ at $f=1.5$ to $\eta_{QP}=0.57$ at $f=2.7$.
\citet{shen_turbulent_2003} found that a slip ratio around $s_r=0.8$
($f=1.2$) is optimal for a body undergoing traveling wave motion of
constant amplitude, in order to reduce separation effects and
turbulence intensity. In our case, however, these results do not
strictly apply because the undulations have a non-constant envelope,
and especially for higher frequency are confined to a small section of
the fish.

%%%%%%%%%%%%%%%%%%%
\subsection{Application to a three-dimensional danio-shaped swimmer}

\begin{figure}
 \centerline{\includegraphics[width=0.6\textwidth]{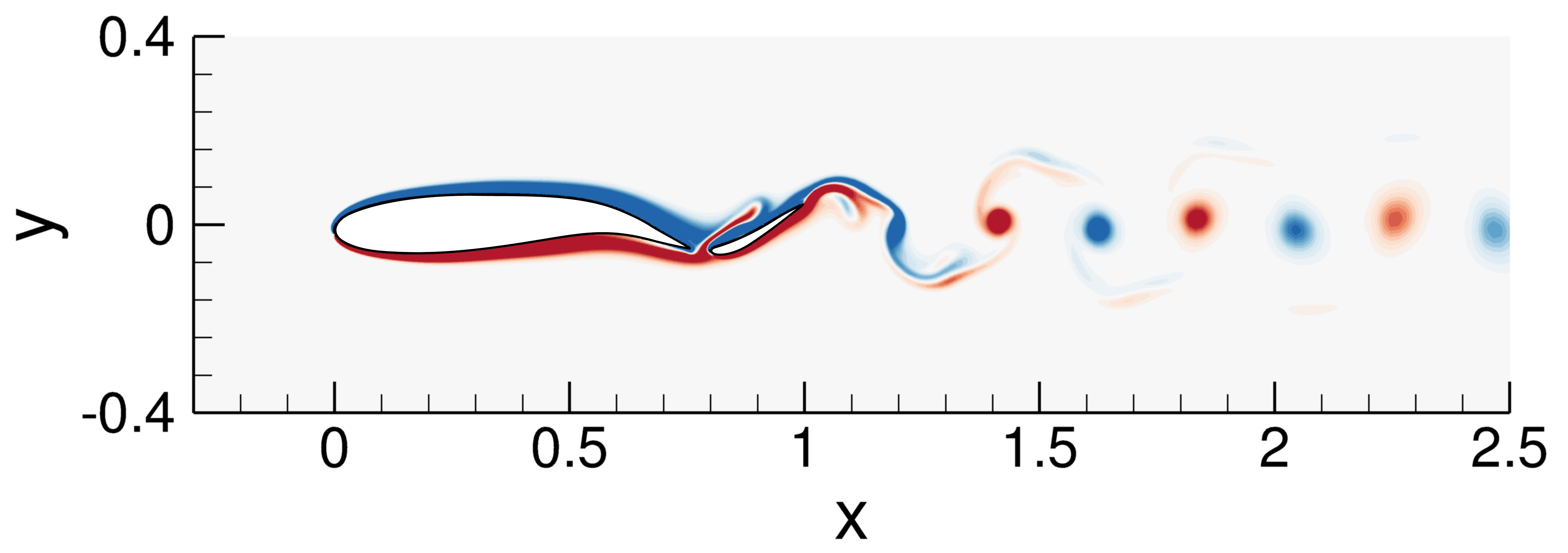}}
  \caption{Snapshot of the vorticity field around a two-dimensional
    fish with a separate tail fin.}
\label{fig:foil_tail}
\end{figure}

We have so far modeled a fish by a two-dimensional fish-like flexible
foil. However, fish have a highly three-dimensional geometry. In
particular, most carangiform and thunniform swimmers are characterized
by a region of reduced depth, around $20\%$ from the trailing edge,
the peduncle. In order to model this region of reduced added mass with
a two-dimensional geometry, it might be more appropriate to model a
fish with a separate foil for the tail, as illustrated in figure
\ref{fig:foil_tail}. The fish model shown in this figure undulates
with the optimized gait at frequency $f=2.4$ identified earlier, and
the performance ($\eta_{QP}=0.54$) is very close to that obtained with
a single foil, indicating that the results are robust to changes in
the geometry.

\begin{figure}
 \centerline{\includegraphics[width=0.75\textwidth]{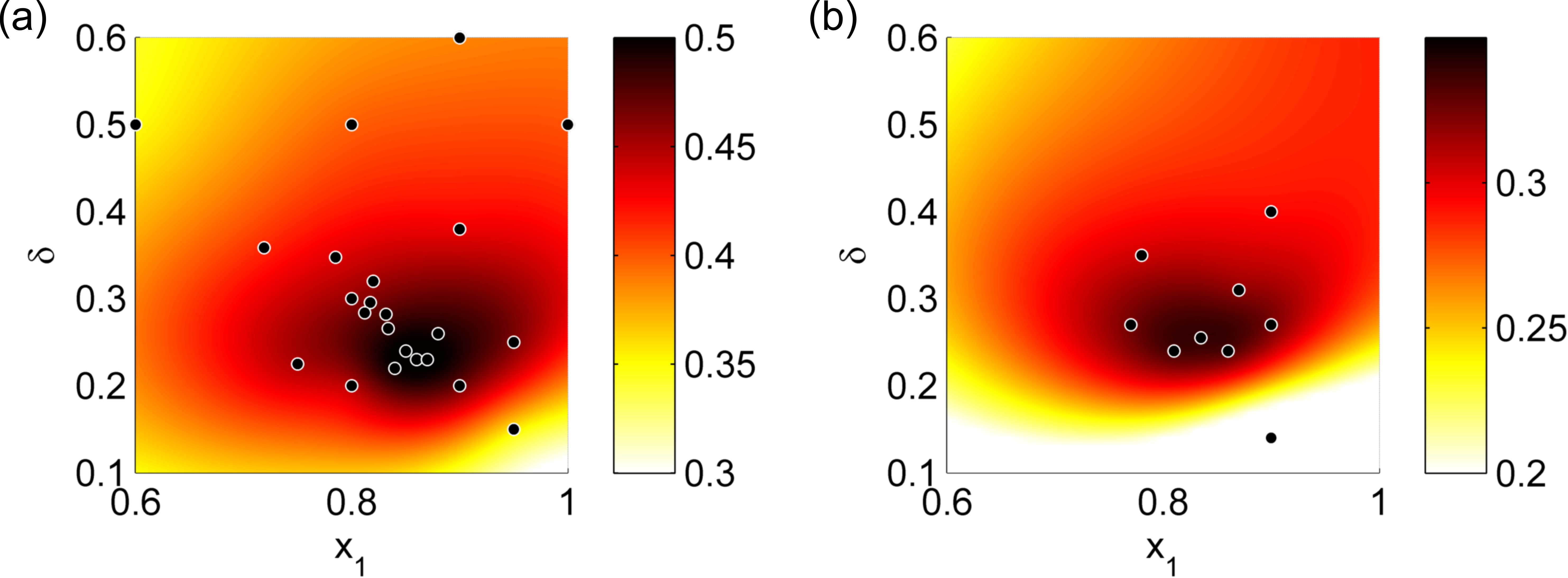}}
  \caption[$\eta _{QP}$ as a function of $x_1$ and $\delta$ near the
    optimum for (a) 2D and (b) 3D geometries with $f=2.4$.]{$\eta
    _{QP}$ as a function of $x_1$ and $\delta$ near the optimum for
    (a) $2D$ and (b) $3D$ geometries with $f=2.4$. The black dots show
    the location of the points that have been used to build the
    thin-plate smoothing spline ($\mathsf{tpaps}$ function in Matlab
    with smoothing parameter $p=0.999$) represented in color.}
\label{fig:optim_3D}
\end{figure}

In the rest of this section, we consider a simplified
three-dimensional fish shape, shown in figure \ref{fig:danio}, which
is based on a giant danio (\emph{Devario aequipinnatu}).  For this
geometry, we fix the undulation frequency to $f=2.4$ and optimize a
Gaussian envelope for quasi-propulsive efficiency (for a fixed
swimming speed $U_s$, we minimize the expanded power
$\overline{P_{in}}$). In figure \ref{fig:optim_3D} we compare how
$\eta_{QP}$ changes with the envelope parameters $x_1$ and $\delta$
for a two-dimensional fish and for the three-dimensional shape. The
efficiency is generally lower with the three-dimensional shape, but
the dependence on $x_1$ and $\delta$ is very similar for both
geometries: the most efficient gaits are for $0.8<x_1<0.9$ and
$0.2<\delta<0.3$ with a sharp decrease in efficiency for
$\delta<0.2$. This shows that, even though three-dimensional effects
reduce the swimming efficiency, most of the conclusions about BCF
swimming drawn from the two-dimensional study extend to
three-dimensional shapes.

The parameters and properties of the optimized gait for $f=2.4$ are
compared to those of the carangiform gait in table \ref{tab:3D}. Like
in the 2D case, the optimization decreases the power consumption by
$50\%$ compared with the carangiform gait. As in 2D, the optimized
gait manages to bring the phase angle between the heave and pitch
motion of the trailing edge close to $90^\circ$, which significantly
reduces the angle of attack. As a result, the optimized gait for the
three-dimensional fish shape have a pitch angle, phase angle and angle
of attack very close to the optimized gait for the two-dimensional
fish. However, since the 3D effects reduce the thrust produced by the
undulating motion, the Strouhal number is higher than in 2D,
especially for the carangiform gait.

\begin{table}
  \begin{center}
  \begin{tabular*}{1\textwidth}{@{\extracolsep{\fill}} ccccccccccc}
  \toprule
      $f$  & $x_1$   & $\delta$ & $a_0$ & $a$ & $\theta_{\max}(^{\circ})$ &  $\alpha_{\max}(^{\circ})$ & $\psi(^{\circ})$ & $St$ & $\overline{C_P}$ &  $\eta_{QP}$\\
      \midrule
      $3.0$ & \multicolumn{2}{c}{carangiform} & $0.099$ & $0.18$ & $34$ & $41$ & $59$ & $0.53$ & $0.035$  & $0.22$ \\
       $2.4$ & $0.84$ & $0.26$ & $0.085$ & $0.18$ & $37$ & $17$ & $87$ & $0.43$ & $0.023$ & $0.34$ \\
       \bottomrule  
  \end{tabular*}
  \caption[Parameters and properties of 3D undulating
    gaits.]{Parameters and properties of 3D undulating
    gaits. Properties are the peak to peak displacement amplitude at
    the trailing edge $a$, maximum pitch angle at the trailing edge
    $\theta_{\max}$, maximum angle of attack $\alpha_{\max}$, heave
    and pitch phase angle $\psi$, Strouhal number $St$, time-averaged
    power coefficient $\overline{C_P}$, and the quasi-propulsive
    efficiency $\eta_{QP}$. The optimized gait at $f=2.4$ is compared
    to the carangiform gait at $f=3$.}
  \label{tab:3D}
  \end{center}
\end{table}

\begin{figure}
 \centerline{\includegraphics[width=0.85\textwidth]{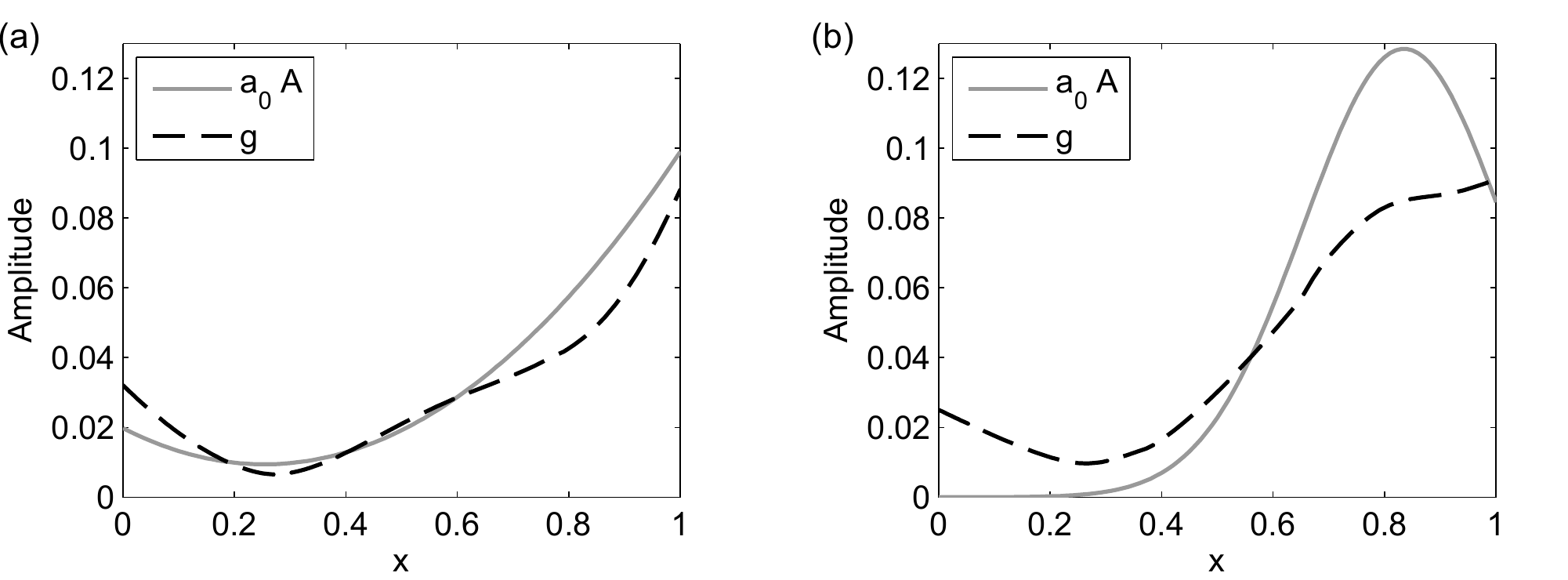}}
  \caption{Prescribed deformation envelope $a_0 A(x)$ and displacement
    envelope $g(x)$ for (a) carangiform gait with $f=3$ and (b)
    optimized gait with $f=2.4$.}
\label{fig:amp_3D}
\end{figure}

\begin{figure}
 \centerline{\includegraphics[width=0.8\textwidth]{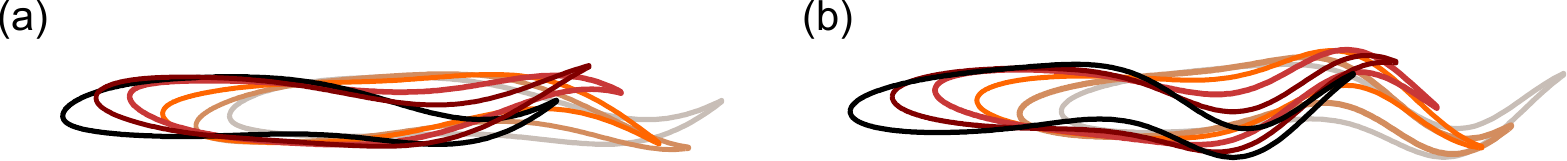}}
  \caption{Superimposed body outlines over one undulation period for
    (a) the carangiform motion and (b) the optimized gait.}
\label{fig:3D_motion}
\end{figure}

Figure \ref{fig:amp_3D} shows the deformation envelope $A(x)$ and the
displacement envelope $g(x)$ for the carangiform gait at $f=3$ and for
the optimized gait.  The superimposed body outlines for the optimized
gait shown in figure \ref{fig:3D_motion}b also look very similar to
the body outlines of the optimized motions in 2D: the deformation of
the tail follows the trajectory of the trailing edge, resulting in an
efficient low angle of attack. The body outlines for the carangiform
motion, on the other hand, show that the pitch of the tail is out of
phase with its velocity (phase angle far from $90^{\circ}$), which
results in a very inefficient gait, with a large angle of attack.
%%$$

\begin{figure}
 \centerline{\includegraphics[width=0.9\textwidth]{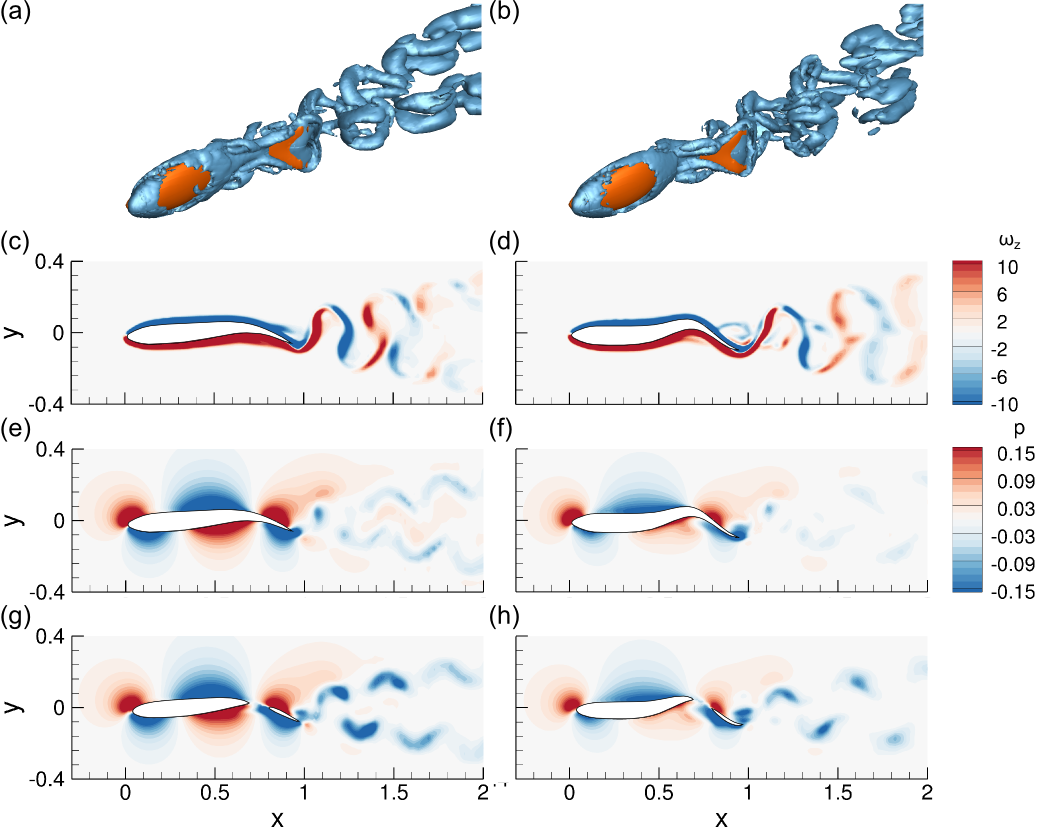}}
  \caption[Snapshots of the flow around a three-dimensional fish with
    a carangiform and optimized gait.]{Snapshots of the flow around a
    three-dimensional fish with (a,c,e,f) a carangiform and (b,d,g,h)
    an optimized gait . (a,b): Three-dimensional vortical structures
    visualized using the $\lambda_2$-criterion; (c,d): $z$ component
    of the vorticity in the $z=0$ plane; (e,f): pressure in the $z=0$
    plane; (g,h): pressure in the $z=0.06$ plane.}
\label{fig:3D_snapshots}
\end{figure}

Finally, we show the flow structure around the 3D fish model for both
gaits in figure \ref{fig:3D_snapshots}. The performance difference
between the two gaits is accompanied by noticeable differences in the
wake structure of the two swimmers. For both gaits, figures
\ref{fig:3D_snapshots}a and \ref{fig:3D_snapshots}b show wakes
comprised of two interconnected vortex loops per cycle, together with
other smaller structures. In particular, the structure in the wake of
the optimized motion is complex, with many vortex tubes interlaced
with each other. Indeed, as can also be seen in the vorticity field at
$z=0$ (figure \ref{fig:3D_snapshots}d), the deformation at the
peduncle is quite large for the optimized gait, resulting in vortex
tubes separating from the main body and then interacting with the
structures shed from the tail.

\citet{borazjani_role_2010} also observed in their 3D simulations
that, for Strouhal number greater than $St=0.3$, the wake structure
observed in 2D, dominated by a single vortex pair (or ring in 3D),
transitions to vortex loops wrapping around each
other. \citet{dong_wake_2006} showed that the same phenomenon happens
to elliptical flapping foils of finite aspect ratio: at low aspect
ratio/large Strouhal number, two vortex rings are shed each cycle. As
the aspect ratio increases or the Strouhal number decreases, the tip
vortices do not merge together any more and the wake consists of
interconnected loops. As the Strouhal number further decreases or the
aspect ratio increases, the three-dimensional effects become even
weaker and the linkage between tip vortices disappears. At this point,
the 3D wake looks similar to the (reverse) \karman vortex street
observed in 2D.

In the carangiform example shown here, the tip vortices merge, while
with the optimized gait, which has a lower Strouhal number and angle
of attack, they do not. At higher Reynolds number, the Strouhal number
would be smaller and a wake similar to that observed in 2D would
probably emerge. Figure \ref{fig:3D_snapshots}c,d shows that, near the
tail, the vorticity in the $z=0$ plane looks very similar to that
behind a 2D foil. However, under the influence of the tip vortices,
the vortex sheets shed by the tail do not evolve into two strong
vortices as in 2D. As a result, whereas the pressure field around the
undulating fish-like shape is very similar to the pressure around an
undulating airfoil, the pressure signature in the wake shown at the
plane $z=0$ is very weak (\ref{fig:3D_snapshots}e,f). However, the
pressure signature in the plane $z=0.06$, just above the peduncle is
much stronger (\ref{fig:3D_snapshots}g,h).

\begin{figure}
 \centerline{\includegraphics[width=0.9\textwidth]{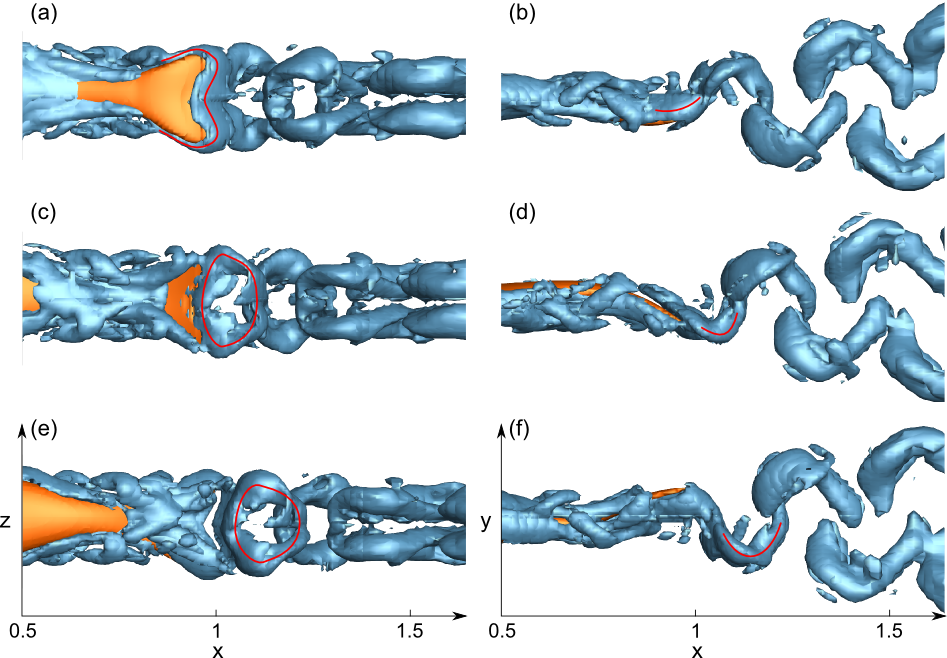}}
  \caption{(a,c,e) Side-view and (b,d,f) top-view of the vortex
    structures at several time-steps for the carangiform gait. (a,b):
    $t/T=0.1\ (\text{mod } 1)$; (c,d): $t/T=0.4\ (\text{mod } 1)$;
    (e,f): $t/T=0.7\ (\text{mod } 1)$. A red line shows the formation
    of a vortex ring.}
\label{fig:3D_Videler_zoom}
\end{figure}

Figure \ref{fig:3D_Videler_zoom} shows a magnified view of the vortex
structures generated by the carangiform motion. A red line shows the
formation of a clear vortex ring at the trailing edge of the tail
between figures \ref{fig:3D_Videler_zoom}a and
\ref{fig:3D_Videler_zoom}c. In figure \ref{fig:3D_Videler_zoom}e, the
vortex ring is fully formed and detached from the tail. Since the
vortex rings are oblique, they produce a large transverse velocity,
which is inefficient and results in waste of energy. We also see a
spanwise narrowing of the vortex rings as they convect downstream, as
also observed in the simulations of \citet{blondeaux_numerical_2005}
and \citet{dong_wake_2006} for a respectively rectangular and
elliptical pitching and heaving foil.

\begin{figure}
 \centerline{\includegraphics[width=0.9\textwidth]{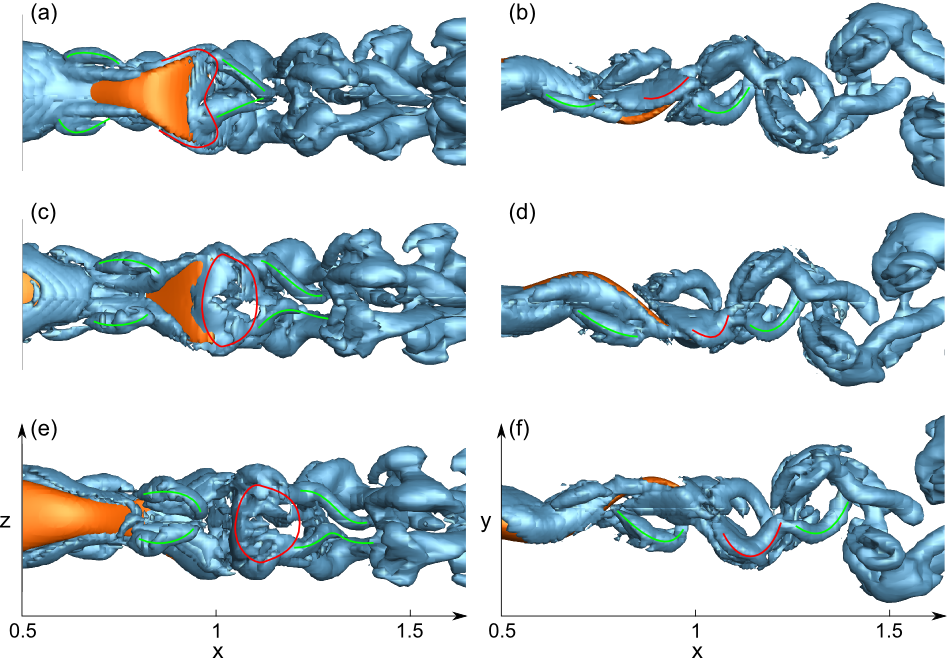}}
  \caption{(a,c,e) Side-view and (b,d,f) top-view of the vortex
    structures at several time-steps for the optimized gait. (a,b):
    $t/T=0.1\ (\text{mod } 1$; (c,d): $t/T=0.4\ (\text{mod } 1$;
    (e,f): $t/T=0.7\ (\text{mod } 1$. A red line shows a vortex shed
    from the tail that never fully develops into a ring, while green
    lines show the vortices shed from the body.}
\label{fig:3D_opt_zoom}
\end{figure}

Figure \ref{fig:3D_opt_zoom} shows a magnified view of the vortex
structures generated by the optimized gait. The structure of the wake
is more intricate than for the carangiform motion. In particular,
instead of one set of interconnected vortex tubes, there are two sets
of tubes, marked in red and green in the figure. The loop marked in
red is the same as observed for the carangiform gait, but at this
lower Strouhal number, it never fully closes into a clearly defined
vortex ring. The tubes marked in green are formed upstream of the tail
and are shed from the body as a result of the large curvature at the
peduncle. The resulting vortex tubes are interlaced with the vortex
loops from the tail with which they have a phase difference of close
to $180^\circ$.

For a three-dimensional fish shape with two-dimensional undulation, as
for a two-dimensional foil, the Strouhal number, pitch angle (or heave
motion), nominal angle of attack and phase angle at the trailing edge
are the key parameters for efficient swimming. The optimization
results in a lower Strouhal number and angle of attack, which reduces
the three-dimensional effects observed behind the non-optimized gait,
such as formation of inefficient oblique vortex ring chains. With the
optimized gait, the production of thrust is also distributed between
the body and the tail, which shedd vortex structures with opposite
phase. It has been shown with turbines, for instance, that
distributing the thrust production (or energy capture) could
significantly improve the efficiency, and it is possible that fish use
the same process. Finally, while we used a simplified fish geometry
with a two-dimensional undulation, fish can also rely on
three-dimensional motion of their dorsal and pectoral fins to save
energy \citep{lauder_fish_2007, drucker_locomotor_2001}.

%%%%%%%%%%%%%%%%%%%%%%%%%%%%%%%%%%%%%%%%%%%%%%%%%%%%%%%%%%%%%%%%%%%%%%%%%
\section{Energy saving in two fish swimming in close proximity \label{sec:2fish}}

The experimental study by \citet{gopalkrishnan_active_1994} and the
theoretical study by \citet{streitlien_efficient_1996} demonstrated
that flapping rigid foils placed within a \karman street can extract
significant energy from the flow through vorticity control.
Subsequently, it was documented that live fish swimming within a
\karman vortex street formed behind a cylinder tend to synchronize
their motion to the oncoming cylinder vortices.  This allows them to
significantly reduce the energy spent to hold station
\citep{liao_karman_2003, liao_fish_2003, akanyeti_kinematic_2013}, or
even generate propulsive force with no input power as evidenced by the
passive propulsion of anaesthesized fish
\citep{beal_passive_2006}. The phenomenon of fish \karman gaiting has
been explained by the faculty of fish to sense and harness the energy
of the vortices.

Since a foil or a fish can extract energy from the vortices in a
\karman street, in principle there is no reason why they could not
extract energy from the vortices in a reverse \karman street.
\citet{boschitsch_propulsive_2014} recently showed that the net
propulsive efficiency of a pitching foil located behind a similarly
pitching foil could be anywhere between $0.5$ and $1.5$ times that of
an isolated foil, depending on the phase. This indicates that the
energy extracted from the vortices in a reverse \karman street more
than compensates for the effect of increased drag caused by the jet
forming in a reverse \karman street (unlike the energetically
beneficial drag wake forming behind a bluff body). Despite the strong
evidence that it is possible to harness the energy of individual
vortices within a reverse \karman street, there is no conclusive
evidence that fish actually do harness this energy. Liao summarizes in
his review of fish swimming in altered flows that “no hydrodynamic or
physiological data exist to evaluate the hypothesis that fish can
increase swimming performance by taking advantage of the wake of other
members” \citep{liao_review_2007}.

Due to the difficulty of experimentally measuring the swimming power
of individual fish in a school, simulations can provide valuable
information to help clarify this issue. Hence, we consider next a pair
of undulating fish-like foils. We have shown in the previous section
that a two-dimensional fish, undulating in open-water, can attain a
quasi-propulsive efficiency of almost $60 \%$ by optimizing its
gait. The goal in this section is to determine whether, by working as
a pair, fish can further reduce the power required to travel at
constant speed $U_s$.

%%%%%%%%%%%%%%%%%%%%%%%%%%%%%
\subsection{Flow in the wake of a self-propelled undulating fish}

\begin{figure}
 \centerline{\includegraphics[width=1\textwidth]{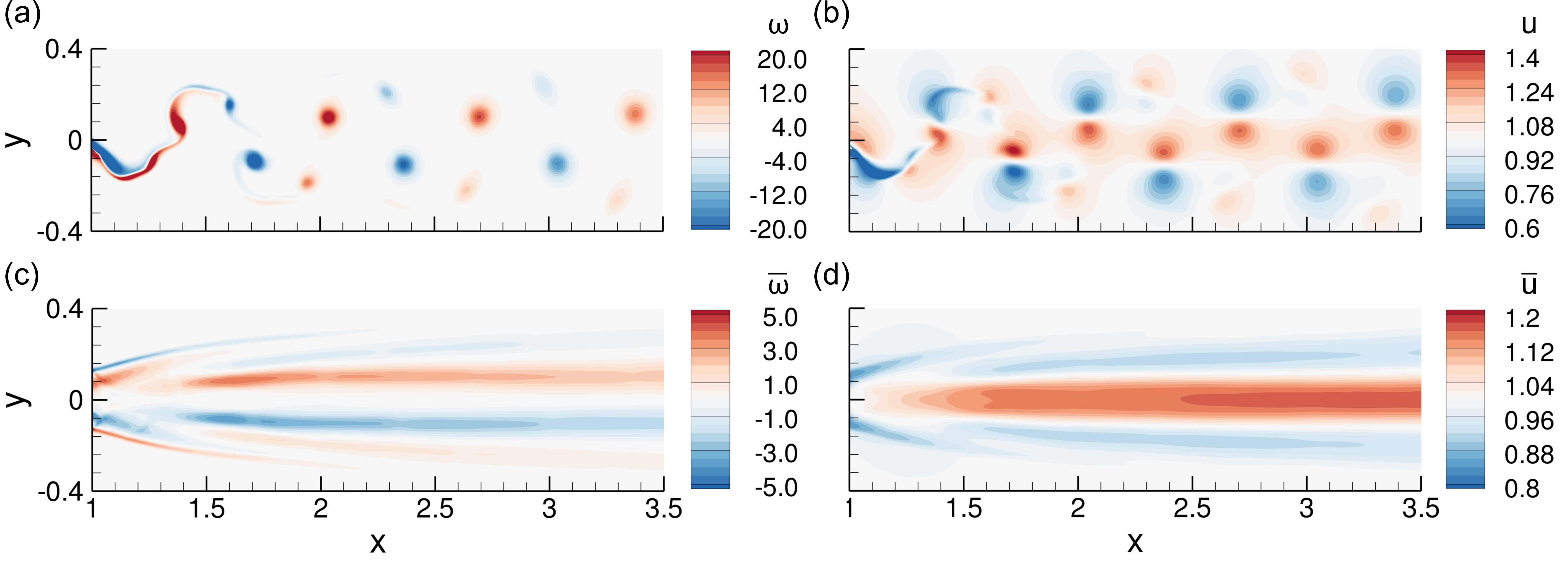}}
  \caption[Wake behind a self-propelled undulating fish for the
    optimized gait with Gaussian envelope and $\lambda=1$ at frequency
    $f=1.5$]{Wake behind a self-propelled undulating fish for the
    optimized gait with Gaussian envelope and $\lambda=1$ at frequency
    $f=1.5$. (a): instantaneous vorticity field; (b): instantaneous
    $x$-velocity field; (c): time-averaged vorticity field; (d):
    time-averaged $x$-velocity field.}
\label{fig:wake}
\end{figure}

The flow in the wake of a self-propelled undulating fish consists of
vortices of alternating sign. The vorticity snapshot in figure
\ref{fig:wake}a shows that the vortices decrease in strength under the
effect of diffusion, but this is a slow process and the wake is
primarily characterized by its periodicity. Figure \ref{fig:wake}b
shows that the vortices are arranged in such a way that the flow along
the $y=0$ axis is faster than the ambient flow (jet-like flow),
whereas the flow away from the centerline moves slower than the
ambient flow; consistent with the momentum-less wake, on average, of a
self-propelled body \citep{triantafyllou_optimal_1993}. As a result,
the time-averaged vorticity field in figure \ref{fig:wake}c is
characterized by four shear layers of alternating sign, resulting in a
jet along the centerline with strips of slowed-down flow on either
side, as shown in figure \ref{fig:wake}d.

\begin{figure}
 \centerline{\includegraphics[width=0.85\textwidth]{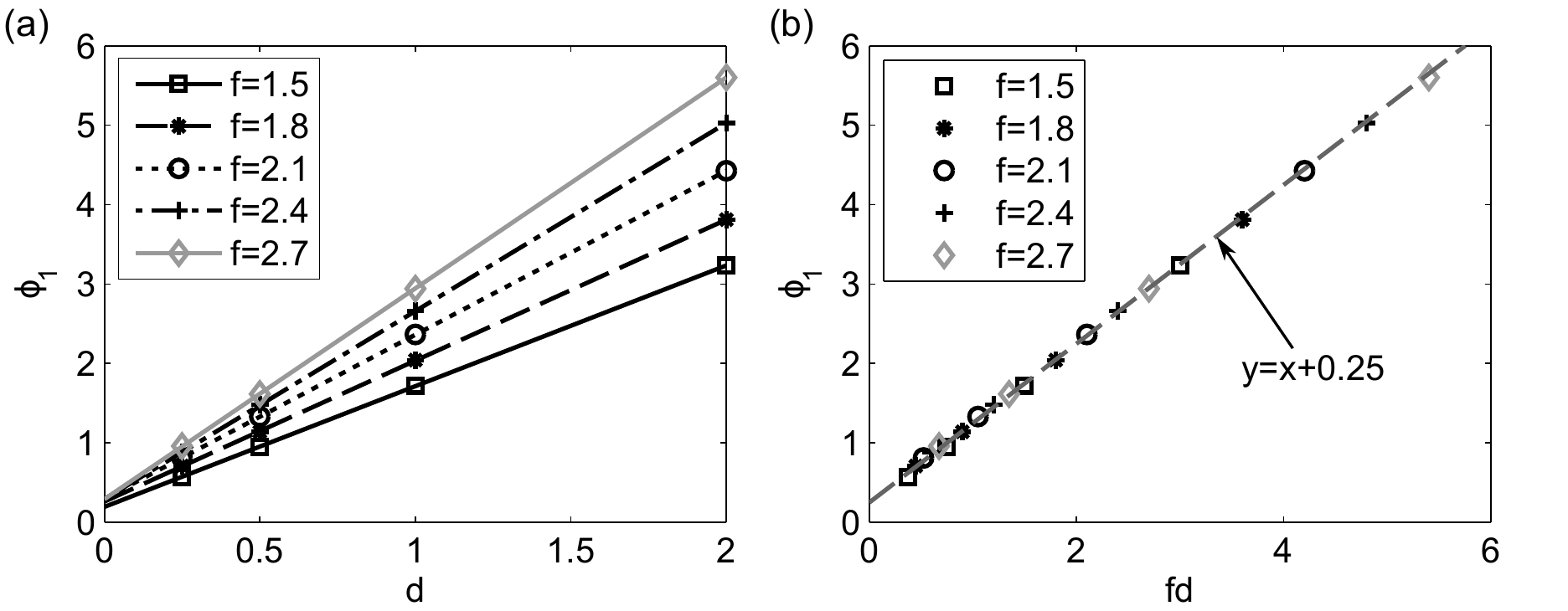}}
  \caption[Vorticity phase in the wake of a self-propelled undulating
    fish.]{Vorticity phase in the wake of self-propelled undulating
    fish as a function of (a) distance and (b) distance times
    frequency. For each frequency, the optimized gait with Gaussian
    envelope is used.}
\label{fig:wake_phase}
\end{figure}

The reverse \karman vortex street behind a self-propelled undulating
fish is characterized by its periodic structure with vortices moving
parallel to the $y=0$ axis in stable formation. The vorticity at
longitudinal distance $d$ from the trailing edge in the wake of an
undulating fish can be modeled as:
\begin{equation}
\omega(d,y,t)=\omega_y(y,t)\,\omega_d(d)\sin \big(2\pi\left(\phi_1(d) -ft\right)  \big), \qquad \phi_1(d)=d/\lambda_w + \phi_w,
\end{equation}
where the frequency $f$ is given by the undulation frequency and the
wavelength $\lambda_w$ and phase $\phi_w$ of the wake need to be
determined. For the five optimized gaits with Gaussian envelope and
$\lambda=1$ presented in \S\,\ref{sec:1fish}, we estimated from the
vorticity field the phase $\phi_1$ at several distances $d$ along the
wake. In figure \ref{fig:wake_phase}a we show the phase $\phi_1$ as a
function of the distance $d$, as well as the least squares linear fit
for each swimming gait. The coefficients for the linear fit are
summarized in table \ref{tab:fit}. For all the gaits, the phase is
essentially proportional to the distance $d$, with a coefficient of
proportionality very close to the undulation frequency $f$. Since
$\lambda_w = f c_w$, where $c_w$ is the speed at which the vortices
travel in the wake, this result shows that the vortices travel at the
same speed as the free-stream. Moreover, for the five gaits
considered, the phase at $d=0$ is around $0.25$, which means that the
vortices are shed by the fish when the trailing edge has maximum
transverse velocity. Finally, we confirm these observations by
plotting $\phi_1$ as a function of $fd$ in figure
\ref{fig:wake_phase}b. Assuming $c_w=1$, the least-squares estimate
($\pm$ standard deviation) of the phase $\phi_w$ is:
\begin{equation}
\phi_w = 0.24 \pm 0.02.
\end{equation}
From now on, $\lambda_w=1/f$ and $\phi_w=0.25$ will be used to
estimate the phase $\phi_1$ encountered by a downstream fish whose
leading edge is located at distance $d$ from the upstream fish.

\begin{table}
  \begin{center}
  \begin{tabular*}{0.8\textwidth}{@{\extracolsep{\fill}} cccccc}
  \toprule
  \multicolumn{3}{c}{Gait parameters} & \multicolumn{2}{c}{$\phi_1$ linear fit} & $c_w=1$ \\
 \cmidrule{1-3}  \cmidrule{4-5}  \cmidrule{6-6}
     $f$ & $x_1$ & $\delta$ & $1/\lambda_w$ & $\phi_w$ & $\phi_1-fd$ \\
     \midrule
     $1.5$ & $0.73$ & $0.52$ & $1.52$ & $0.19$ & $0.22$\\
     $1.8$ & $0.77$ & $0.36$ & $1.77$ & $0.26$ & $0.24$\\
     $2.1$ & $0.81$ & $0.28$ & $2.07$ & $0.30$ & $0.26$\\
     $2.4$ & $0.87$ & $0.23$ & $2.37$ & $0.29$ & $0.26$\\
     $2.7$ & $0.88$ & $0.21$ & $2.65$ & $0.29$ & $0.25$\\
       \bottomrule  
  \end{tabular*}
  \caption[Parameters of the gaits used in the wake vorticity phase
    estimate and fitted phase and wavelength for the vorticity in the
    wake.]{Parameters of the gaits used in the wake vorticity phase
    estimate and fitted phase and wavelength for the vorticity in the
    wake. An estimate of the phase $\phi_w$ assuming a phase speed
    $c_w=\lambda_w/f=1$ is also provided.}
  \label{tab:fit}
  \end{center}
\end{table}

Whereas the pressure signature in the wake shown at the plane $z=0$ is
very weak (\ref{fig:3D_snapshots}e,f), the pressure signature in the
plane $z=0.06$, just above the peduncle is much stronger
(\ref{fig:3D_snapshots}g,h), and could still be used by a downstream
fish to reduce its swimming energy.

%%%%%%%%%%%%%%%%%%%%%%%%%%%%%
\subsection{Effect of phase and distance for two fish in-line \label{sec:2fish_dist}}

\begin{figure}
 \centerline{\includegraphics[width=0.9\textwidth]{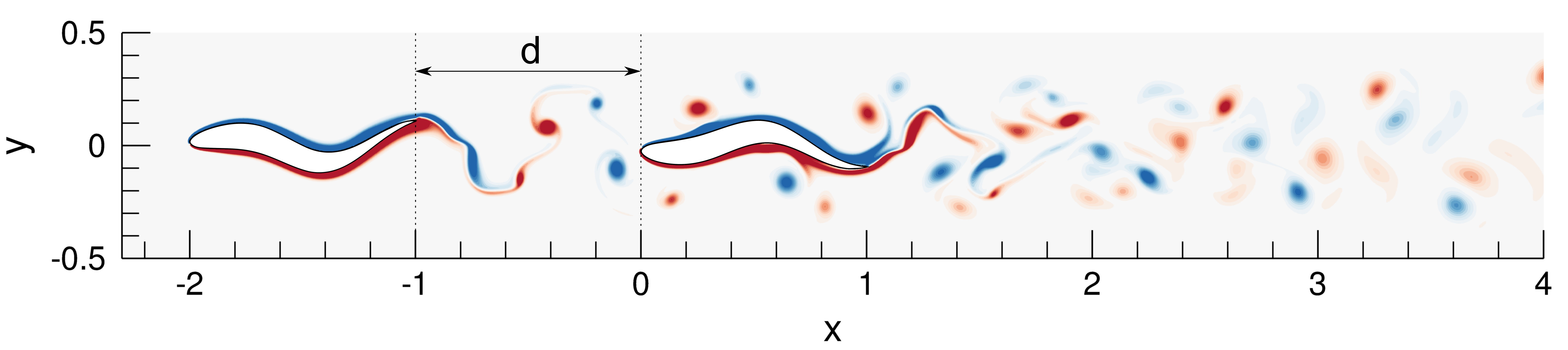}}
  \caption[Snapshot of the vorticity field for two fish undulating at
    $f=1.5$ with separation distance $d=1$ and optimal phase $\Delta
    \phi = 0.83$.]{Snapshot of the vorticity field for two fish
    undulating at $f=1.5$ with separation distance $d=1$ and optimal
    phase $\Delta \phi = 0.83$ at time $t/T-\phi = 0.25\ (\text{mod }
    1)$. The color axis is the same as in figure
    \ref{fig:opt_gauss_vort}.}
\label{fig:2fish_vort1}
\end{figure}

Here we consider two fish-like foils following each other and
undulating at frequency $f=1.5$ with the optimized gait for this
frequency, as illustrated in figure \ref{fig:2fish_vort1}. The
amplitude of undulation, $a_0$, is adjusted independently for each
fish to ensure that both fish are in a stable position and produce
zero net thrust on average.  We vary the distance $d$ between the
trailing edge of the upstream fish and the leading edge of the
downstream fish, as well as $\phi$, the phase of the downstream fish
motion as defined in Eq. \ref{eq:motion}. An important parameter will
be $\Delta \phi$, the phase difference between the motion of the
downstream fish leading edge and the vortices it encounters: $\Delta
\phi = \phi-\phi_1(d)$. In order to measure the impact of the pair
configuration on each fish, we define $R(C_P)$ (resp. $R(a_0)$), the
ratio of the power coefficient (resp. amplitude) in the pair
configuration over the power coefficient (resp. amplitude) for the
corresponding gait in open water.

\begin{figure}
 \centerline{\includegraphics[width=0.85\textwidth]{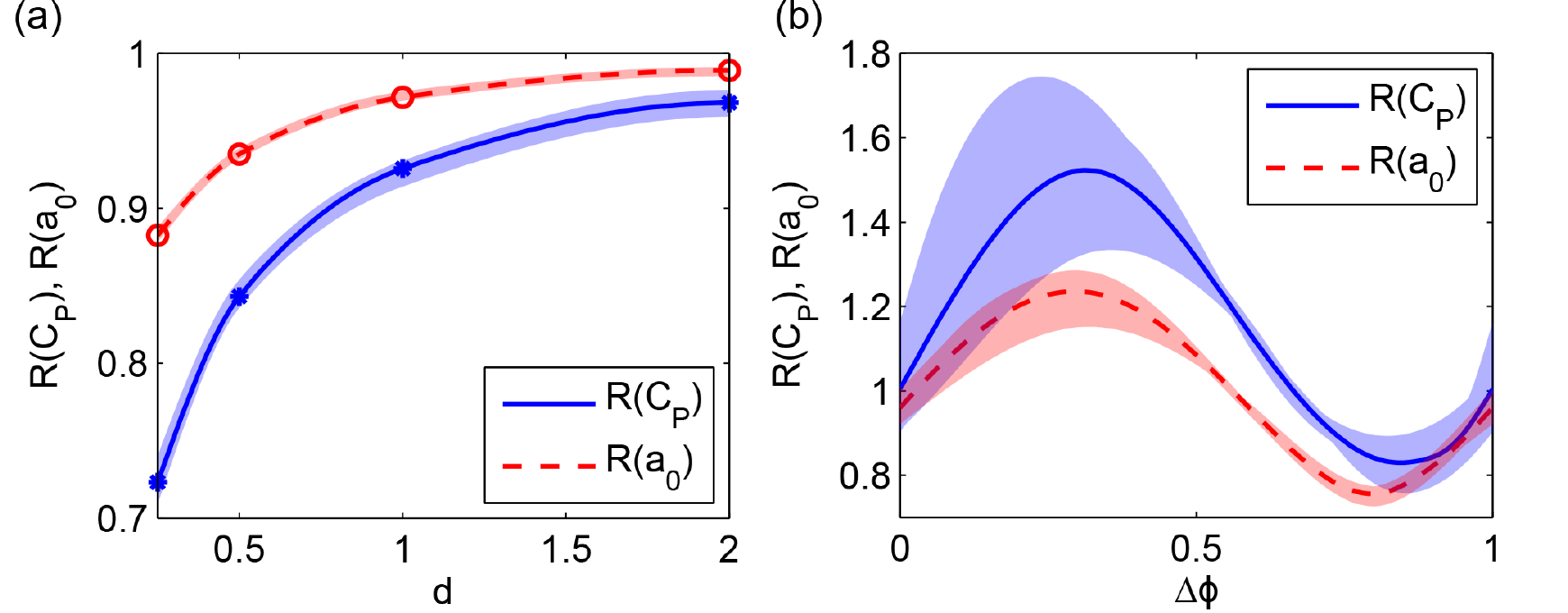}}
  \caption[Time-averaged power coefficient $\overline{C_P}$ and
    amplitude $a_0$ for (a) the upstream fish as a function of
    distance $d$ and (b) the downstream fish as a function of phase
    $\Delta \phi$.]{Time-averaged power coefficient $\overline{C_P}$
    and amplitude $a_0$ for (a) the upstream fish as a function of
    ditance $d$ and (b) the downstream fish as a function of phase
    $\Delta \phi$. The solid (resp. dashed) line marks the average
    value with respect to the phase (resp. distance) and the shaded
    area indicates the range of values reached across the various
    distances (resp. phases).}
\label{fig:2fish_dist2}
\end{figure}

Both fish can benefit from swimming in pair, but the trends are very
different. The swimming power and amplitude of the upstream fish is
virtually independent of the phase of the downstream fish, as shown in
figure \ref{fig:2fish_dist2}a. For large separations $d$, the
downstream fish does not impact the upstream fish whose efficiency is
then almost the same as in open-water. However, as the downstream fish
gets close ($d<0.5$), the high pressure around the leading edge of the
downstream fish `pushes' the upstream foil, regardless of their phase
difference. As a result, the upstream fish can reduce its swimming
amplitude, expending less power than it would in open-water. At
$d=0.25$, the undulation amplitude is reduced by $10\%$, resulting in
$28\%$ energy saving, corresponding to a quasi-propulsive efficiency
(based on the towed drag on open-water) of $\eta_{QP}=69\%$.

For the downstream fish, the situation is very different. Even when
the upstream fish is several chord lengths ahead, the downstream fish
encounters its wake. The performance of the downstream fish is
determined by its interaction with the vortices of the wake. It
appears from \ref{fig:2fish_dist2}b that the swimming power of the
downstream fish depends primarily on the phase difference $\Delta
\phi$ between its undulation and the encountered vortices. Regardless
of the distance $d$, the swimming power of the downstream fish is low
if $\Delta \phi$ is between $0.7$ and $1$, and it is high if $\Delta
\phi$ is between $0.1$ and $0.5$. Like for the upstream fish, the
reduced swimming power results from a reduced undulation amplitude
$a_0$, but despite a more significant reduction in amplitude ($27\%$
for $\Delta \phi=0.8$), the power reduction does not exceed that of
the upstream fish. For the downstream fish, a maximum energy saving of
$24\%$ is reached at $\Delta \phi = 0.85$, corresponding to an
efficiency of $\eta_{QP}=65\%$.

\begin{figure}
 \centerline{\includegraphics[width=0.75\textwidth]{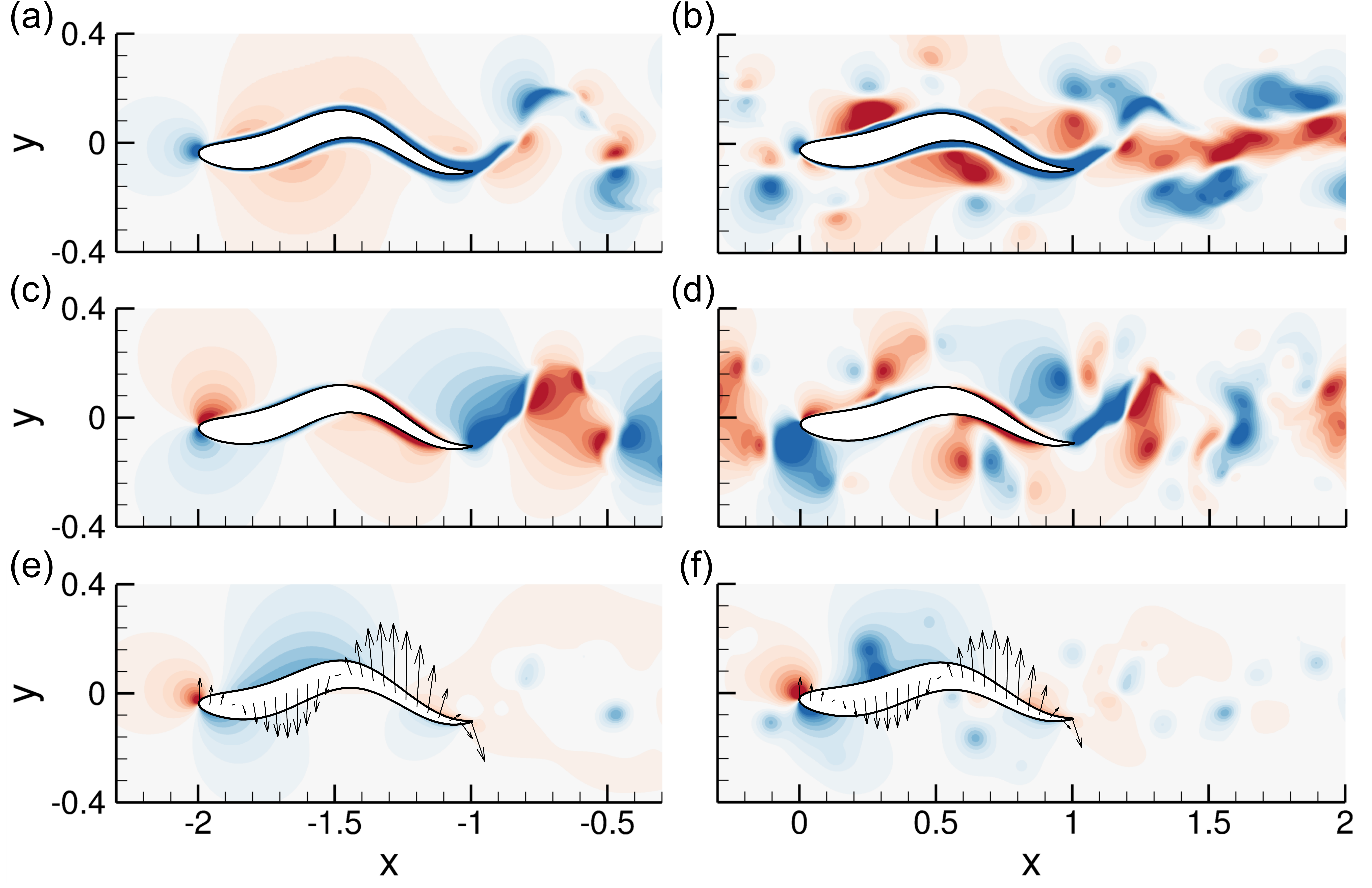}}
  \caption[Snapshot of the velocity and pressure field for two fish
    undulating at $f=1.5$ with separation $d=1$.]{Snapshot of the
    velocity and pressure field for two fish undulating at $f=1.5$
    with separation $d=1$ and optimal phase $\Delta \phi =
    0.83$. (a,b) $x$-velocity; (c,d): $y$-velocity; (e,f): pressure
    and arrows showing the velocity of the fish. (a,c,e): upstream
    fish at $t/T = 0.25\ (\text{mod } 1)$; (b,d,f): downstream fish at
    $t/T-\phi = 0.25\ (\text{mod } 1)$. The same color axis as in
    figure \ref{fig:opt_gauss_press} is used for the pressure, and the
    same as in figure \ref{fig:wake}b is used for the velocity
    (centered in $0$ for the $y$-velocity).}
\label{fig:2fish_press1}
\end{figure}

Figure \ref{fig:2fish_vort1} shows the vorticity field around the two
fish undulating with frequency $f=1.5$ at distance $d=1$ for the phase
$\Delta \phi=0.83$ that minimizes the swimming power of the downstream
fish. At $t/T-\phi=0.25$, the downstream fish approaches the negative
vortex (at $x=-0.1$ on its left) as it is turning its ``head''
(leading edge) to the right. This acceleration of the head causes a
low pressure on the left side of the head, as shown in figure
\ref{fig:2fish_press1}e. Due to its position, the approaching negative
vortex causes an increase in the longitudinal velocity, as shown in
figure \ref{fig:2fish_press1}b, which results in an increased
stagnation pressure (figure \ref{fig:2fish_press1}f). However, this
vortex also generates a large transverse velocity with negative sign,
as shown in figure \ref{fig:2fish_press1}d. As a result, the effects
of the head motion are amplified by the incoming vortex, displacing
the stagnation point downstream on the right side and deepening the
low pressure on the left side (figure \ref{fig:2fish_press1}f). While
the energy required by the fish to rotate its head is increased, the
drag is decreased, despite the faster flow encountered by the fish.

At the same time, the positive vortex located at $x=0.2$ on the right
side of the fish thickens the boundary layer and significantly
accelerates the flow in a region where the fish undulation already
accelerates it (figure \ref{fig:2fish_press1}a,b). This interaction
between the vortex and the fish results in a very large pressure drop
around $x=0.3$ that also contributes to the reduction in drag while
increasing the swimming power. The vortices are convected downstream
at a speed which is substantially lower than the phase speed of the
fish deformation. Further downstream, the distance between the
vortices and the fish increases, and their interaction becomes
weaker. At the trailing edge, the phase between the vortices and the
fish motion is close to $\pi$, such that the positive vortex reaches
$x=1$ as the trailing edge of the fish is at its leftmost
position. This vortex will be shed just upstream of the same sign
vortex shed by the downstream fish. The resulting wake configuration
is unstable and it takes several body lengths for the wake to
reorganize into two pairs of opposite sign vortices per cycle.

The results presented in this section are consistent with the
experimental results from thrust producing rigid pitching foils in an
in-line configuration \citep{boschitsch_propulsive_2014}. We found
that for small separation distance the propulsive efficiency of the
upstream fish is greatly improved. We also showed that the efficiency
of the downstream fish only weakly depends on the separation distance,
the primary parameter being the phase difference between the wake from
the upstream fish and the undulating motion of the downstream fish. If
the undulation amplitude was fixed, the downstream fish would
experience an increased drag and decreased power for $0 \leq \Delta
\phi \leq 0.5$, whereas it would experience a decreased drag and
increase power for $0.5 \leq \Delta \phi \leq 1$. For a self-propelled
fish, the energetic benefits of a reduced amplitude resulting from a
reduced drag overcome the power increase caused by the vortices.

%%%%%%%%%%%%%%%%%%%%%%%%%%%%%
\subsection{Effect of frequency for two undulating fish in-line}

\begin{figure}
 \centerline{\includegraphics[width=0.75\textwidth]{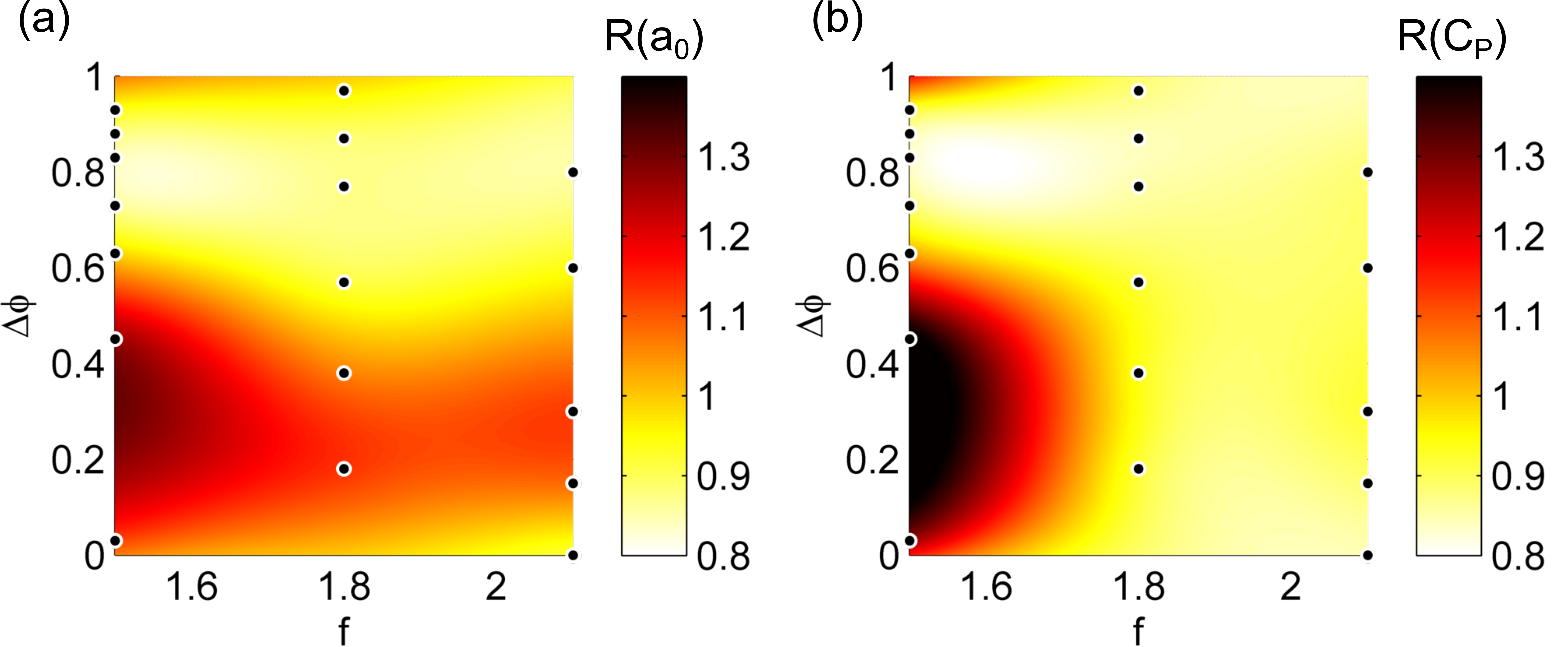}}
  \caption[Ratio of (a) amplitude, and (b) power coefficient, as a
    function of frequency $f$ and phase $\Delta \phi$ for two fish
    swimming in-line.]{Ratio of (a) amplitude and (b) power
    coefficient as a function of frequency $f$ and phase $\Delta \phi$
    for two fish swimming in-line at distance $d=1$.  The ratios are
    defined with respect to the corresponding gait in open water.  The
    black dots show the location of the points that have been used to
    build the thin-plate smoothing spline ($\mathsf{tpaps}$ function
    in Matlab with smoothing parameter $p=0.999$) represented in
    color.}
\label{fig:2fish_f}
\end{figure}

Next, we fix the distance between the two fish to $d=1$ and vary their
undulation frequency. For frequencies $f=[1.5,\, 1.8,\,2.1]$, their
optimized Gaussian envelope is used, for which the parameters are
summarized in table \ref{tab:fit}.  Figure \ref{fig:2fish_f} shows
that most of the conclusions drawn in \S\,\ref{sec:2fish_dist} still
hold as the undulation frequency is increased. While the upstream fish
is mostly unaffected by the presence and phase of the downstream fish,
the undulation amplitude of the downstream fish is largest for $0 \leq
\Delta \phi \leq 0.5$ and smallest for $0.5 \leq \Delta \phi \leq
1$. However, the correlation between amplitude and power coefficient
is not as strong any more, and the exact value of the optimal phase
depends on the frequency. For instance, at $f=2.1$, the amplitude is
minimum for $\Delta \phi = 0.85$, but the power coefficient is minimum
for $\Delta \phi = 0$. Whereas with $f=1.5$ most phases result in an
increased amplitude and power coefficient, with $f=1.8$ and $f=2.1$,
the amplitude and power coefficient of the downstream fish never
exceed that of the upstream fish. Therefore, at these frequencies, it
is always beneficial to swim in the wake of an undulating fish,
despite the increased average velocity of the encountered flow.

\begin{figure}
 \centerline{\includegraphics[width=0.8\textwidth]{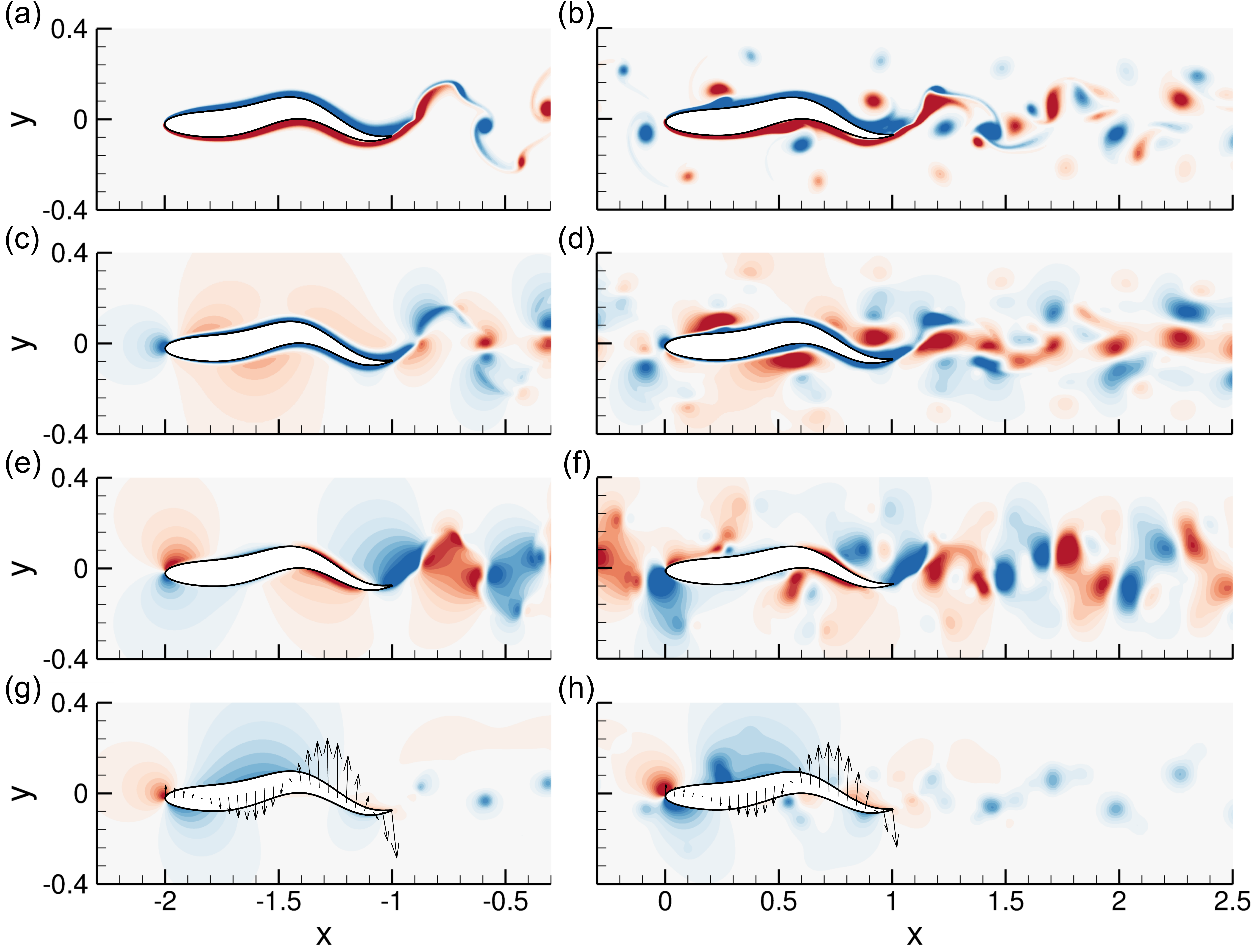}}
  \caption[Snapshot of the vorticity, velocity and pressure field for
    two fish undulating at $f=1.8$ with separation distance $d=1$ and
    optimal phase $\Delta \phi = 0.87$.]{Snapshot of the vorticity,
    velocity and pressure field for two fish undulating at $f=1.8$
    with separation distance $d=1$ and optimal phase $\Delta \phi =
    0.87$. (a,b) vorticity; (c,d) $x$-velocity; (e,f): $y$-velocity;
    (g,h): pressure and arrows showing the velocity of the
    fish. (a,c,e,g): upstream fish at $t/T = 0.25\ (\text{mod } 1)$;
    (b,d,f,h): downstream fish at $t/T-\phi = 0.25\ (\text{mod }
    1)$. The same color axis as in figure \ref{fig:opt_gauss_press} is
    used for the pressure, and the same as in figures \ref{fig:wake}a
    and b for vorticity and velocity (centered in $0$ for the
    $y$-velocity).}
\label{fig:2fish_f18}
\end{figure}

At frequency $f=1.8$, the results for the optimal phase, shown in
figure \ref{fig:2fish_f18}, are very similar to those described in the
previous section for $f=1.5$, with the vortices from the upstream fish
reinforcing the effect of the body undulation. However, the wake at
this frequency is narrower; therefore the vortices are closer to the
fish and they lose more strength through their interaction with the
boundary layer. Moreover, since the distance between two consecutive
vortices is proportional to the undulation frequency, the vortices are
spaced closer to each other. The resulting wake is dominated by two
single vortices shed by the downstream fish, each accompanied by
weaker vortices of opposite sign from the upstream fish.  This is
identical to the destructive vortex interaction mode of a flapping
foil within a \karman street in \citet{gopalkrishnan_active_1994},
which has been associated with increased efficiency.

\begin{figure}
 \centerline{\includegraphics[width=0.8\textwidth]{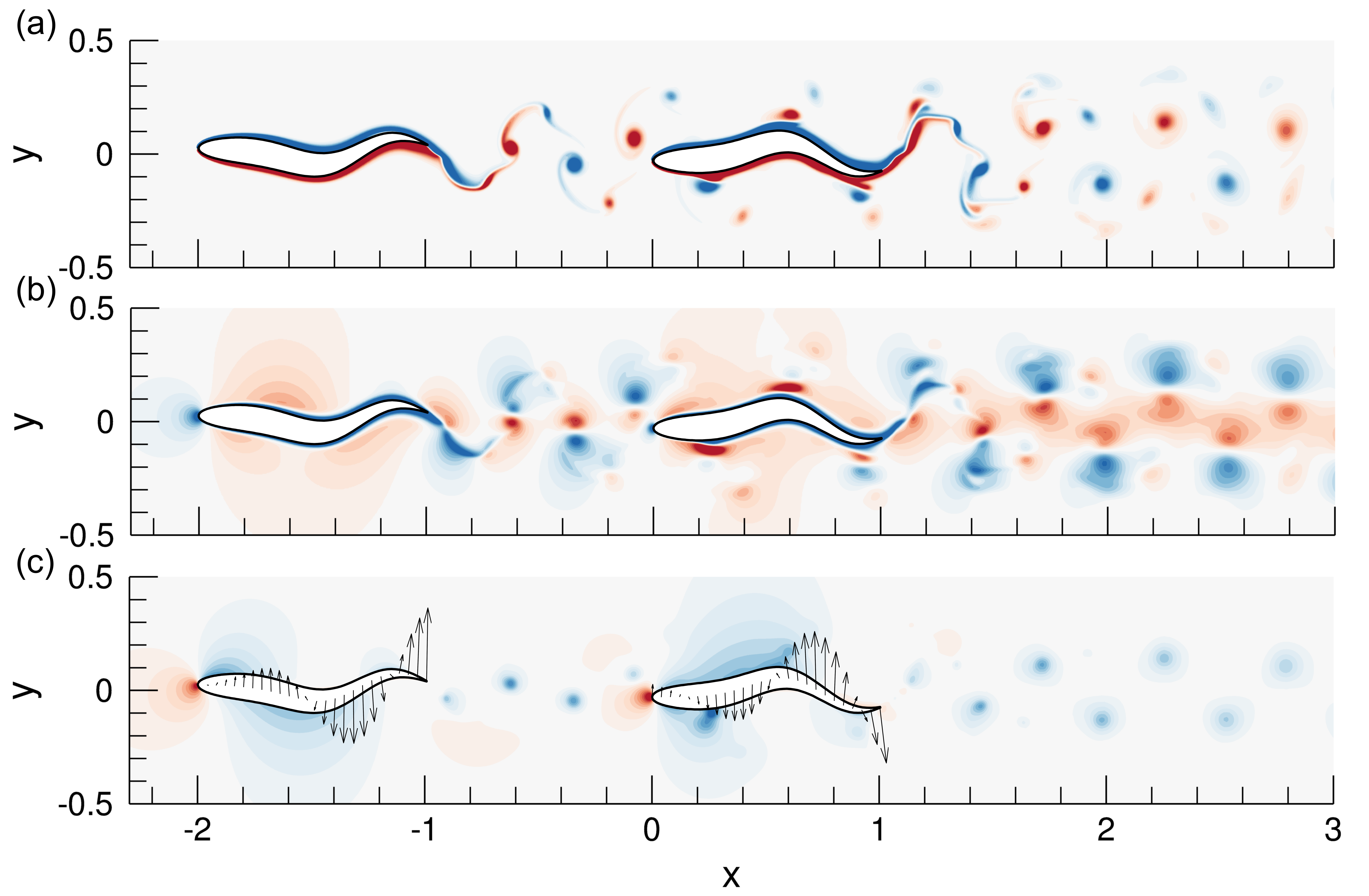}}
  \caption[Snapshots of (a) vorticity, (b) $x$-velocity, and (c)
    pressure field for two fish undulating at $f=1.8$ with separation
    distance $d=1$ and phase $\Delta \phi = 0.38$.]{Snapshots of (a)
    vorticity, and (b) pressure field for two fish undulating at
    $f=1.8$ with separation distance $d=1$ and phase $\Delta \phi =
    0.38$ at $t/T-\phi = 0.25\ (\text{mod } 1)$. The same color axis
    as in figure \ref{fig:wake}a is used for the vorticity and the
    same as figure \ref{fig:opt_gauss_press} for the pressure.}
\label{fig:2fish_f18_worst}
\end{figure}

Figure \ref{fig:2fish_f18_worst} illustrates the cases of the
downstream fish undulating with the phase providing the worse swimming
efficiency for $f=1.8$. In this configuration, the vortices from the
upstream fish counteract the effects of the undulating motion of the
downstream fish. As the fish turns its head to the right, displacing
the stagnation point to the right and causing a low pressure on the
left side of the head, the positive $y$-velocity caused by the
approaching positive vortex has the opposite effect. The high velocity
regions caused by the vortices along the fish correspond to low
velocity regions from the undulating motion. Finally, when the
vortices from the upstream fish reach the trailing edge, they merge
with the same sign vortices from the downstream fish. The resulting
wake is very stable and is a classical reverse \karman vortex street
much wider than the one behind a single fish.  This corresponds
precisely to the constructive interaction mode of a flapping foil
within a \karman street in \citet{gopalkrishnan_active_1994}, which
has been shown to have reduced efficiency.

\begin{figure}
 \centerline{\includegraphics[width=0.8\textwidth]{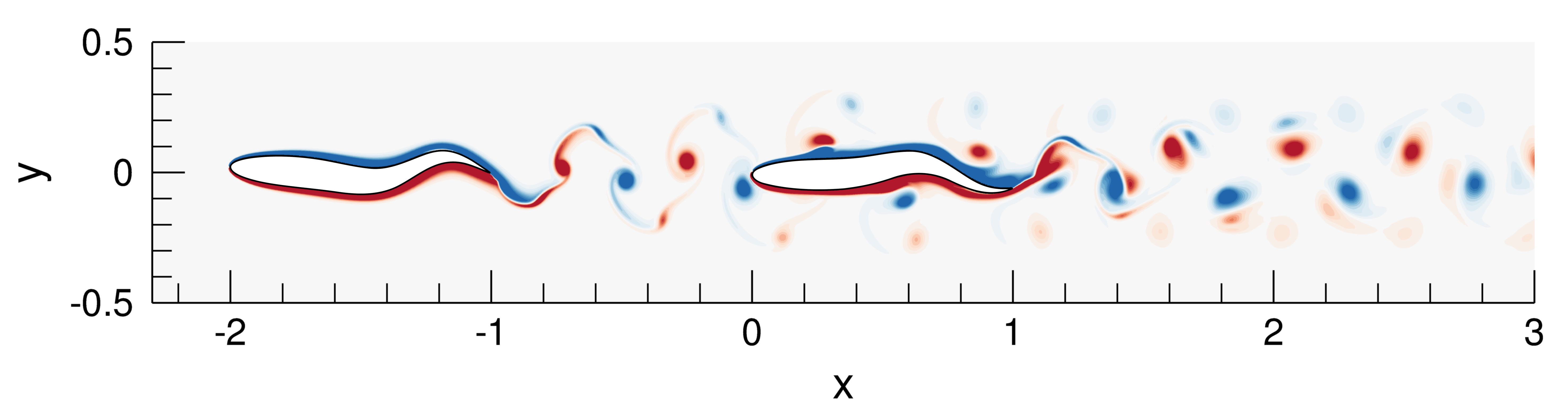}}
  \caption[Snapshot of the vorticity field for two fish undulating at
    $f=2.1$ with separation distance $d=1$ and phase $\Delta \phi =
    0$.]{Snapshot of the vorticity field for two fish undulating at
    $f=2.1$ with separation distance $d=1$ and phase $\Delta \phi =
    0$. The same color axis is used as in figure \ref{fig:wake}a.}
\label{fig:2fish_f21}
\end{figure}

As the frequency increases further, the vortices from the upstream
fish lose even more energy through interaction with the boundary layer
of the downstream fish; therefore for each period of oscillation the
wake behind the two fish contains a pair of single vortices, as shown
in figure \ref{fig:2fish_f21}. Moreover, since the efficiency of the
downstream fish mostly depends on the phase of the leading edge with
respect to the arrival of the upstream fish reverse \karman street
vortices, the phase of the trailing edge with respect to the incoming
vortices in the optimal configuration changes with frequency.

For all the frequencies considered here, a self-propelled fish can
save energy by undulating behind another self-propelled fish
undulating at the same frequency, reaching efficiencies close to
$\eta_{QP}=70\%$. By properly phasing its motion with respect to the
incoming vortex street, the vortices can reinforce the effect of the
undulation. Whereas for a fixed amplitude this phase would result in
an increased swimming power, the reduction in drag results in an
overall decreased swimming power.

%%%%%%%%%%%%%%%%%%%%%%%%
\subsection{Fish undulating in the reduced velocity region of the wake}

We have so far considered the case of a pair of fish swimming in an
in-line configuration. Since our fish model has a feedback controller
ensuring its stability in $y$, it is also possible to impose an
asymmetric configuration with an offset in the $y$ direction. Indeed,
according to Weihs' theory \citep{weihs_hydromechanics_1973}, the only
way for a fish to save energy in a school is to swim in the region of
reduced velocity located between two wakes. We have already shown that
a fish can save energy by swimming directly in the wake of another
self-propelled fish and will now investigate if additional savings are
possible by swimming in areas of reduced flow velocity. Figure
\ref{fig:wake}d shows that, even with a single fish upstream, the flow
on either side of the wake is slower than the free stream: for $f=1.5$
the average flow is slowest at $y= \pm 0.17$. With the downstream fish
offset from the upstream fish by $\Delta y=0.17$, we vary the phase
difference $\Delta \phi$ in order to see if the downstream fish can
also save energy when swimming at this location.

\begin{figure}
 \centerline{\includegraphics[width=0.85\textwidth]{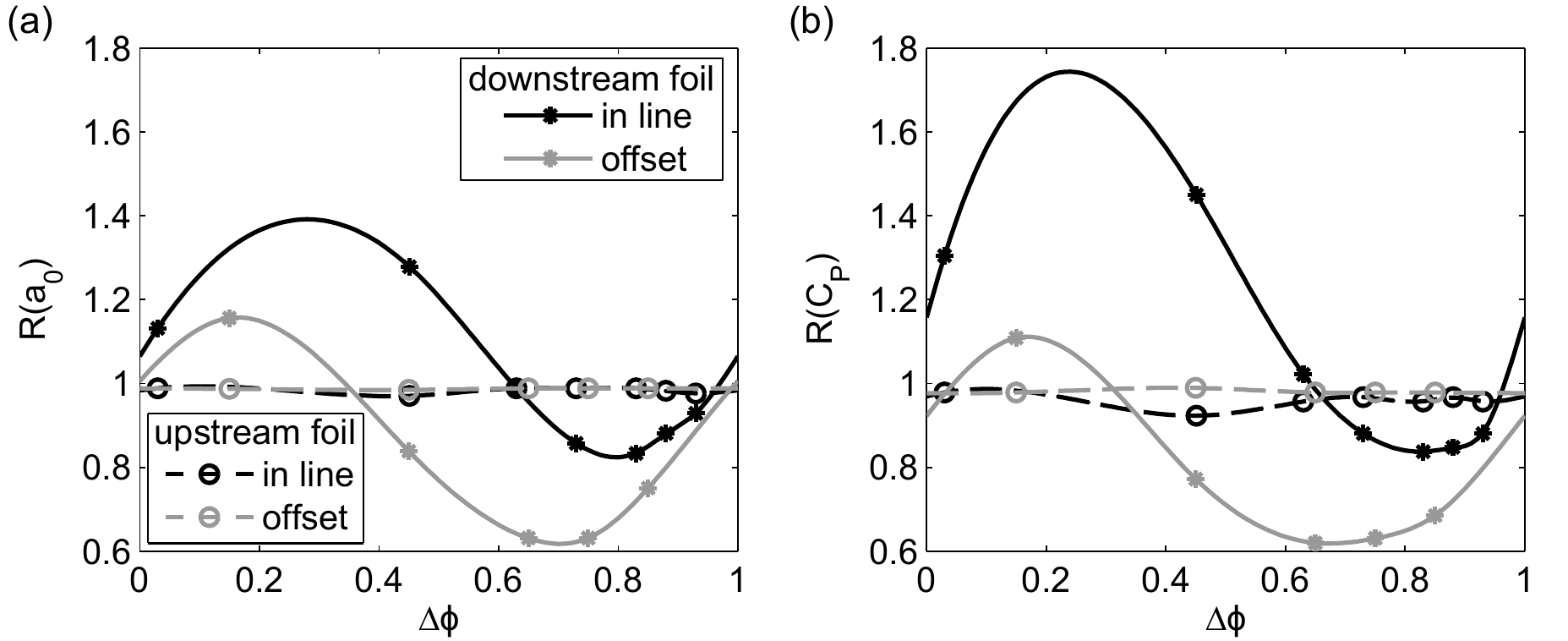}}
  \caption{Ratio of undulation amplitude $a_0$ and time-averaged power
    coefficient $\overline{C_P}$ as a function of phase for two fish
    undulating at $f=1.5$ with longitudinal separation distance $d=1$.
    In-line fish and fish at an offset $\Delta y=0.17$ are compared.}
\label{fig:2fish_y017_CP}
\end{figure}

Figure \ref{fig:2fish_y017_CP} shows that, even when the downstream
fish is offset from the vortex street, its swimming performance
greatly varies with the phase. However, it is easier for the fish to
save energy in this region of reduced flow velocity than directly
behind the upstream fish. Directly behind the upstream fish, only
$30\%$ of the phases result in energy savings, and by using the
unsteadiness of the wake, the quasi-propulsive efficiency at $f=1.5$
can be brought up from $50\%$ to $60\%$. When undulating in the region
of reduced flow velocity, it is easier to save energy since over
$70\%$ of the phases result in energy savings. The energy savings can
even be very large since $\eta_{QP}=80\%$ is possible for $\Delta \phi
= 0.65$.

\begin{figure}
 \centerline{\includegraphics[width=0.8\textwidth]{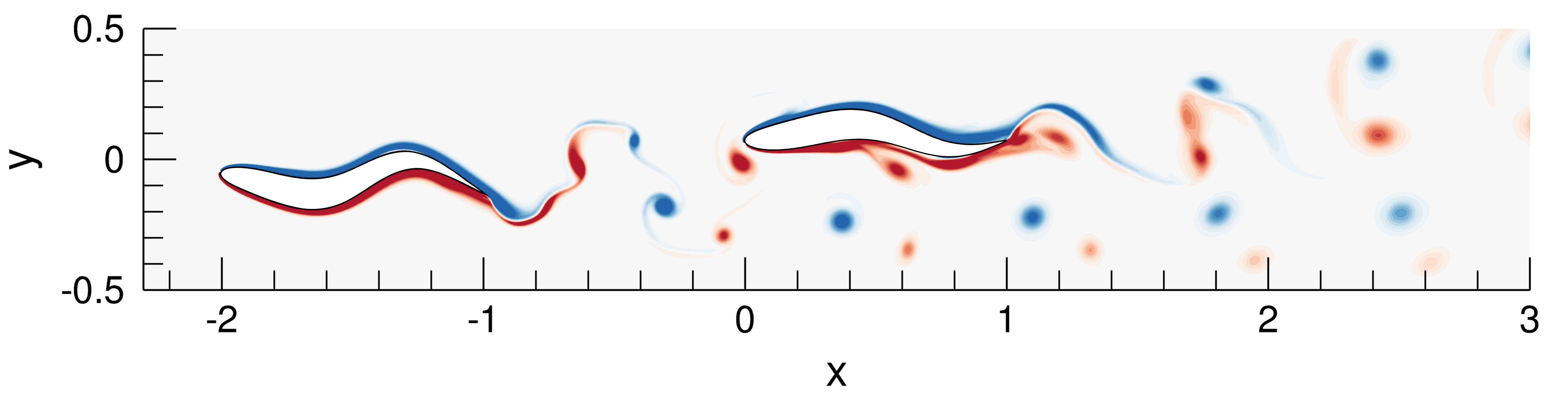}}
  \caption[Snapshot of the vorticity field for two fish undulating at
    $f=1.5$ with longitudinal separation distance $d=1$, transverse
    separation $\Delta y=0.17$ and optimal phase $\Delta \phi =
    0.65$.]{Snapshot of the vorticity field for two fish undulating at
    $f=1.5$ with longitudinal separation distance $d=1$, transverse
    separation $\Delta y=0.17$ and optimal phase $\Delta \phi = 0.65$
    at time $t/T-\phi = 0.1\ (\text{mod } 1)$. The color axis is the
    same as in figure \ref{fig:opt_gauss_vort}.}
\label{fig:2fish_vort_y017}
\end{figure}

\begin{figure}
 \centerline{\includegraphics[width=0.8\textwidth]{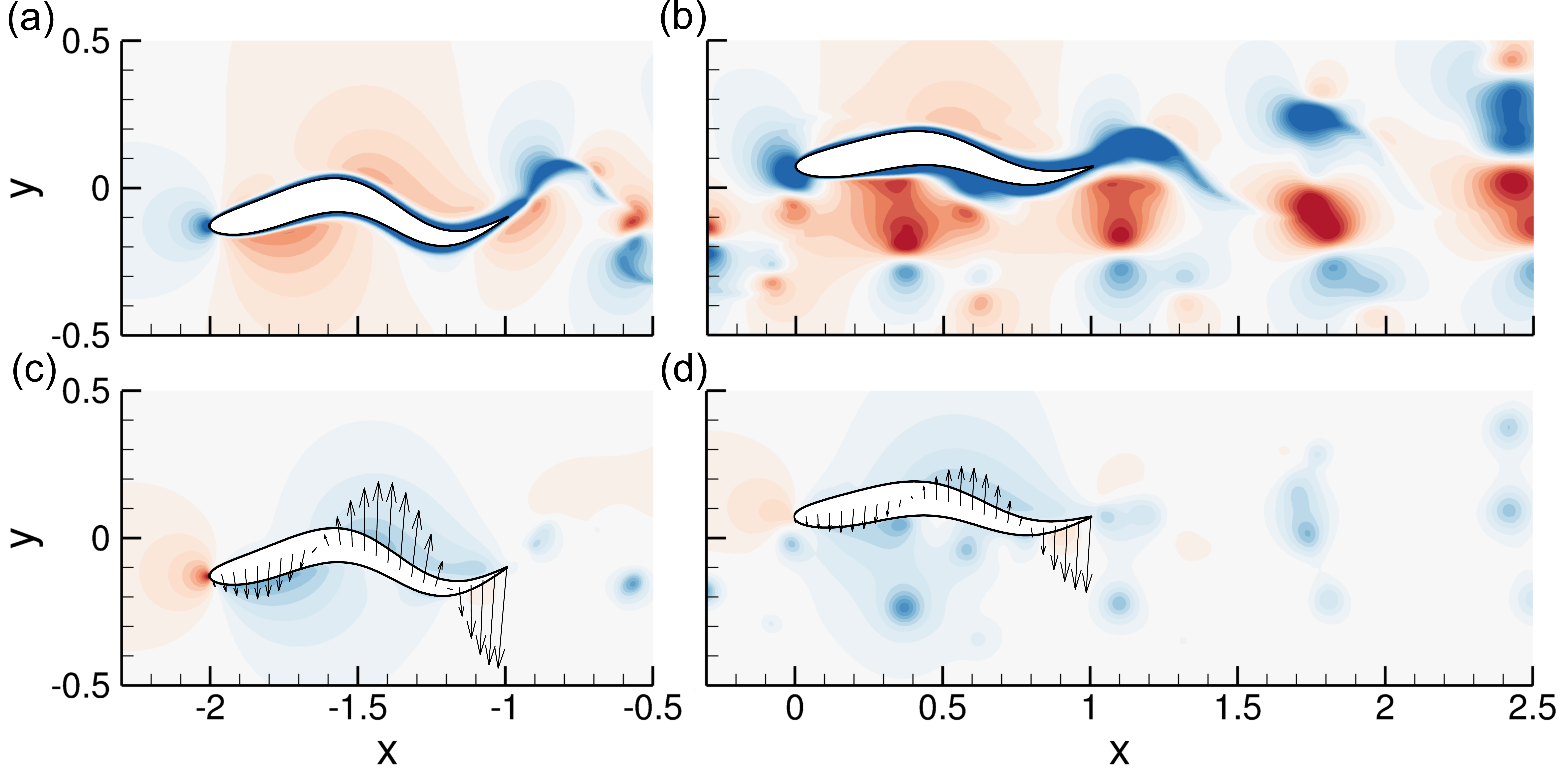}}
  \caption[Snapshot of the $x$-velocity and pressure field for two
    fish undulating at $f=1.5$ with longitudinal separation distance
    $d=1$, transverse separation $dy=0.17$ and optimal phase $\Delta
    \phi = 0.65$.]{Snapshot of the (a,b) $x$-velocity and (c,d)
    pressure field for two fish undulating at $f=1.5$ with
    longitudinal separation distance $d=1$, transverse separation
    $dy=0.17$ and optimal phase $\Delta \phi = 0.65$. (a,c): upstream
    fish, $t/T = 0.1\ (\text{mod } 1)$; (b,d): downstream fish, $t/T
    -\phi = 0.1\ (\text{mod } 1)$.The same color axis as in figure
    \ref{fig:opt_gauss_press} is used for the pressure, and the same
    as in figure \ref{fig:wake}b for the velocity.}
\label{fig:2fish_velpress_y017}
\end{figure}

Figure \ref{fig:2fish_vort_y017} shows that, at the optimal phase, the
leading edge of the downstream fish reaches its leftmost position at
the same time as it reaches a positive vortex. Figure
\ref{fig:2fish_velpress_y017}b shows that the leading edge of the
downstream fish exactly passes through the region where the
longitudinal velocity is smallest. As a result, the stagnation
pressure is greatly reduced (figure
\ref{fig:2fish_velpress_y017}d). Moreover, the region of accelerated
flow between the negative vortex and the fish ($x=0.4$) reinforces the
accelerated region caused by the undulation, which we showed earlier
is beneficial.

%%%%%%%%%%%%%%%%%%%%%%%%%%%%%%%%%%%%%%%%%%%%%%%%%%%%%%%%%%%%%%%%%%%%%%%%%
\section{Discussion}
%%%%%%%%%%%%%%%%%%%%%
\subsection{Optimal BCF propulsion and the role of fish shape}

It is interesting to note that, as the peak of the Gaussian describing
the fish bending motion becomes sharper, the curvature imposed by the
envelope in the peduncle section becomes much larger than the
curvature caused by the traveling wave. This is particularly
noticeable for $f=2.7$, at which frequency the undulations are mostly
restricted to what would be the peduncle and tail sections for a fish.

The increase of the curvature in the peduncle region corresponding
with a decreased stride length and increased swimming frequency is
essential, allowing the instantaneous deformation of the fish to match
the trailing edge trajectory (Figure
\ref{fig:body_motion_gauss}). This serves to avoid the efficiency loss
associated with a large angle of attack at the tail, as well as to
keep the boundary layer attached to the fish. Indeed, the boundary
layer remains attached to the fish as previously observed for waves
traveling faster than the free stream \citep{taneda_visual_1977,
  shen_turbulent_2003}, and a reverse \karman vortex street forms in
the wake, consistent with previous studies of efficient thrust
production in oscillating foils \citep{triantafyllou_wake_1991,
  anderson_oscillating_1998}. The width and wavelength of the reverse
\karman vortex street decreases with increasing undulation frequency,
and secondary small vortices develop at low frequency.

The total lateral displacement of a live swimming fish is maximum at
the trailing edge, but, once recoil is substracted, it becomes
apparent that the body deformation is largest around the peduncle. We
show that efficiency improves when the envelope of the body
deformation is largest upstream of the trailing edge, because this
creates a part of the body that is capable of acting as a caudal fin
(and has roughly the length of the caudal fin). The high curvature of
the peduncle region allows the (equivalent) caudal fin area to pitch
independently from the motion of the body, in order to control the
timing of trailing edge vortex shedding.

Fish employing the carangiform and thunniform swimming mode generally
have a narrow peduncle, and our results suggest that there is function
associated with this form. The narrow peduncle allows for sharp
bending of the tail even at an angle with respect to the body
deformation (providing a discontinuous slope), such as in the optimal
motion at high frequency. Tunas, for example, have a special anatomy
at the peduncle that allows powerful tendons to actuate the tail with
substantial torque. This allows for the independent control of the
tail in manipulating vorticity formed upstream of the tail along the
body \citep{zhu_three-dimensional_2002}, or from externally generated
vortices. \citet{wolfgang_near-body_1999} demonstrated experimentally
and numerically that high flexibility of the peduncle region allows
for the caudal fin to precisely redirect vorticity shed upstream for
optimal propulsion. In the 2D simulations conducted here the high
curvature in the peduncle region serves the same purpose in allowing
the fish shed vortices optimally.

Similar to \citep{borazjani_role_2010}, we show that the optimal
kinematics is mostly independent of body shape. Indeed, even a
two-dimensional geometry can help assess the energetic performance of
swimming kinematics.

%%%%%%%%%%%%%%%%%%%%%%%%
\subsection{Proposed schooling theory and comparison with Weihs' theory}

To our knowledge, the only existing hydrodynamic theory of schooling
has been proposed by \citet{weihs_hydromechanics_1973}. This theory
provides useful insight based on time- averaged flow considerations,
but does not factor in the potential benefits of energy extraction
from the vortices. According to Weihs, a fish swimming directly behind
another fish would experience higher relative velocity and would
therefore have to spend extra energy. On the contrary, a fish swimming
between two adjacent fish wakes would experience a reduced relative
speed, allowing it to save energy. This strategy is known as flow
refuging \citep{liao_review_2007} or drafting.

As we have shown in this paper it is possible for a fish to save
energy regardless of whether it swims directly behind another fish or
at a lateral offset that allows it to benefit from a reduced flow
velocity. The phase difference between its undulation and the wake
vortices, $\Delta \phi$, determines whether its drag is reduced or
enhanced. When swimming directly behind another fish, where the
averaged flow is faster than the free-stream, the fish cannot save
energy through drafting, and must therefore capture energy from
individual vortices in order to save energy. We show that by using the
transverse velocity of the individual vortices to amplify the pressure
effects of the undulating motion, the fish can reduce its
drag. Conversely, swimming in the region of the wake where the flow
is, on average, slower, does not guarantee a reduced drag. However, we
observed that it is possible to reduce drag and save energy by
undulating in the region of reduced velocity than directly behind
another fish, due to the possibility of taking advantage of both
drafting and individual vortex energy capture. Up to 81\%
quasi-propulsive efficiency can be reached for a fish undulating with
proper phase in the region of reduced flow velocity, compared with a
maximum of 66\% for a fish swimming directly behind another fish
(Table \ref{tab:2fish}).

\begin{table}
  \begin{center}
  \begin{tabular*}{1\textwidth}{@{\extracolsep{\fill}} cccccc}
  \toprule
    $f$  & $d$   & $\Delta y$ & $\Delta \phi$ (mod $1$) & $\eta_{QP}$(upstream) & $\eta_{QP}$(downstream) \\
      \midrule
      $1.5$ & $1$ & $0$ & $0.83$ & $0.52$ &  $0.60$ \\
      $1.8$ & $1$ & $0$ & $0.84$ & $0.55$ & $0.61$ \\
      $2.1$ & $1$ & $0$ & $0.00$ & $0.56$ & $0.62$ \\
      $1.5$ & $0.25$ & $0$ & $0.83$ & $0.67$ & $0.66$ \\
      $1.5$ & $1$ & $0.17$ & $0.65$ & $0.51$ & $0.81$ \\
       \bottomrule  
  \end{tabular*}
  \caption[Efficiency for a pair of undulating fish in various
    advantageous configurations.]{Efficiency for a pair of undulating
    fish in various advantageous configurations. The undulation
    frequency $f$, longitudinal separation $d$, transverse distance
    $\Delta y$ and the phase difference $\Delta \phi$ between the
    leading edge of the downstream fish and the vortices in the wake
    of the upstream fish are considered.}
  \label{tab:2fish}
  \end{center}
\end{table}

The energy contained in individual vortices can be harnessed in
several ways. For a flapping foil in a \karman vortex street, it is
well known that the efficiency increases if the foil vortices
destructively interact with the oncoming vortices, while the thrust
can be enhanced if the foil vortices constructively interact with the
oncoming vortices \citep{gopalkrishnan_active_1994,
  streitlien_efficient_1996}. For a fish swimming within a wake,
constructive and destructive interaction can occur between the
oncoming wake vortices and vortices emanating from the boundary layer
of the fish, as well as between the wake vortices and the vortices
shed at the fish's trailing edge. Indeed, at the tail of the
downstream fish, we observed that constructive interactions between
the oncoming wake vortices and the vortices shed at the trailing edge
were associated with reduced efficiency, while destructive
interactions correlated with increased efficiency.

The interactions between the fish body and the oncoming vortices can
result in enhanced thrust and/or improved efficiency. We show that the
downstream fish can reduce its drag by consistently turning its head
in a manner that employs the oncoming vortex flow to increase the
transverse velocity across the head, amplifying the pressure field
created at the head. While this increases the power consumed by the
fish to rotate its head, the pressure drag at the head is decreased
substantially to result in an overall improvement to the efficiency.

While reduced drag implies a reduced undulation amplitude for
open-water self-propelled swimming, the correlation with energy saving
is not as straightforward within a school, because the vortices impact
both the drag and the swimming power. Since the quasi-propulsive
efficiency is defined as $RU_s/\bar{P}_{in}$, the power consumed must
be reduced for an increased efficiency. However, it can be directly or
indirectly reduced. Directly, the power is reduced when vortices along
the body of the fish exert force in the direction the fish is
oscillating, thereby doing work on the fish. Indirectly, the vortices
can help to reduce the overall drag on the fish, therefore reducing
the amount of work the fish must perform to self-propel. If the
undulation amplitude was kept constant, phases $0 \leq \Delta \phi
\leq 0.5$ would result in an increased drag and decreased power, and
the reverse would apply to phases $0.5 \leq \Delta \phi \leq 1$. The
energy benefits of a reduced amplitude generally more than compensate
for the increased swimming power, such that drag reduction tends to
result in power reduction. However, the phase resulting in the
smallest amplitude usually does not coincide with the optimal
one. This suggests that multiple mechanisms are important for the
efficiency of the downstream foil. In particular, the drag is mostly
governed by the interaction between the head of the fish and the
vortices, whereas the power is mostly governed by the interaction
between these vortices and the tail, where the transverse velocities
are much larger. The exact value of the optimal phase, therefore,
depends on the undulation frequency and the gait.

In summary, a fish undulating in a vortex street cannot be considered
as a rigid body with a propeller, located inside a jet. Regardless of
the exact location of the fish in the vortex street, constructive
interactions between the undulating body and the individual vortices
can result in enhanced thrust, while destructive interactions result
in increased swimming power. The exact value of the optimal phase
depends on the gait details, but in general the drag reduction
configurations are the most advantageous, and it is easier to reduce
drag when undulating in a region of averaged reduced flow velocity,
even in an asymmetric configuration.

%%%%%%%%%%%%%%%%%%%%%%%%%%%%%%%%%%%%%%%%%%%%%%%%%%%%%%%%%%%%%%%%%%%%%%%%%
\section{Summary and Conclusions}

We established through 2D and 3D numerical simulation the conditions
for optimal propulsion in undulatory fish swimming, first for a single
self-propelled fish, and then for a pair of identically shaped
self-propelled fish moving in-line or at an offset, separated axially
by a distance $d$.

First, we considered the problem of optimal propulsion of an
undulating, self-propelled fish-like body, fully accounting for linear
and angular recoil.  We employed 2D simulations to conduct an
extensive parametric search and then by employing targeted 3D
simulations we established that properties found in 2D are
qualitatively similar to those for 3D simulations.

In summary, the assumptions we employed in order to render the study
feasible are:
\begin{enumerate}[labelwidth=0.6cm,labelindent=16pt, leftmargin=1cm, align=left]
\item \noindent Simulations were conducted at a Reynolds number of
  5000.
\item \noindent For the 2D simulations, we modeled the fish body as an
  undulating NACA0012, while a danio-shaped body was used for 3D
  simulations. We model the main body of the fish and its caudal fin
  but not the other fins or details on the body such as other fins,
  finlets, eyes and other protrusions, and scales.
\item \noindent The motion of the fish consists of a steady axial
  translation at a prescribed speed, and a lateral body deformation in
  the form of a traveling wave of constant frequency $f$ and
  wavelength $\lambda=1$. Free axial rigid body motion as well as
  lateral and angular recoil were permitted and studied for their
  effect on propulsive efficiency.
\item \noindent The bending envelope was chosen to be either a
  quadratic or a Gaussian function of the length along the fish. For
  both functions, two parameters were sufficient to specify the shape
  of the envelope, with a third parameter $a_0$ to control the
  amplitude of the envelope. The total displacement envelope of live
  carangiform and anguilliform swimmers can be approximated by a
  quadratic function, but, after subtraction of the recoil, the
  envelope of body deformation is best approximated by a Gaussian
  function.
\item \noindent To maximize the quasi-propulsive efficiency (i.e.~to
  minimize the energy expended), the two parameters used to specify
  the shape for each envelope (equations 3.1 and 3.2) were varied
  using an optimization scheme.
\item \noindent A PID controller adjusts the amplitude of oscillation
  $a_0$ until self-propulsion is obtained, and additionally maintains
  the heading of the fish by adjusting the camber.
\end{enumerate}
  
The conclusions for optimal 2D swimming are as follows:
\begin{enumerate}[labelwidth=0.6cm,labelindent=16pt, leftmargin=1cm, align=left]
\item \noindent As with rigid flapping foils, the Strouhal number,
  phase angle between heave and pitch at the trailing edge, and
  nominal angle of attack are the principal parameters affecting the
  efficiency of propulsion.
\item \noindent Angular recoil has a significant impact on the
  efficiency of propulsion and hence must always be accounted for.
\item \noindent A Gaussian envelope enables a body deformation with
  high curvature in the region where the peduncle of the fish would
  be. This effectively allows the caudal fin to pitch independently
  with respect to the peduncle motion, and this extra degree of
  freedom allows for the control of flow patterns forming upstream of
  that position. Hence, whereas the convex profile traditionnaly used
  to model carangiform swimming provides quasi-propulsive efficiency
  of around 40\%, an optimized profile results in efficiency of 57\%.
\end{enumerate}

For 3D swimming of a danio-shaped fish, which explicitly models the
peduncle and caudal fin, the optimality conditions were observed to be
very close qualitatively to those of 2D swimming, although the
efficiency was consistently lower:
\begin{enumerate}[labelwidth=0.6cm,labelindent=16pt, leftmargin=1cm, align=left]
\item \noindent Angular recoil has significant impact on efficiency,
  as in 2D swimming.
\item \noindent Similarly to finite aspect ratio rigid flapping foils,
  the Strouhal number and maximum angle of attack are principal
  parameters affecting efficiency: For higher values of the Strouhal
  number the wake is also found to bifurcate, from a single row of
  connected vortex rings to a double row of vortex rings, resulting in
  reduced efficiency, as well as a complex three-dimensional structure
  in the wake.  As expected, optimization reduces the Strouhal number
  to be closer to the optimal range, and hence the effect of wake
  bifurcation is also reduced.  Within our parametric space, the
  Strouhal number was not reduced to values close to 2D swimming, and
  hence the effect of a bifurcating wake was not totally eliminated.
  It is expected that by, for example, increasing the span of the
  caudal fin, which reduces the required thrust coefficient, further
  increase in efficiency is possible.
\item \noindent The sharp curvature of the envelope of body
  deformation at the peduncle affects efficiency significantly.  The
  efficiency increases from 22\% for a convex imposed motion to 35\%
  for an optimized body deformation with large deformation around the
  peduncle; both at Reynolds number 5000.
\item \noindent Heave and pitch motion at the trailing edge is close
  to $90^{\circ}$, as for a 2D swimming fish.  The parametric
  dependence of the envelope shape is also qualitatively similar
  (Figure \ref{fig:amp_3D}).
\end{enumerate}

The resulting wake has a periodic 3D structure with coherent vortices
that another fish can use to save energy by properly timing its
motion. However, the three dimensional flow around a fish is far more
complex than the flow around a two-dimensional flow. Since the
three-dimensional effects mostly result in a loss of efficiency, the
optimization reduces these effects while distributing the production
of thrust between the body and the tail (resulting in
$\eta_{QP}=34\%$).

Turning to the 2D swimming of two identical, self-propelled fish in
close proximity, the geometric models and assumptions employed for a
single fish optimization were used, under the following additional
conditions: Both fish swim at the same speed and at a constant
distance; the downstream fish is either directly behind the upstream
fish, or at a lateral offset.  The amplitude of motion of each fish is
adjusted separately to achieve self-propulsion, and their kinematics
are also optimized separately for expended energy.  The power of each
fish is compared with the power required when swimming alone.  The
following results are obtained:
\begin{enumerate}[labelwidth=0.6cm,labelindent=16pt, leftmargin=1cm, align=left]
\item \noindent The upstream fish may benefit from the mere presence
  of the downstream fish for very short relative distances. For
  example, a 28\% energy saving at a distance of $d=0.25$ is achieved,
  but as the distance increases this is quickly reduced.
\item \noindent The downstream fish can benefit energetically even at
  axial distances equal to several times the body length, and in both
  the in-line position and in an offset position. This proves that
  energy saving is achieved through interaction with individual
  vortices as opposed to taking advantage of reduced oncoming flow
  (drafting), because in the in-line position the downstream fish is
  in a region of averaged increased relative velocity, which would be
  expected to cause an increase in drag.
\item \noindent The axial force on the downstream fish is mostly
  affected by the interaction between the head of the fish and the
  oncoming vortices, whereas the power required to sustain the
  undulating motion is affected by the interaction between these
  vortices and the body motion downstream from the head. The critical
  parameter for efficiency is the phasing between the head motion and
  the arrival of the vortices, since in general, a reduced drag
  results in a greater power reduction.
\item \noindent For the in-line arrangement, when fixing the relative
  distance to one body length, $d = 1$, and varying the frequency, the
  efficiency of the downstream fish can increase to 66\% for the
  optimal phase between the head motion and the arrival of the
  upstream fish vortices.
\item \noindent For the offset arrangement, at $\Delta y = 0.17$,
  efficiency increases further, as the downstream fish also exploits
  the reduction in oncoming velocity. Still, an optimized phase is
  required, which makes it possible the reach a quasi-propulsive
  efficiency 81\%, even though the fish can only interact with every
  other vortex produced by the upstream fish. It is expected that the
  efficiency of the downstream fish would increase further if there
  were two fish in the front, spaced by $\Delta y = 0.34$ and
  perfectly synchronized.
\item \noindent Although the upstream fish vortices interact with the
  downstream fish over its entire length, as described above, it is
  remarkable that at the tail of the downstream fish the upstream and
  downstream fish vortices interact following the rules of
  \citet{gopalkrishnan_active_1994}, viz.~constructive interaction
  results in reduced efficiency, while destructive interaction
  provides increased efficiency.
\end{enumerate}

Hence, we can conclude that swimming power can be reduced by swimming
in a group for any position of the downstream fish; and for the
upstream fish when positioned at close distances from the downstream
fish. For the downstream fish, it can improve its thrust by
interacting with oncoming vortices, and since reduced drag also
reduces the power required to swim, this results in an increase of
efficiency. On the contrary, bad timing leads to enhanced drag and
swimming power.  The schooling theory by
\citet{weihs_hydromechanics_1973} predicts that a fish swimming
directly behind another fish would experience increased drag and have
to expend more power than in open-water.  Wee show here that an
additional consideration must be made on energy capture from the
oncoming vortices, which depends on the phasing of the undulating
motion with respect to the vortex street.  When swimming in an offset
location, energy savings can be maximized by simultaneously extracting
energy from individual vortices and taking advantage of reduced
oncoming flow velocity.

%%%%%%%%%%%%%%%%%%%%%%%%%%%%%%%%%
\section*{Acknowledgements}
The authors wish to acknowledge support from the Singapore-MIT
Alliance for Research and Technology through the CENSAM Program, and
from the MIT Sea Grant Program.

%%%%%%%%%%%%%%%%%%%%%%%%%%%%%%%%%%%%%%%%%%%%%%%%%%%%%%%%%%%%%%%%%%
\appendix
%%%%%%%%%%%%%%%%%%%%%%%%%%%%%%%%%

\section{Numerical method validation \label{sec:valid}}

Problems previously studied with BDIM include ship flows and flexible
wavemaker flows \citep{weymouth_advancements_2006}, shedding of
vorticity from a rapidly displaced foil \citep{wibawa_global_2012},
and a cephalopod-like deformable jet-propelled body
\citep{weymouth_ultra-fast_2013}. In \citet{maertens_accurate_2015} we
have demonstrated the ability of BDIM to handle several moving bodies
and generalized the original method to accurately simulate the flow
around streamlined foils at Reynolds numbers on the order of $\Rey =
10^4$.

In order to validate the code for simulating undulating foils, the
force and power resulting from a fully imposed kinematics are compared
with results reported in the literature.  Finally, a convergence study
and sensitivity analysis on a self-propelled undulating foil are
performed.

\subsection{Undulating NACA0012 with fully imposed kinematics}

\begin{figure}
 \centerline{\includegraphics[width=0.6\textwidth]{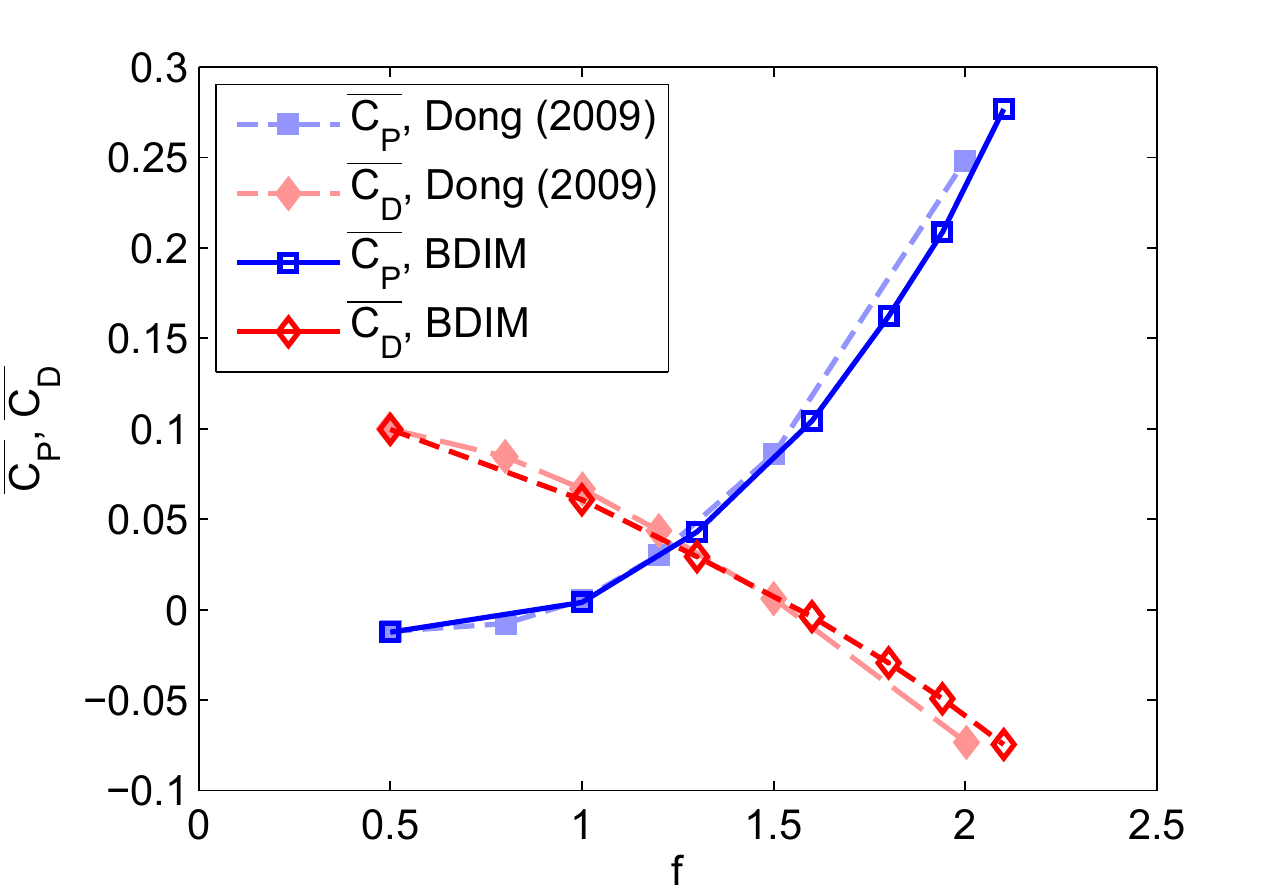}}
  \caption[Time-averaged drag and power coefficients for an undulating
    NACA0012 as a function of frequency.]{Time-averaged drag and power
    coefficients for an undulating NACA0012 as a function of
    frequency, compared with values from
    \citet{dong_characteristics_2007}.}
\label{fig:Dong}
\end{figure}

Using a fully imposed carangiform undulation:
 \begin{equation}
 h(x,t) = \big(0.1-0.0825(x-1)+0.1625(x^2-1)\big) \sin \big( 2\pi ( x - ft )\big) ,
\end{equation} 
the undulation frequency is varied from $f=0.5$ to $f=2$ and the
resulting time-averaged force and power coefficients are compared to
the values from \citet{dong_characteristics_2007} in Figure
\ref{fig:Dong}. Note that in these simulations the kinematics is fully
imposed, not allowing for recoil.

Similarly to \citet{dong_characteristics_2007}, we find that the
average power coefficient, slightly negative at $f=0.5$, increases to
around $0.25$ at $f=2$, and that the drag is positive for $f < 1.6$
and negative for $f > 1.6$. The good agreement between our method and
the results from \citet{dong_characteristics_2007} serves as a
validation of the force and power calculation routines for an
undulating foil.

%%%%%%%%%%%%%%%%%%%
\subsection{Self-propelled undulating NACA0012}

We now ensure that the simulation results presented here are
independent of the grid parameters. In this section we consider the
carangiform motion with frequency $f=2.1$.

\begin{figure}
 \centerline{\includegraphics[width=0.9\textwidth]{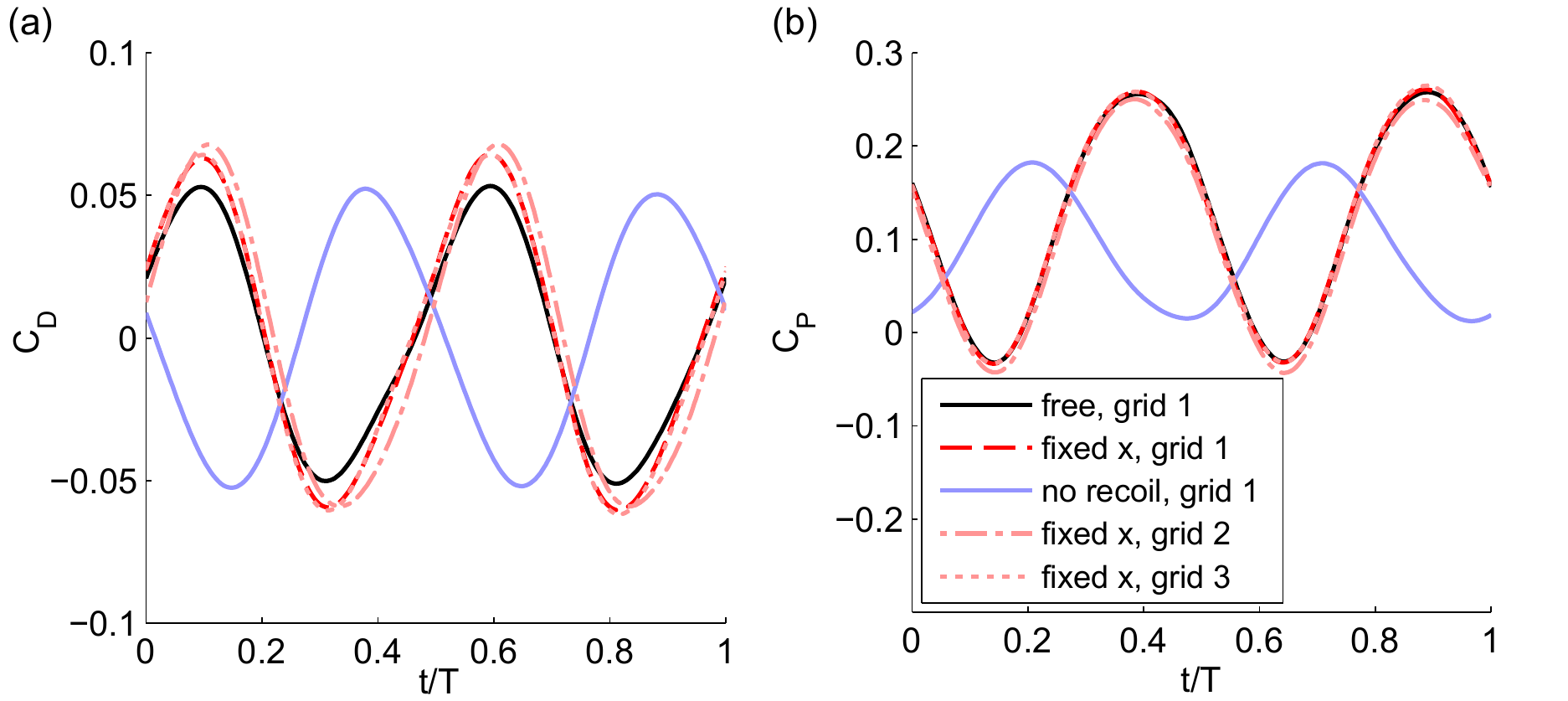}}
  \caption[(a) Drag and (b) pressure coefficient on an undulating
    NACA0012 with carangiform motion at $f=1/T=2.1$.]{(a) Drag and (b)
    pressure coefficient on an undulating NACA0012 with carangiform
    motion at $f=1/T=2.1$. Various grids and constraints are
    compared. Grid $2$ is twice as fine as grid $1$, while the
    computational domain of grid $3$ is twice as large as that of grid
    $1$.}
\label{fig:conv}
\end{figure}

Figure \ref{fig:conv} shows the evolution of power and drag
coefficients during an undulation period $T$=1/f for various
configurations. By comparing the free undulation and the fixed $x$
case, we first notice that fixing the $x$ location of the foil does
not impact the power, confirming the observations from
\citet{bale_energy_2014}. The amplitude of the drag oscillations are a
bit larger for the case with fixed $x$ location, as would be expected,
but this does not impact any of the results discussed in this
paper. On the other hand, precluding all recoil completely changes the
phase and amplitude of the power and drag coefficients. Figure
\ref{fig:conv} also shows that the power and drag coefficients
estimated on grid $1$ (introduced in \S\,\ref{sec:numerical}) are very
close to those estimated on a grid twice as fine (grid $2$, $\dd x=\dd
y=1/320$) and a grid twice as large (grid $3$, $x \in [-12,\,14]$, $y
\in[-4, \,4]$). Table \ref{tab:undul} summarizes the mean and maximum
power, maximum drag, and undulation amplitude $a_0$ for all these
cases.

These results confirm that, while fixing the $x$ location of the foil
will not impact our swimming efficiency estimates, the foil should be
let free to heave and pitch. Therefore, a foil fixed in $x$, free to
heave and pitch under the influence of the hydrodynamic forces will be
used throughout this chapter.  Moreover, the estimates on grid $1$
being very close to those on a finer and larger grid, grid $1$ ($5$
points across the boundary layer) will be used for the optimization
procedures with a fish in open-water, whereas grid $2$ ($10$ points
across the boundary layer) will be used for visualization and for a
swimming pair.

\begin{table}
  \begin{center}
  \begin{tabular*}{0.8\textwidth}{@{\extracolsep{\fill}} lcccc}
  \toprule
      Case  & $\overline{C_P}$   & $(C_P-\overline{C_P})_{\max}$ &  $(C_D)_{\max}$ & $a_0$  \\
      \midrule
       free, grid $1$   & $0.124$ & $0.153$ & $0.054$ & $0.100$\\
       fixed $x$, grid $1$   & $0.125$ & $0.155$ & $0.065$ & $0.100$ \\
       no recoil, grid $1$  & $0.093$ & $0.087$ & $0.054$ & $0.065$\\
       fixed $x$, grid $2$  & $0.112$ & $0.165$ & $0.068$ & $0.097$ \\
       fixed $x$, grid $3$ & $0.125$ & $0.156$ & $0.065$ & $0.099$\\
       \bottomrule  
  \end{tabular*}
  \caption{Mean and maximum amplitude of power coefficient, amplitude
    of drag coefficient and undulation amplitude for a NACA0012 with
    carangiform amplitude at $f=2.1$ and $0$ drag.}
  \label{tab:undul}
  \end{center}
\end{table}

%%%%%%%%%%%%%%%%%%%%%%%%%%%%%%%%%%%%%%%%%%%%%%%
\section{Feedback controller \label{sec:PID}}

In steady state, the time-averaged velocity of a swimming fish is
constant and the mean forces on the swimmer are $0$. In order to
ensure that the system converges toward a steady state in which the
swimming velocity is the prescribed velocity $U_s$, we designed a
proportional-integral-derivative (PID) controller that adjusts the
thrust by tuning the amplitude of the swimming gait $a_0$. If the fish
is fully self-propelled, the time-averaged linear momentum in $x$ is
used as feedback (referred to as displacement control).

Since the amplitude of the oscillations in $v_c^x$ is very small, in
most cases we actually fix the fish in $x$ in order to reduce the PID
convergence time. In this case (referred to as force control), the
time-averaged drag is used as feedback. However, it is important to
let the fish move freely in heave and pitch under the effect of the
hydrodynamic forces. In order to ensure stability of the fish in heave
and pitch, the time-averaged linear momentum in $y$ is used as the
input to a PID controller that tunes the camber parameter $C$ of the
$y_1(x)$ function defined in Eq. \ref{eq:camber}.

For a self-propelled fish with flapping frequency $f=1/T$, we define
the error as:
\begin{equation}
\vec{e}(t_{n}) = f \sum_{k=n_0}^{n-1} m \vec{v}_c(t_k) (t_{k+1}-t_k),
\end{equation}
where $n_0$ is the first index $k$ such that $t_k\geq t_{n}-T$.  If
the $x$ motion is fixed and force control is used, $F_h^x$ replaces $m
v_c^x$ in the calculation of $e^x$.

The integral of the error is calculated as:
\begin{equation}
\vec{e}_i(t_n) = \sum_{k = 0}^n \vec{e}(t_k) (t_{k+1}-t_k),
\end{equation}
and its derivative is:
\begin{equation}
\vec{e}_d(t_n) = f \left[\frac{\big( t_{n_0}-(t_n-T) \big) 
\vec{e}(t_{n_0-1}) + \big( (t_n-T)-t_{n_0-1} \big) 
\vec{e}(t_{n_0})} {t_{n_0}-t_{n_0-1}}\right] .
\end{equation}
At the beginning of each time step, the parameters $a_0$ from equation
\ref{eq:motion} and $C$ from Eq. \ref{eq:camber} are updated as:
\begin{subequations}
\begin{empheq}[left=\empheqlbrace]{align}
& a_0(t_{n}) = \max \big[ a_0(t_n)+ (t_{n}-t_{n-1})
\big( K_p^x e^x(t_n) + K_i^x e_i^x(t_n) 
+K_d^x e_d^x(t_n) \big) , \, 0 \big], \\
& C(t_n) = - \big( K_p^y e^y(t_n) + K_i^y e_i^y(t_n) +K_d^y e_d^y(t_n) \big) ,
\end{empheq}
\end{subequations}
where $e^x$ and $e^y$ denote respectively the $x$ and $y$ components
of the error vector $\vec{e}$.

The gain coefficients used in this study are 
\begin{align}
& \text{for force control in } x: \qquad & K_p^x = 5, \quad K_i^x =
  5,\ \, \quad K_d^x = 5,\ \\
& \text{for displacement control in } x: \qquad & K_p^x = 5, \quad
  K_i^x = 1, \quad K_d^x = 100,\\
& \text{for displacement control in } y: \qquad & K_p^y = 8, \quad
  K_i^y = 10, \quad K_d^y = 12,
\end{align}

%%%%%%%%%%%%%%%%%%%%%%%%%%%%%%%%%%%%%%%%%%%%%%%%%%%%%%%%%%%%%%%%%%%%%%%%%
 \bibliographystyle{jfm}
\bibliography{swim}

\begin{thebibliography}{74}
\expandafter\ifx\csname natexlab\endcsname\relax\def\natexlab#1{#1}\fi

\bibitem[Abrahams \& Colgan(1987)]{abrahams_fish_1987}
{\sc Abrahams, Mark~V. \& Colgan, Patrick~W.} 1987 Fish schools and their
  hydrodynamic function: a reanalysis. {\em Environ Biol Fish\/} {\bf 20}~(1),
  79--80.

\bibitem[Akanyeti \& Liao(2013)]{akanyeti_kinematic_2013}
{\sc Akanyeti, Otar \& Liao, James~C.} 2013 A kinematic model of {Kármán}
  gaiting in rainbow trout. {\em J Exp Biol\/} p. jeb.093245.

\bibitem[Anderson {\em et~al.\/}(1998)Anderson, Streitlien, Barrett \&
  Triantafyllou]{anderson_oscillating_1998}
{\sc Anderson, J.~M., Streitlien, K., Barrett, D.~S. \& Triantafyllou, M.~S.}
  1998 Oscillating foils of high propulsive efficiency. {\em Journal of Fluid
  Mechanics\/} {\bf 360}, 41--72.

\bibitem[Bainbridge(1961)]{bainbridge_problems_1961}
{\sc Bainbridge, Richard} 1961 Problems of fish locomotion. In {\em Symp.
  {Zool}. {Soc}. {Lond}\/}, , vol.~5, pp. 13--32.

\bibitem[Bale {\em et~al.\/}(2014)Bale, Hao, Bhalla \&
  Patankar]{bale_energy_2014}
{\sc Bale, Rahul, Hao, Max, Bhalla, Amneet Pal~Singh \& Patankar, Neelesh~A.}
  2014 Energy efficiency and allometry of movement of swimming and flying
  animals. {\em PNAS\/} p. 201310544.

\bibitem[Beal {\em et~al.\/}(2006)Beal, Hover, Triantafyllou, Liao \&
  Lauder]{beal_passive_2006}
{\sc Beal, D.~N., Hover, F.~S., Triantafyllou, M.~S., Liao, J.~C. \& Lauder,
  G.~V.} 2006 Passive propulsion in vortex wakes. {\em Journal of Fluid
  Mechanics\/} {\bf 549}, 385--402.

\bibitem[Bergmann {\em et~al.\/}(2014)Bergmann, Iollo \&
  Mittal]{bergmann_effect_2014}
{\sc Bergmann, Michel, Iollo, Angelo \& Mittal, Rajat} 2014 Effect of caudal
  fin flexibility on the propulsive efficiency of a fish-like swimmer. {\em
  Bioinspiration \& Biomimetics\/} .

\bibitem[Blondeaux {\em et~al.\/}(2005)Blondeaux, Fornarelli, Guglielmini,
  Triantafyllou \& Verzicco]{blondeaux_numerical_2005}
{\sc Blondeaux, Paolo, Fornarelli, Francesco, Guglielmini, Laura,
  Triantafyllou, Michael~S. \& Verzicco, Roberto} 2005 Numerical experiments on
  flapping foils mimicking fish-like locomotion. {\em Physics of Fluids
  (1994-present)\/} {\bf 17}~(11), 113601.

\bibitem[Borazjani \& Sotiropoulos(2008)]{borazjani_numerical_2008}
{\sc Borazjani, Iman \& Sotiropoulos, Fotis} 2008 Numerical investigation of
  the hydrodynamics of carangiform swimming in the transitional and inertial
  flow regimes. {\em J Exp Biol\/} {\bf 211}~(10), 1541--1558.

\bibitem[Borazjani \& Sotiropoulos(2010)]{borazjani_role_2010}
{\sc Borazjani, I. \& Sotiropoulos, F.} 2010 On the role of form and kinematics
  on the hydrodynamics of self-propelled body/caudal fin swimming. {\em J Exp
  Biol\/} {\bf 213}~(1), 89--107.

\bibitem[Boschitsch {\em et~al.\/}(2014)Boschitsch, Dewey \&
  Smits]{boschitsch_propulsive_2014}
{\sc Boschitsch, Birgitt~M., Dewey, Peter~A. \& Smits, Alexander~J.} 2014
  Propulsive performance of unsteady tandem hydrofoils in an in-line
  configuration. {\em Physics of Fluids\/} {\bf 26}~(5), 051901.

\bibitem[Breder(1926)]{breder_locomotion_1926}
{\sc Breder, Charles~Marcus} 1926 The locomotion of fishes. {\em Zoologica\/}
  {\bf 4}, 159--297.

\bibitem[Carling {\em et~al.\/}(1998)Carling, Williams \&
  Bowtell]{carling_self-propelled_1998}
{\sc Carling, J., Williams, T.~L. \& Bowtell, G.} 1998 Self-propelled
  anguilliform swimming: simultaneous solution of the two-dimensional
  navier-stokes equations and {Newton}'s laws of motion. {\em J Exp Biol\/}
  {\bf 201}~(23), 3143--3166.

\bibitem[Connell \& Yue(2007)]{connell_flapping_2007}
{\sc Connell, Benjamin S.~H. \& Yue, Dick K.~P.} 2007 Flapping dynamics of a
  flag in a uniform stream. {\em Journal of Fluid Mechanics\/} {\bf 581},
  33--67.

\bibitem[Deng {\em et~al.\/}(2013)Deng, Xu, Chen, Dai, Wu \&
  Tian]{deng_numerical_2013}
{\sc Deng, Hong-Bin, Xu, Yuan-Qing, Chen, Duan-Duan, Dai, Hu, Wu, Jian \& Tian,
  Fang-Bao} 2013 On numerical modeling of animal swimming and flight. {\em
  Comput Mech\/} {\bf 52}~(6), 1221--1242.

\bibitem[Dong \& Lu(2007)]{dong_characteristics_2007}
{\sc Dong, Gen-Jin \& Lu, Xi-Yun} 2007 Characteristics of flow over traveling
  wavy foils in a side-by-side arrangement. {\em Physics of Fluids\/} {\bf
  19}~(5), 057107--057107--11.

\bibitem[Dong {\em et~al.\/}(2006)Dong, Mittal \& Najjar]{dong_wake_2006}
{\sc Dong, H., Mittal, R. \& Najjar, F.~M.} 2006 Wake topology and hydrodynamic
  performance of low-aspect-ratio flapping foils. {\em Journal of Fluid
  Mechanics\/} {\bf 566}, 309.

\bibitem[Drucker \& Lauder(2001)]{drucker_locomotor_2001}
{\sc Drucker, Eliot~G. \& Lauder, George~V.} 2001 Locomotor function of the
  dorsal fin in teleost fishes: experimental analysis of wake forces in
  sunfish. {\em Journal of Experimental Biology\/} {\bf 204}~(17), 2943--2958.

\bibitem[Eldredge(2006)]{eldredge_numerical_2006}
{\sc Eldredge, Jeff~D.} 2006 Numerical simulations of undulatory swimming at
  moderate {Reynolds} number. {\em Bioinspir. Biomim.\/} {\bf 1}~(4), S19.

\bibitem[Eloy(2013)]{eloy_best_2013}
{\sc Eloy, Christophe} 2013 On the best design for undulatory swimming. {\em
  Journal of Fluid Mechanics\/} {\bf 717}, 48--89.

\bibitem[Förster {\em et~al.\/}(2007)Förster, Wall \&
  Ramm]{forster_artificial_2007}
{\sc Förster, Christiane, Wall, Wolfgang~A. \& Ramm, Ekkehard} 2007 Artificial
  added mass instabilities in sequential staggered coupling of nonlinear
  structures and incompressible viscous flows. {\em Computer Methods in Applied
  Mechanics and Engineering\/} {\bf 196}~(7), 1278--1293.

\bibitem[Gazzola {\em et~al.\/}(2014)Gazzola, Argentina \&
  Mahadevan]{gazzola_scaling_2014}
{\sc Gazzola, Mattia, Argentina, Médéric \& Mahadevan, L.} 2014 Scaling
  macroscopic aquatic locomotion. {\em Nat Phys\/} {\bf advance online
  publication}.

\bibitem[Gero(1952)]{gero_hydrodynamic_1952}
{\sc Gero, D.~R.} 1952 The hydrodynamic aspects of fish propulsion. {\em Fish
  propulsion\/} {\bf 1601}, 1--32, 32 p. : ill. ; 24 cm.

\bibitem[Ginneken {\em et~al.\/}(2005)Ginneken, Antonissen, Müller, Booms,
  Eding, Verreth \& Thillart]{ginneken_eel_2005}
{\sc Ginneken, Vincent~van, Antonissen, Erik, Müller, Ulrike~K., Booms,
  Ronald, Eding, Ep, Verreth, Johan \& Thillart, Guido van~den} 2005 Eel
  migration to the {Sargasso}: remarkably high swimming efficiency and low
  energy costs. {\em J Exp Biol\/} {\bf 208}~(7), 1329--1335.

\bibitem[Gopalkrishnan {\em et~al.\/}(1994)Gopalkrishnan, Triantafyllou,
  Triantafyllou \& Barrett]{gopalkrishnan_active_1994}
{\sc Gopalkrishnan, R., Triantafyllou, M.~S., Triantafyllou, G.~S. \& Barrett,
  D.} 1994 Active vorticity control in a shear flow using a flapping foil. {\em
  Journal of Fluid Mechanics\/} {\bf 274}, 1--21.

\bibitem[Gray(1933)]{gray_studies_1933}
{\sc Gray, J.} 1933 Studies in {Animal} {Locomotion} {I}. {The} {Movement} of
  {Fish} with {Special} {Reference} to the {Eel}. {\em Journal of Experimental
  Biology\/} {\bf 10}~(1), 88--104.

\bibitem[Harper \& Blake(1990)]{harper_fast-start_1990}
{\sc Harper, David~G. \& Blake, Robert~W.} 1990 Fast-{Start} {Performance} of
  {Rainbow} {Trout} {Salmo} {Gairdneri} and {Northern} {Pike} {Esox} {Lucius}.
  {\em J Exp Biol\/} {\bf 150}~(1), 321--342.

\bibitem[Ijspeert(2014)]{ijspeert_biorobotics_2014}
{\sc Ijspeert, Auke~J.} 2014 Biorobotics: {Using} robots to emulate and
  investigate agile locomotion. {\em Science\/} {\bf 346}~(6206), 196--203.

\bibitem[Johnson(n.d.)]{nlopt}
{\sc Johnson, Steven~G.} n.d. The {NLopt} nonlinear-optimization package,
  {http://ab-initio.mit.edu/nlopt}.

\bibitem[Kern \& Koumoutsakos(2006)]{kern_simulations_2006}
{\sc Kern, Stefan \& Koumoutsakos, Petros} 2006 Simulations of optimized
  anguilliform swimming. {\em J Exp Biol\/} {\bf 209}~(24), 4841--4857.

\bibitem[Killen {\em et~al.\/}(2012)Killen, Marras, Steffensen \&
  McKenzie]{killen_aerobic_2012}
{\sc Killen, Shaun~S., Marras, Stefano, Steffensen, John~F. \& McKenzie,
  David~J.} 2012 Aerobic capacity influences the spatial position of
  individuals within fish schools. {\em Proc Biol Sci\/} {\bf 279}~(1727),
  357--364.

\bibitem[Lauder \& Madden(2007)]{lauder_fish_2007}
{\sc Lauder, George~V. \& Madden, Peter G.~A.} 2007 Fish locomotion: kinematics
  and hydrodynamics of flexible foil-like fins. {\em Experiments in Fluids\/}
  {\bf 43}~(5), 641--653.

\bibitem[Liao(2007)]{liao_review_2007}
{\sc Liao, James~C} 2007 A review of fish swimming mechanics and behaviour in
  altered flows. {\em Philosophical Transactions of the Royal Society B:
  Biological Sciences\/} {\bf 362}~(1487), 1973 --1993.

\bibitem[Liao {\em et~al.\/}(2003{\natexlab{{\em a\/}}})Liao, Beal, Lauder \&
  Triantafyllou]{liao_fish_2003}
{\sc Liao, J.~C., Beal, D.~N., Lauder, G.~V. \& Triantafyllou, M.~S.}
  2003{\natexlab{{\em a\/}}} Fish exploiting vortices decrease muscle activity.
  {\em Science\/} {\bf 302}~(5650), 1566--1569.

\bibitem[Liao {\em et~al.\/}(2003{\natexlab{{\em b\/}}})Liao, Beal, Lauder \&
  Triantafyllou]{liao_karman_2003}
{\sc Liao, James~C., Beal, David~N., Lauder, George~V. \& Triantafyllou,
  Michael~S.} 2003{\natexlab{{\em b\/}}} The {\textbackslash}karman gait: novel
  body kinematics of rainbow trout swimming in a vortex street. {\em J Exp
  Biol\/} {\bf 206}~(6), 1059--1073.

\bibitem[Lighthill(1960)]{lighthill_note_1960}
{\sc Lighthill, M.~J.} 1960 Note on the swimming of slender fish. {\em Journal
  of Fluid Mechanics\/} {\bf 9}~(02), 305--317.

\bibitem[Liu {\em et~al.\/}(2011)Liu, Yu \& Tong]{liu_flow_2011}
{\sc Liu, Geng, Yu, Yong-Liang \& Tong, Bing-Gang} 2011 Flow control by means
  of a traveling curvature wave in fishlike escape responses. {\em Phys. Rev.
  E\/} {\bf 84}~(5), 056312.

\bibitem[Maertens {\em et~al.\/}(2015)Maertens, Triantafyllou \&
  Yue]{maertens_efficiency_2015}
{\sc Maertens, A.~P., Triantafyllou, M.~S. \& Yue, D. K.~P.} 2015 Efficiency of
  fish propulsion. {\em Submitted to Bioinspiration and biomimetics\/} Under
  review.

\bibitem[Maertens \& Weymouth(2015)]{maertens_accurate_2015}
{\sc Maertens, Audrey~P. \& Weymouth, Gabriel~D.} 2015 Accurate
  {Cartesian}-grid simulations of near-body flows at intermediate {Reynolds}
  numbers. {\em Computer Methods in Applied Mechanics and Engineering\/} {\bf
  283}, 106--129.

\bibitem[Marras {\em et~al.\/}(2014)Marras, Killen, Lindström, McKenzie,
  Steffensen \& Domenici]{marras_fish_2014}
{\sc Marras, Stefano, Killen, Shaun~S., Lindström, Jan, McKenzie, David~J.,
  Steffensen, John~F. \& Domenici, Paolo} 2014 Fish swimming in schools save
  energy regardless of their spatial position. {\em Behav Ecol Sociobiol\/} pp.
  1--8.

\bibitem[Partridge \& Pitcher(1979)]{partridge_evidence_1979}
{\sc Partridge, B.~L. \& Pitcher, T.~J.} 1979 Evidence against a hydrodynamic
  function for fish schools. {\em Nature\/} {\bf 279}~(5712), 418--419.

\bibitem[Peng \& Zhu(2009)]{peng_energy_2009}
{\sc Peng, Zhangli \& Zhu, Qiang} 2009 Energy harvesting through flow-induced
  oscillations of a foil. {\em Physics of Fluids\/} {\bf 21}~(12), 123602.

\bibitem[Pitcher(1986)]{pitcher_functions_1986}
{\sc Pitcher, Tony~J.} 1986 Functions of {Shoaling} {Behaviour} in {Teleosts}.
  In {\em The {Behaviour} of {Teleost} {Fishes}\/} (ed. Tony~J. Pitcher), pp.
  294--337. Springer US.

\bibitem[Portugal {\em et~al.\/}(2014)Portugal, Hubel, Fritz, Heese, Trobe,
  Voelkl, Hailes, Wilson \& Usherwood]{portugal_upwash_2014}
{\sc Portugal, Steven~J., Hubel, Tatjana~Y., Fritz, Johannes, Heese, Stefanie,
  Trobe, Daniela, Voelkl, Bernhard, Hailes, Stephen, Wilson, Alan~M. \&
  Usherwood, James~R.} 2014 Upwash exploitation and downwash avoidance by flap
  phasing in ibis formation flight. {\em Nature\/} {\bf 505}~(7483), 399--402.

\bibitem[Powell(2009)]{powell_bobyqa_2009}
{\sc Powell, Michael~JD} 2009 The {BOBYQA} algorithm for bound constrained
  optimization without derivatives. {\em Cambridge NA Report NA2009/06,
  University of Cambridge, Cambridge\/} .

\bibitem[Read {\em et~al.\/}(2003)Read, Hover \&
  Triantafyllou]{read_forces_2003}
{\sc Read, D.A., Hover, F.S. \& Triantafyllou, M.S.} 2003 Forces on oscillating
  foils for propulsion and maneuvering. {\em Journal of Fluids and
  Structures\/} {\bf 17}~(1), 163--183.

\bibitem[van Rees {\em et~al.\/}(2013)van Rees, Gazzola \&
  Koumoutsakos]{van_rees_optimal_2013}
{\sc van Rees, Wim~M., Gazzola, Mattia \& Koumoutsakos, Petros} 2013 Optimal
  shapes for anguilliform swimmers at intermediate {Reynolds} numbers. {\em
  Journal of Fluid Mechanics\/} {\bf 722}, null--null.

\bibitem[Reid {\em et~al.\/}(2009)Reid, Hildenbrandt, Padding \&
  Hemelrijk]{reid_flow_2009}
{\sc Reid, Daniel A.~P., Hildenbrandt, H., Padding, J.~T. \& Hemelrijk, C.~K.}
  2009 Flow around fishlike shapes studied using multiparticle collision
  dynamics. {\em Phys. Rev. E\/} {\bf 79}~(4), 046313.

\bibitem[Reid {\em et~al.\/}(2012)Reid, Hildenbrandt, Padding \&
  Hemelrijk]{reid_fluid_2012}
{\sc Reid, Daniel A.~P., Hildenbrandt, H., Padding, J.~T. \& Hemelrijk, C.~K.}
  2012 Fluid dynamics of moving fish in a two-dimensional multiparticle
  collision dynamics model. {\em Phys. Rev. E\/} {\bf 85}~(2), 021901.

\bibitem[Rios \& Sahinidis(2013)]{rios_derivative-free_2013}
{\sc Rios, Luis~Miguel \& Sahinidis, Nikolaos~V.} 2013 Derivative-free
  optimization: a review of algorithms and comparison of software
  implementations. {\em J Glob Optim\/} {\bf 56}~(3), 1247--1293.

\bibitem[Sefati {\em et~al.\/}(2013)Sefati, Neveln, Roth, Mitchell, Snyder,
  MacIver, Fortune \& Cowan]{sefati_mutually_2013}
{\sc Sefati, Shahin, Neveln, Izaak~D., Roth, Eatai, Mitchell, Terence R.~T.,
  Snyder, James~B., MacIver, Malcolm~A., Fortune, Eric~S. \& Cowan, Noah~J.}
  2013 Mutually opposing forces during locomotion can eliminate the tradeoff
  between maneuverability and stability. {\em PNAS\/} {\bf 110}~(47),
  18798--18803.

\bibitem[Sfakiotakis {\em et~al.\/}(1999)Sfakiotakis, Lane \&
  Davies]{sfakiotakis_review_1999}
{\sc Sfakiotakis, Michael, Lane, D.M. \& Davies, J.B.C.} 1999 Review of fish
  swimming modes for aquatic locomotion. {\em IEEE Journal of Oceanic
  Engineering\/} {\bf 24}~(2), 237--252.

\bibitem[Shen {\em et~al.\/}(2003)Shen, Zhang, Yue \&
  Triantafyllou]{shen_turbulent_2003}
{\sc Shen, Lian, Zhang, Xiang, Yue, Dick K.~P. \& Triantafyllou, Michael~S.}
  2003 Turbulent flow over a flexible wall undergoing a streamwise travelling
  wave motion. {\em Journal of Fluid Mechanics\/} {\bf 484}, 197--221.

\bibitem[Shirgaonkar {\em et~al.\/}(2009)Shirgaonkar, MacIver \&
  Patankar]{shirgaonkar_new_2009}
{\sc Shirgaonkar, Anup~A., MacIver, Malcolm~A. \& Patankar, Neelesh~A.} 2009 A
  new mathematical formulation and fast algorithm for fully resolved simulation
  of self-propulsion. {\em Journal of Computational Physics\/} {\bf 228}~(7),
  2366--2390.

\bibitem[Stefanini {\em et~al.\/}(2012)Stefanini, Orofino, Manfredi, Mintchev,
  Marrazza, Assaf, Capantini, Sinibaldi, Grillner, Wallén \&
  Dario]{stefanini_novel_2012}
{\sc Stefanini, C., Orofino, S., Manfredi, L., Mintchev, S., Marrazza, S.,
  Assaf, T., Capantini, L., Sinibaldi, E., Grillner, S., Wallén, P. \& Dario,
  P.} 2012 A novel autonomous, bioinspired swimming robot developed by
  neuroscientists and bioengineers. {\em Bioinspir. Biomim.\/} {\bf 7}~(2),
  025001.

\bibitem[Streitlien {\em et~al.\/}(1996)Streitlien, Triantafyllou \&
  Triantafyllou]{streitlien_efficient_1996}
{\sc Streitlien, Knut, Triantafyllou, George~S. \& Triantafyllou, Michael~S.}
  1996 Efficient foil propulsion through vortex control. {\em AIAA Journal\/}
  {\bf 34}~(11), 2315--2319.

\bibitem[Taneda(1977)]{taneda_visual_1977}
{\sc Taneda, Sadatoshi} 1977 Visual study of unsteady separated flows around
  bodies. {\em Progress in Aerospace Sciences\/} {\bf 17}, 287--348.

\bibitem[Tokić \& Yue(2012)]{tokic_optimal_2012}
{\sc Tokić, Grgur \& Yue, Dick K.~P.} 2012 Optimal shape and motion of
  undulatory swimming organisms. {\em Proc. R. Soc. B\/} {\bf 279}~(1740),
  3065--3074.

\bibitem[Triantafyllou {\em et~al.\/}(1993)Triantafyllou, Triantafyllou \&
  Grosenbaugh]{triantafyllou_optimal_1993}
{\sc Triantafyllou, G.~S., Triantafyllou, M.~S. \& Grosenbaugh, M.~A.} 1993
  Optimal {Thrust} {Development} in {Oscillating} {Foils} with {Application} to
  {Fish} {Propulsion}. {\em Journal of Fluids and Structures\/} {\bf 7}~(2),
  205--224.

\bibitem[Triantafyllou \& Triantafyllou(1995)]{triantafyllou_efficient_1995}
{\sc Triantafyllou, Michael~S. \& Triantafyllou, George~S.} 1995 An {Efficient}
  {Swimming} {Machine}. {\em Scientific American\/} {\bf 272}, 64--70.

\bibitem[Triantafyllou {\em et~al.\/}(1991)Triantafyllou, Triantafyllou \&
  Gopalkrishnan]{triantafyllou_wake_1991}
{\sc Triantafyllou, M.~S., Triantafyllou, G.~S. \& Gopalkrishnan, R.} 1991 Wake
  mechanics for thrust generation in oscillating foils. {\em Physics of Fluids
  A: Fluid Dynamics (1989-1993)\/} {\bf 3}~(12), 2835--2837.

\bibitem[Tytell(2004)]{tytell_hydrodynamics_2004}
{\sc Tytell, Eric~D.} 2004 The hydrodynamics of eel swimming {II}. {Effect} of
  swimming speed. {\em J Exp Biol\/} {\bf 207}~(19), 3265--3279.

\bibitem[Tytell \& Lauder(2004)]{tytell_hydrodynamics_2004-1}
{\sc Tytell, Eric~D. \& Lauder, George~V.} 2004 The hydrodynamics of eel
  swimming {I}. {Wake} structure. {\em J Exp Biol\/} {\bf 207}~(11),
  1825--1841.

\bibitem[Videler(1993)]{videler_fish_1993}
{\sc Videler, J.~J.} 1993 {\em Fish {Swimming}\/}. Springer.

\bibitem[Videler \& Hess(1984)]{videler_fast_1984}
{\sc Videler, J.~J. \& Hess, F.} 1984 Fast {Continuous} {Swimming} of {Two}
  {Pelagic} {Predators}, {Saithe} ({Pollachius} {Virens}) and {Mackerel}
  ({Scomber} {Scombrus}): a {Kinematic} {Analysis}. {\em J Exp Biol\/} {\bf
  109}~(1), 209--228.

\bibitem[Webb(1971)]{webb_swimming_1971}
{\sc Webb, P.~W.} 1971 The {Swimming} {Energetics} of {Trout} {II}. {Oxygen}
  {Consumption} and {Swimming} {Efficiency}. {\em J Exp Biol\/} {\bf 55}~(2),
  521--540.

\bibitem[van Weerden {\em et~al.\/}(2014)van Weerden, Reid \&
  Hemelrijk]{van_weerden_meta-analysis_2014}
{\sc van Weerden, J~Fransje, Reid, Daniel A~P \& Hemelrijk, Charlotte~K} 2014 A
  meta-analysis of steady undulatory swimming. {\em Fish Fish\/} {\bf 15}~(3),
  397--409.

\bibitem[Weihs(1973)]{weihs_hydromechanics_1973}
{\sc Weihs, D.} 1973 Hydromechanics of {Fish} {Schooling}. {\em Nature\/} {\bf
  241}~(5387), 290--291.

\bibitem[Weymouth {\em et~al.\/}(2006)Weymouth, Dommermuth, Hendrickson \&
  Yue]{weymouth_advancements_2006}
{\sc Weymouth, G.~D., Dommermuth, D.~G., Hendrickson, K. \& Yue, D. K.-P.} 2006
  Advancements in {Cartesian}-grid methods for computational ship
  hydrodynamics. In {\em 26th {Symposium} on {Naval} {Hydrodynamics}, {Rome},
  {Italy}, 17–22 {September} 2006\/}. Rome, Italy.

\bibitem[Weymouth \& Triantafyllou(2013)]{weymouth_ultra-fast_2013}
{\sc Weymouth, G.~D. \& Triantafyllou, M.~S.} 2013 Ultra-fast escape of a
  deformable jet-propelled body. {\em J. Fluid Mech.\/} {\bf 721}, 367--385.

\bibitem[Wibawa {\em et~al.\/}(2012)Wibawa, Steele, Dahl, Rival, Weymouth \&
  Triantafyllou]{wibawa_global_2012}
{\sc Wibawa, M.~S., Steele, S.~C., Dahl, J.~M., Rival, D.~E., Weymouth, G.~D.
  \& Triantafyllou, M.~S.} 2012 Global vorticity shedding for a vanishing wing.
  {\em J. Fluid Mech.\/} {\bf 695}, 112--134.

\bibitem[Wolfgang {\em et~al.\/}(1999)Wolfgang, Anderson, Grosenbaugh, Yue \&
  Triantafyllou]{wolfgang_near-body_1999}
{\sc Wolfgang, M.~J., Anderson, J.~M., Grosenbaugh, M.~A., Yue, D.~K. \&
  Triantafyllou, M.~S.} 1999 Near-body flow dynamics in swimming fish. {\em
  Journal of Experimental Biology\/} {\bf 202}~(17), 2303--2327.

\bibitem[Zhu \& Shoele(2008)]{zhu_propulsion_2008}
{\sc Zhu, Qiang \& Shoele, Kourosh} 2008 Propulsion performance of a
  skeleton-strengthened fin. {\em J Exp Biol\/} {\bf 211}~(13), 2087--2100.

\bibitem[Zhu {\em et~al.\/}(2002)Zhu, Wolfgang, Yue \&
  Triantafyllou]{zhu_three-dimensional_2002}
{\sc Zhu, Q., Wolfgang, M.~J., Yue, D. K.~P. \& Triantafyllou, M.~S.} 2002
  Three-dimensional flow structures and vorticity control in fish-like
  swimming. {\em Journal of Fluid Mechanics\/} {\bf 468}, 1--28.

\end{thebibliography}

 \end{document}